\documentclass{aa}

\usepackage{graphicx}
\usepackage{txfonts}
\usepackage{natbib}
\usepackage{longtable}
\usepackage[colorlinks=true,linkcolor=blue,citecolor=blue,urlcolor=blue]{hyperref}
\bibpunct{(}{)}{;}{a}{}{,}

\begin{document}

\title{The HARPS search for southern extra-solar planets
}

\subtitle{XLI. A dozen planets around the M dwarfs GJ~3138, GJ~3323, GJ~273, 
  GJ~628, and GJ~3293
\thanks{Based on observations made with the HARPS instrument 
  on the ESO 3.6 m telescope under the program 
  IDs 180.C-0886(A), 183.C-0437(A), and 191.C-0873(A) at Cerro La Silla (Chile).}
\textsuperscript{,}\thanks{Radial velocity data (full Tables A.1--A.5) are only 
available at the CDS via anonymous ftp to 
cdsarc.u-strasbg.fr (130.79.128.5)\newline
or via\newline
http://cdsarc.u-strasbg.fr/viz-bin/qcat?J/A+A/602/A88
}}

\authorrunning{Astudillo-Defru et al.}
\titlerunning{Planets orbiting the M dwarfs GJ~3138, 
GJ~3323, GJ~273, GJ~628, and GJ~3293}

\author{N. Astudillo-Defru \inst{1, 2}, T. Forveille\inst{2}, X. Bonfils\inst{2}, 
  D. S\'egransan\inst{1}, F. Bouchy\inst{1}, X. Delfosse\inst{2},   
  C. Lovis\inst{1}, M. Mayor\inst{1}, F. Murgas\inst{2}, 
  F. Pepe\inst{1}, N. C. Santos\inst{3, 4}, 
  S. Udry\inst{1}, A. W\"unsche\inst{2}}

\institute{Observatoire de Gen\`eve, Universit\'e de Gen\`eve, 
    51 ch. des Maillettes, 1290 Sauverny, Switzerland\newline
    \email{nicola.astudillo@unige.ch}
  \and Univ. Grenoble Alpes, CNRS, IPAG, F-38000 Grenoble, France 
  \and Instituto de Astrof\'isica e Ci\^encias do Espa\c{c}o, Universidade do Porto, CAUP,
  Rua das Estrelas, PT4150-762 Porto, Portugal 
  \and Departamento de F\'isica e Astronomia, Faculdade de Ci\^encias, 
    Universidade do Porto, Portugal 
}

\date{Received 28 November 2016 / Accepted 13 March 2017\\}

\abstract
{ Low-mass stars are currently the best targets when searching for rocky 
  planets in the habitable zone of their host star.
  Over the last 13 years, precise radial velocities measured with the HARPS 
  spectrograph have identified over a dozen super-Earths and Earth-mass 
  planets (m$ \sin i \leq$ 10~M$_\oplus$) around M~dwarfs, with a well-understood 
  selection function. This well-defined sample provides information on their frequency 
  of occurrence and on the distribution of their orbital parameters, and 
  therefore already constrains our understanding of planetary formation. 
  The subset of these low-mass planets that were found within the habitable 
  zone of their host star also provide prized targets for future searches of  atmospheric 
  biomarkers.
}
{We are working to extend this planetary sample to lower masses and longer
  periods through dense and long-term monitoring of the radial velocity
  of a small M~dwarf sample.
}
{We obtained large numbers of HARPS spectra for the  M~dwarfs GJ\,3138, 
  GJ\,3323, GJ\,273, GJ\,628,  and GJ\,3293, from which we derived radial velocities 
  (RVs) and spectroscopic activity indicators. We searched  for 
  variabilities, periodicities, Keplerian modulations, and correlations,
  and attribute the radial-velocity variations to combinations of planetary 
  companions and stellar activity.}
{We detect 12 planets, 9 of which  are new with masses ranging from 1.17 to 10.5~M$_\oplus$. 
These planets have relatively short orbital periods (P$<$40 d), except for two that have 
periods of 217.6 and 257.8 days. Among these systems, GJ~273 harbor two planets with masses 
close to the  Earth's. With a distance of only 3.8 parsec, GJ~273 is the second 
nearest known planetary system -- after Proxima Centauri -- with a planet orbiting the 
circumstellar habitable zone.
}
{}


\maketitle
\section{Introduction}
\label{sec:introduction}
M dwarfs lie at the bottom of the main sequence, and consequently are 
small, cool, and intrinsically faint. These physical characteristics 
provide many advantages when looking for smaller, cooler, and fainter planets. 
For orientation, an Earth-mass planet in the habitable zone (HZ)  of a 
0.2 M$_\odot$ M dwarf, produces a radial velocity (RV) wobble that is 
over an order of magnitude larger than that of the Earth generates in the Sun. 
When caught in transit, this Earth-size planet decreases the flux 
of the M dwarf (with a 0.2R$_\odot$ radius) by 25 times as much 
as the Earth does when it crosses the Sun. This makes planets around 
M dwarfs easier to detect, and even more importantly easier to characterize. 
M dwarfs with transiting planets -- like GJ436 
\citep{2004ApJ...617..580B, 2007A&A...471L..51G}, GJ1214 \citep{2009Natur.462..891C}, 
GJ3470 \citep{2012A&A...546A..27B}, GJ1132 \citep{2015Natur.527..204B}, and K2-3 
\citep{2015ApJ...804...10C, 2015A&A...581L...7A} -- are therefore very popular 
targets in the exoplanet characterization community. Our HARPS 
program has made significant contributions to the detection, 
confirmation, and/or characterization of all of them. Some of the 
more numerous HARPS detections that do not transit may become amenable 
to characterization once future instruments are able to to resolve 
them angularly, for example the Mid-infrared E-ELT Imager and Spectrograph (METIS) 
at the E-ELT\citep [][]{2015A&A...576A..59S}.

In this paper, we report a total of twelve planet detections in five planetary 
systems. The five stars, GJ~3138, GJ~3323, GJ~273, GJ~628, and GJ~3293, were observed 
with the HARPS spectrograph as part of our long-term effort to search for 
planets 
around bright, nearby M~dwarfs. Section~\ref{sec:stellar_properties} summarizes 
the stellar properties of each host star, while 
Section~\ref{sec:analysis_from_harps} briefly describes our  data set and 
how both radial velocities and spectroscopic indices are measured from 
the HARPS spectra. In Section~\ref{sec:searching} we identify periodic 
signals and ascribe them to either planets or activity-induced variations. 
Finally, Sect~\ref{sec:Conclusion} presents our conclusions.

\section{Stellar properties}
\label{sec:stellar_properties}

\begin{table*}[t]
\caption{ Stellar properties and parameters.}
\normalsize
\begin{tabular*}{\hsize}{@{\extracolsep{\fill}}llllll}

\hline\hline
\noalign{\smallskip}
& GJ~3138 & GJ~3323 & GJ~273 & GJ~628 & GJ~3293\\
\noalign{\smallskip}
\hline
\noalign{\smallskip}
Spectral Type$^{(1)}$ & M0 & M4 & M3.5 & M3.5 & M2.5 \\
$\alpha$ (J2000) & $02^h 09^m 10.9^s$ & $05^h 01^m 57.5^s$ & $07^h 27^m 24.49^s$ & $16^h 30^m 18.1^s$ & $04^h 28^m 35.6^s$ \\
$\delta$ (J2000) & $-16^\circ 20^\prime 22.5^{\prime\prime}$ & $-06^\circ 56^\prime 45.9^{\prime\prime}$ & $+05^\circ 13^\prime 32.8^{\prime\prime}$ & $-12^\circ 39^\prime 45.3^{\prime\prime}$ & $-25^\circ 10^\prime 16^{\prime\prime}$ \\
V$^{(2)}$ & 10.98 & 12.22& 9.872 &  10.03 & 11.962 \\
J$^{(3)}$ & $8.076\pm0.019$ & $7.617\pm0.032$ & $5.714\pm0.032$ & $5.950\pm0.024$ & $8.362\pm0.024$ \\
H$^{(3)}$ & $7.412\pm0.031$ & $7.065\pm0.046$ & $5.219\pm0.063$ & $5.373\pm0.040$ & $7.749\pm0.038$ \\
K$_S^{(3)}$ & $7.246\pm0.016$ & $6.736\pm0.024$ & $4.857\pm0.023$ & $5.075\pm0.024$ & $7.486\pm0.033$ \\
$\pi\; [mas]^{(4, 5)}$ & $33.44\pm2.17$ & $187.92\pm1.26$ & $262.98\pm1.39$ & $232.98\pm1.60$ & $55\pm9$ \\
$M_V$ & $8.6\pm0.13$ & $13.59\pm0.01$ & $12.26\pm0.01$ & $11.87\pm0.01$ & $10.66\pm0.31$ \\
$M_K$ & $4.87\pm0.13$ & $8.11\pm0.03$ & $6.96\pm0.03$ & $6.91\pm0.03$ & $6.19\pm0.31$ \\
$BC_K$ & $2.62\pm0.05$ & $2.72\pm0.07$ & $2.68\pm0.08$ & $2.71\pm0.06$ & $2.71\pm0.08$ \\
L [L$_{Sun}$]$^{(6)}$ & $0.045$ & $0.0027$ & $0.0088$ & $0.0102$ & $0.022$ \\
 & $\pm0.007$ & $\pm0.0066$ & $\pm0.0066$ &  $\pm0.0066$  & $\pm0.0066$ \\
M [M$_\odot$]$^{(7)}$ & 0.681 & 0.164 & 0.29 &  0.294 & 0.42 \\
R [R$_\odot$]$^{(6)}$ & $0.50\pm0.03$ & $0.119\pm0.027$ & $0.293\pm0.027$ &  $0.307\pm0.027$ & $0.404\pm0.027$ \\
T$_{eff}$ [K]$^{(6)}$ & $3717\pm49$ & $3159\pm49$ & $3382\pm49$ &  $3342\pm49$ & $3466\pm49$ \\
$[Fe/H]^{(8)}$ & $-0.30\pm0.12$ & $-0.27\pm0.09$ & $0.09\pm0.17$ &  $-0.09\pm0.09$ & $0.02\pm0.09$ \\
$\mu_\alpha \;[mas/yr]^{(4, 9, 2)}$ & $517.58\pm1.98$ &  $-550.2\pm8$ & $572.51\pm1.50$ &  $-94.81\pm2.30$ & $-87\pm5$ \\
$\mu_\delta \;[mas/yr]^{(4, 9, 2)}$ & $78.39\pm1.63$ & $-533.3\pm8$ & $-3693.51\pm0.96$ &  $-1183.43\pm1.74$ & $-475\pm5$ \\
$dv_r/dt \; [m/s/yr]$ & $0.188\pm0.014$ & $0.073\pm0.001$ & $1.221\pm0.007$ &  $0.139\pm0.001$  & $0.097\pm0.018$\\
Rad. Vel [km/s]$^{(10)}$ & 13.60$\pm$0.50 & 42.45$\pm$0.50 & 18.41$\pm$0.50 &  -21.04$\pm$0.50 & 13.30$\pm$0.50\\
U [km/s] & 63.5$\pm$14.4 & 23.60$\pm$0.04 & -15.72$\pm$0.09 &  12.74$\pm$0.02 & -27.29$\pm$17.1 \\
V [km/s] & -39.1$\pm$6.0 & -17.97$\pm$0.02 & -65.83$\pm$0.05 &  -21.00$\pm$0.01 & -25.9$\pm$6.6 \\
W [km/s] & 11.2$\pm$2.9 & -35.99$\pm$0.05 & -17.27$\pm$0.01 &  -20.56$\pm$0.01  & -22.2$\pm$23.1\\
Dyn. Pop. & YD or YD/OD & YD/OD  & YD/OD &  YD  &  YD or YD/OD\\
Multiplicity$^{(11, 12)}$ & -- & No  & No &  No  &  -- \\
$HZ_{In}\;S/S_\oplus^{(13,\ conservative,\ 1M_\oplus)}$ & 0.937 & 0.925 & 0.929 & 0.928 & 0.931 \\
$HZ_{In}\;S/S_\oplus^{(13,\ conservative,\ 5M_\oplus)}$ & 1.006 & 0.993 & 0.997 & 0.996 & 0.999 \\
$HZ_{In}\;S/S_\oplus^{(14,\ [Fe/H]=-0.5,-,0.3,0.3,0.3)}$ & 1.69 & -- & 1.64 & 1.63 & 1.68 \\
$HZ_{Out}\;S/S_\oplus^{(13,\ conservative,\ 1M_\oplus)}$ & 0.254 & 0.237 & 0.243 & 0.242 & 0.246 \\
$\log(R^\prime_{HK}$)$^{(15)}$ & -4.855 & -4.839 & -5.560&  -5.523 &  -5.114 \\
$P_{Rot.}$ [days]$^{(15)}$ & 34 & 33 & 99 &  93  & 50 \\
$P_{Rot.}$ [days]$^{(16)}$ & 42.5 & -- & -- &  95  & 41 \\
\noalign{\smallskip}
\hline
\noalign{\smallskip}
\end{tabular*}

\small
\textbf{Notes.} (1) \citet{1996AJ....112.2799H}; (2) \citet{2014MNRAS.443.2561G}; 
(3) \citet{2003yCat.2246....0C}; 
(4) \citet{2007A&A...474..653V}; (5) \citet{2006AJ....132.2360H};
(6) \citet{2012ApJ...757..112B}; (7) \citet{2000A&A...364..217D};
(8) \citet{2013A&A...551A..36N}; 
(9) \citet{2012yCat.1322....0Z}; 
(10) This work. The adopted uncertainty for the HARPS absolute radial velocity is 0.5 km/s;
(11) \citet{2015MNRAS.449.3160R};
(12) \citet{2015MNRAS.449.2618W};
(13) \citet{2013ApJ...765..131K,2014ApJ...787L..29K}; 
(14) \citet{2016ApJ...819...84K}
(15) From \citet{2017A&A...600A..13A}, where the typical error is 8.7\%;
(16) This work. From the periodogram of activity indices. 
On the dynamical population we use the abbreviations YD for young disk and OD for old disk.
\label{tab:targetsProperties}
\end{table*}

GJ~3138, GJ~3323, GJ~273, GJ~628, and GJ~3293 are part of our HARPS sample of 
M~dwarfs because their distance is $d<11$ pc, their $V$~band magnitude 
$V<14$~mag, and their rotational velocity $v\,sini<6.5\ km/s$ 
\citep{2013A&A...549A.109B}. We retrieved their spectral types, photometry,
coordinates, proper motions, parallaxes, and multiplicity from the literature; the   full
references are listed  in Table~\ref{tab:targetsProperties}. The secular acceleration 
($dv/dr$) is calculated following Eq. 2 from \citet{2009A&A...505..859Z}. We computed 
their luminosities using the \citet{2001ApJ...548..908L} bolometric corrections. We 
estimated their physical parameters using calibrations by \citet{2000A&A...364..217D} 
for stellar mass, \citet{2012ApJ...757..112B} for stellar radius and effective temperature, 
and \citet{2013A&A...551A..36N} for metallicity. We computed their UVW galactic 
velocities using the \citet{1987AJ.....93..864J} conventions and assigned them 
to dynamical population following the \citet{1992ApJS...82..351L} prescription. 
We computed the limit of the habitable zone of each star 
with the \citet{2013ApJ...765..131K,2014ApJ...787L..29K} online 
calculator\footnote{\url{http://depts.washington.edu/naivpl/sites/default/files/hz.shtml}}, 
 and adopted the new inner limit of the  HZ from \citet{2016ApJ...819...84K}. 
This new inner limit incorporates 3D global climate models (GCM) in tidally locked 
exoplanets around cool stars. We quantified their magnetic activity by measuring 
their Mount Wilson S~index \citep{1968ApJ...153..221W,1978PASP...90..267V} 
on their average HARPS spectra. 
We converted the S~index into $R'_{HK}$ values following the procedure  
described in \citet{2017A&A...600A..13A}. With $\log(R'_{HK})$ between $-$4.78 
and $-$5.60, all five stars are magnetically quiet or, at most, very moderately 
active. While GJ\,628 has been catalogued as a BY-Dra variable 
\citep{2012AJ....143....2N}, its photometric variability is only moderate 
since the dispersion of its V band flux is just 16$\pm$7 mmag 
\citep{2015AJ....150....6H}. 
As discussed below, we find its variations consistent with a magnetic cycle.
Table~\ref{tab:targetsProperties} summarizes these stellar properties. 

\section{Data}
\label{sec:analysis_from_harps}

We obtained spectra of the five targets with HARPS \citep{2003Msngr.114...20M},
a fiber-fed cross-dispersed echelle spectrograph  at the 3.6m telescope at
ESO/La Silla Observatory (Chile). The instrument has a resolving power of 
R$\sim$115,000 and almost full wavelength coverage between 380~nm and 
690~nm. It is enclosed in a vacuum vessel and temperature-stabilized  
so as to reach a long-term radial-velocity precision $<$ 1 m/s. Calibrations 
are performed daily with a ThAr lamp \citep{2007A&A...468.1115L} and, 
since 2010, additionally with a white-light-illuminated Fabry-P\'erot 
\'etalon \citep{2010SPIE.7735E..4XW}. Using both of  these calibrations, the 
sub-m/s drift of the instrument can be measured by illuminating a 
second fiber with a calibration lamp, while the first fiber 
collects the light of the star. The offset between the simultaneous 
and daily calibrations can be subtracted from the radial velocity 
measurement of the star to improve its precision to a few tens of 
cm/s  \citep[see Sect. 2.6 in][]{2016PASP..128f6001F}. We did not initially rely 
on this simultaneous calibration for our program, because some very strong ThAr 
lines spill from the calibration fiber into the stellar spectra to degrade our
measurements of some important spectroscopic diagnostics in the 
blue-part of M dwarf spectra. The Fabry-P\'erot calibration source
has much more uniform line intensities, and therefore does not
significantly contaminate the stellar spectrum. Starting in  2012 
we thus changed our observing strategy and now use simultaneous calibration
with the Fabry-P\'erot source for stars brighter than V=11 mag. 
For fainter targets, we use the second fiber to 
monitor the sky background. 

Octagonal fibers are used in the entrance of a spectrograph to stabilize 
the illumination and to decrease the effect of imperfect 
guiding or seeing variations \citep{2011SPIE.8151E..15P}. The HARPS fiber link 
was upgraded with such a fiber on May 28, 2015 \citep{2015Msngr.162....9C}, 
stabilizing its line-spread function but introducing a zero point in our 
measurement series.

\subsection{Radial velocities}

The HARPS Data Reduction Software (DRS) provides a quasi-real-time
estimate of the radial velocity through cross-correlation (CCF) of the
stellar spectrum with a software mask 
\citep{1979VA.....23..279B,2002A&A...388..632P}. This measurement,
although a very good first estimate, uses a fixed template across M~ subtypes 
and discards the information contained in the many blended and weak lines 
were rejected at its construction. It therefore has a suboptimal signal-to-noise ratio, 
and we recompute improved velocities off-line through a
few steps \citep{2015A&A...575A.119A}. 
We first align all spectra of a given star to a common 
frame using RV$_{CCF}$ as a first guess, and co-add them into a high 
signal-to-noise stellar template. We then produce a high signal-to-noise 
template of the telluric absorption, subtracting the average stellar
template contribution from each observed spectrum and co-adding the
residuals in the laboratory frame.  The stellar spectrum 
moves on the CCD according to the projection of the radial velocity 
of the Earth toward the star, but the telluric lines are fixed on the 
detector. The stellar and telluric spectra can thus be disentangled. 
Armed with these two templates, we measure a radial velocity from 
each observed spectrum as the Doppler shift that maximizes the 
likelihood between this spectrum and the high signal-to-noise 
template, using the telluric template to mask out the pixels most 
affected by telluric lines. Extracting the RV through maximum-likelihood 
optimization against a high signal-to-noise template was  proposed 
long ago \citep[e.g.,][]{1997MNRAS.284..265H,2006MNRAS.371.1513Z}, and independent
implementations were applied to HARPS data by 
\citet[][in Fourier space]{2000A&A...358L..59C,2005A&A...443..337G} 
and by \citet{2012ApJS..200...15A}.

Compared to the online pipeline, this reprocessing reduces the error bars
by typically 20 to 30\%. Since the HARPS upgrade with an introduction of a 
section of octagonal fiber altered the line spread function, 
we opted to process the {\it pre-} and {\it post-}upgrade epochs 
independently. For each star we thus produce {\it pre-} and {\it post-}upgrade 
templates, and we adjust a zero point offset when analyzing the time series.

\subsection{Activity proxies}

\label{subsec:actind}

Because stellar activity can confuse planet searches, we need to monitor 
its strength and its variation on  timescales that range from the stellar 
rotation period (a few to 100 days) to the length of a magnetic cycle 
(a few hundred days to a dozen years). We therefore monitor the shape 
of the spectral lines and measure spectroscopic indices in our HARPS 
spectra, and we also use publicly available photometry. 

\subsubsection{Spectroscopic indices}
\label{subsubsec:actind}

\begin{table}[t]
\centering
\caption{\small Wavelengths for the passbands defining H$\beta$, H$\gamma$, and Na~D 
spectroscopic activity proxies.}

\label{tab:waveBands}
\begin{tabular*}{\hsize}{@{\extracolsep{\fill}}lccc}

\hline\hline
\noalign{\smallskip}

Index & C & V & R  \\
      & $[\lambda_i,\;\lambda_f]\;[\AA]$      & $[\lambda_i,\;\lambda_f]\;[\AA]$     & $[\lambda_i,\;\lambda_f]\;[\AA]$       \\
\noalign{\smallskip}
\hline
\noalign{\smallskip}
H$\beta$  & [4861.04, 4861.60] & [4855.04, 4860.04] & [4862.6, 4867.2]  \\
H$\gamma$ & [4340.16, 4340.76] & [4333.60, 4336.80] & [4342.00, 4344.00]\\
Na D      & [5889.70, 5890.20] & [5860.00, 5870.00] & [5904.00, 5908.00] \\
          & [5895.67, 5896.17] &  &  \\
\noalign{\smallskip}
\hline

\end{tabular*}
\end{table}

The HARPS DRS provides the full width at half maximum (FWHM) and the 
bisector span of the cross-correlation function \citep{2001A&A...379..279Q}, 
which are both  useful probes of stellar activity. We also measure from the
spectra the chromospheric emission in the \ion{Ca}{ii} H\&K lines,
and  in the H$\alpha$, H$\beta$,  H$\gamma$, and Na~D 
lines \citep{1968ApJ...153..221W,1978ApJ...226..144G,1982ApJ...260..670L}. 
Our measurements of the \ion{Ca}{ii} H\&K follow 
\citet[][S-index]{1978PASP...90..267V} and those of H$\alpha$ 
\citet{2011A&A...534A..30G}. For the other lines that are sensitive 
to stellar activity, we compute the ratio of the integrated flux in 
the line (C) over that in two control bands (V, R):

\begin{equation}
\label{eq:indices}
H \beta = {C_\beta \over {V_\beta+R_\beta}} \; ; \;\;
H \gamma = {C_\gamma \over {V_\gamma+R_\gamma}} \; ; \;\;
Na D = {{C1_{Na}+C2_{Na}} \over {V_{Na}+R_{Na}}}
,\end{equation}
where the wavelength passbands are listed in Table~\ref{tab:waveBands}.

\subsubsection{Photometry}
\label{subsubsec:asas}
Since the contrast between spots and the rest of the photosphere increases
at bluer wavelength, we use photometry in the optical rather than near-infrared
band, and retrieved photometric time series of the five targets from the 
All Sky Automated Survey 
\citep[ASAS\footnote{\url{http://www.astrouw.edu.pl/asas}},][]{1997AcA....47..467P}. 
Since M dwarfs are relatively faint objects in this band, the ASAS photometry 
through the smallest syntheticapertures provide the highest quality measurements 
by minimizingthe sky background.

We systematically searched for spot induced periodicities for the 
whole photometric  time series and  for each seasonal observations. 
The seasonal analysis helps in the detection of the stellar rotation period 
assuming that active regions evolve and last for an unknown period of time.

\section{Data analysis}
\label{sec:searching}
\subsection{Methods}
\label{subsec:RV_analysis}

We applied the same method for the five stars, searching the time series 
for variability and periodicities, and modeling the periodic signals 
as Keplerian orbits. In an effort to disentangle activity signals
from planets, we compare the detected periods with the stellar 
rotation period and we search all RV signals for a counterpart in 
the time series of the activity indicators.

Using the CCF computed by the HARPS DRS, we start by building a 
representative measurement uncertainty for each star that accounts for 
the photon noise \citep{2001A&A...374..733B} and 
the $0.60\ ms^{-1}$  calibration error \citep[see Fig. 6 in][]{2016PASP..128f6001F}, 
but not for the 
stellar jitter. We quadratically add the two noise components for each 
epoch, and compute the typical expected error 
$\sigma_i$ as a weighted average of these epoch uncertainties. 
Prior to the RV analysis, we subtract the secular acceleration (see 
Table~\ref{tab:targetsProperties}). 
The observed dispersions of the pipeline RV measurements of GJ~3138, GJ~3323, 
 GJ~273, GJ~628, and GJ~3293 stand above this expected error by factors of respectively 
1.5, 1.5, 3.2, 3.7, and 3.1. 

This motivated the extraction of the offline RVs described 
above, which we searched for periodicities by computing a generalized Lomb-Scargle (GLS)
periodogram \citep{2009A&A...496..577Z}. Our GLS normalization choice is
such that a GLS power of 1.0 means that a sine wave is a perfect description
of the data, while a power of 0.0 means that a constant model is an equally 
good fit. To translate GLS power values to false alarm probabilities (FAP), 
we use bootstrap Monte Carlo simulations, creating 10,000 synthetic 
time series by shuffling the original velocity values between the 
observation dates. We compute a periodogram for each bootstrapped
data set, and record the distribution of the maximum value of the
bootstrapped periodograms. We obtain the FAP for a given GLS power
as the fraction of the bootstrapped periodograms that had a higher 
maximum value. The window function of the data sets is analyzed in order 
to identify aliasing or periodicities due to our temporal sampling.

When we identify a periodic signal with a FAP under 1\%, we adjust 
a Keplerian to the velocities and look for additional periodic signals 
in the fit residuals, iterating until no significant signal is detected.

To protect against overfitting the data with a more complex model 
than they warrant, we compute the Bayesian information criterion 
\citep[BIC: minus twice the Schwarz criterion,][]{schwarz1978} for
models with increasing numbers of Keplerian components. The BIC 
is widely used for model comparison because it is easily computed\footnote{BIC=-2 (log maximized likelihood)+ \newline 
+number of parameters (log number of measurements)}; however, the BIC's reliability 
as an approximation of the logarithm of the (more rigorous but harder to 
derive) Bayes factor  depends on the sample size \citep{10.2307/2291091} 
and quantifies the evidence in 
favor (or not) of a model over the null hypothesis model $H_0$ (generally 
the simplest model). 

We adjusted the (multi-)Keplerian models using the \textrm{\small YORBIT} 
\citep{2011A&A...535A..54S} code, which efficiently explores the model 
parameter space through a genetic algorithm. After model convergence,
we estimated the parameter uncertainties with a Markov chain Monte Carlo 
algorithm. As noted above, all models include a free zero-point offset 
between the velocities acquired before and after the HARPS fibers 
upgrade \citep{2015Msngr.162....9C}.

\subsection{GJ~3138}
\label{sec:GJ3138_analysis}

\subsubsection{Periodicity analysis}

We acquired 199 radial velocities of GJ~3138, of which 156 were obtained prior 
to the mid-2015 HARPS fibers upgrade and 43 afterwards. We discarded one 
measurement (epoch BJD=2457258.81) which is a strong outlier. The 
root mean square dispersion of these velocities is $\sigma_{(O-C)}=2.87\ ms^{-1}$, 
well above their expected uncertainty of $\sigma_i=1.82\ ms^{-1}$. 

\begin{figure}[t]
\centering
\includegraphics[scale=0.47]{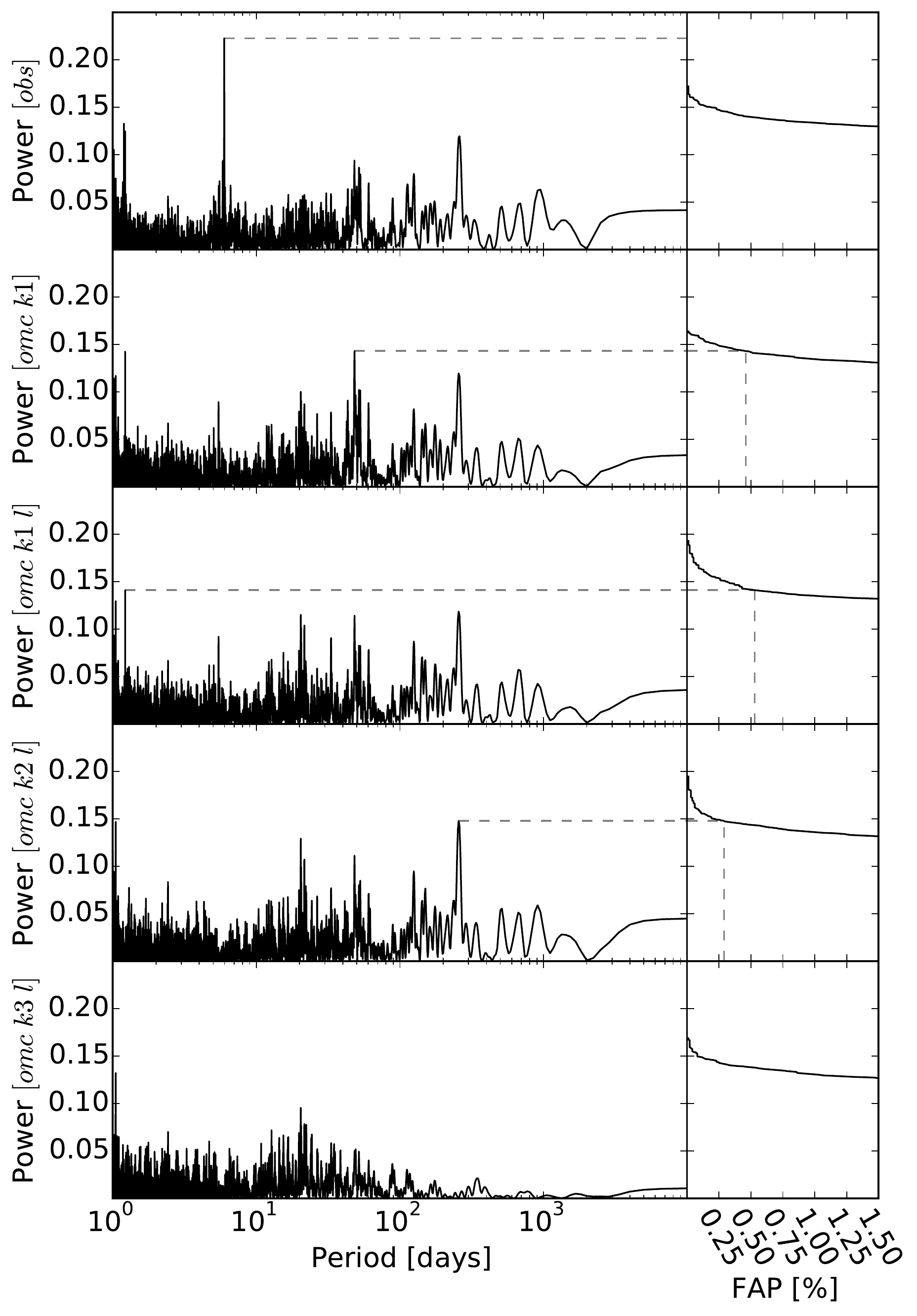}
\includegraphics[scale=0.47]{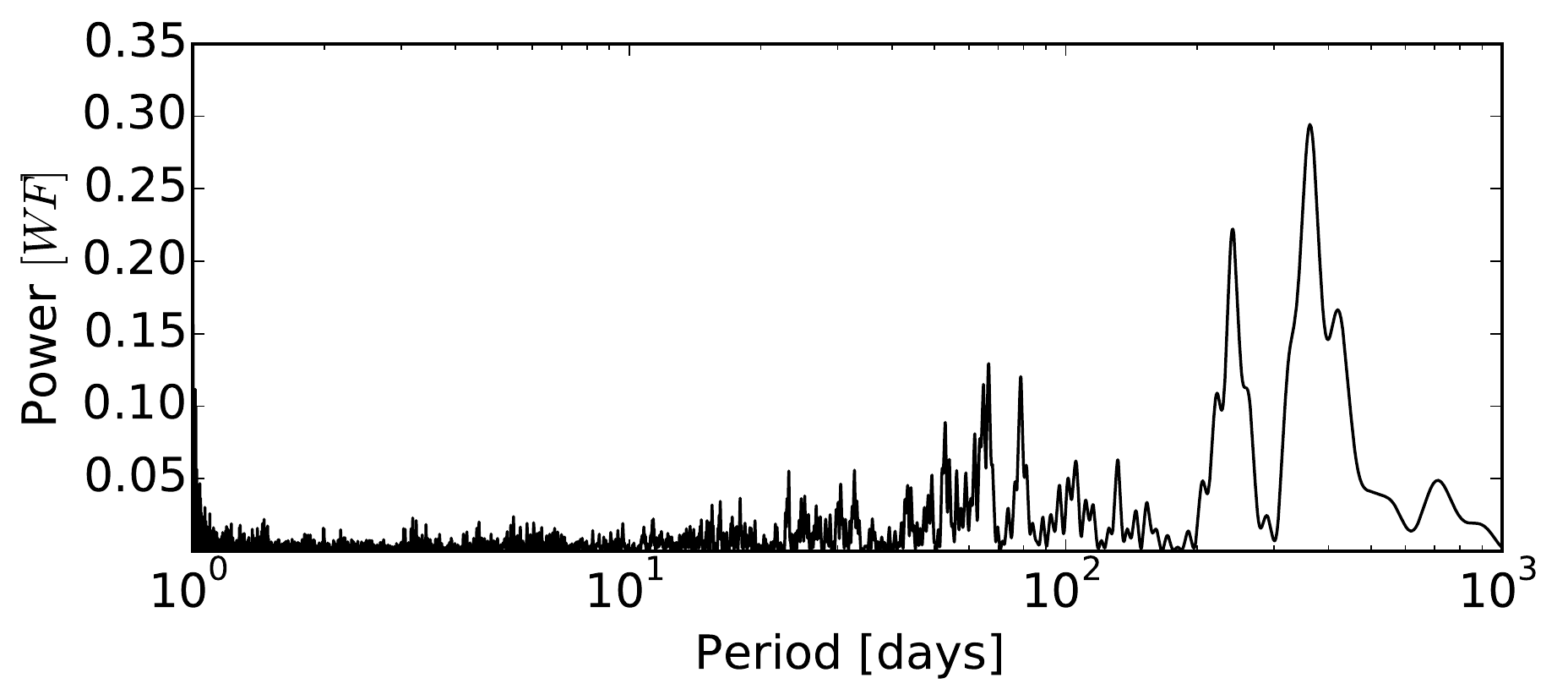}
\caption{\small \textit{Top:} Periodograms (left columns) of the GJ~3138 radial velocities. We 
subsequently subtract a 
model to fit data (raw RVs on top) until no clear signal remains on the residues (bottom). 
The $k$ in the y-axis label depicts a  Keplerian model (single or multiple), while the $l$ 
represents the linear model between RVs and H$\alpha$ (see text). In each step we modeled 
the signal highlighted with the horizontal dashed line. The right column represents the FAP 
derived from  the bootstrap analysis. Each analyzed peak has a FAP under 0.75\%. 
\textit{Bottom:} Window function exhibiting peaks located 
at 365 days, 240 days, 66 day, and 79 days.}
\label{fig:GJ3138_RVs_period}
\end{figure}

Our iterative periodogram analysis detects four significant periodicities 
(Fig.~\ref{fig:GJ3138_RVs_period}). The periodogram of the radial velocities shows a 
strong peak around a period of  6.0~days, with a power far higher than any found in the 10 000 
reshuffled data sets. Its FAP  is thus well below $0.01\%$. 
The residuals of a fit of the corresponding Keplerian model (model {\it k1}) have
periodogram peaks at periods of 48.2, 1.2, and 257 days, with FAPs of respectively 
0.46$\%$, 0.49$\%$, and 4.72$\%$. In the next section, we show that the peak at 48.2 days 
 is likely due to stellar activity since after subtraction of the other periodic 
radial velocity signals we observe a strong correlation with the H$\alpha$ flux. 
Instead of modeling this 48.2-day periodicity as a Keplerian signal, we use the 
RV-H$\alpha$ correlation derived in the next section to describe it. After 
subtraction of both that activity signal and the first planet, the strongest 
periodogram peak is at 1.2 days and has a 0.53\% FAP. We adjust a two-Keplerian
model (model {\it k2}) to the activity-corrected radial velocity time series 
(a RV time series now corrected from the RV-H$\alpha$ correlation). In a fourth 
iteration, the periodogram of the residuals exhibits a significant power excess at 
a period of 257 days, with a FAP=0.29$\%$. A fifth iteration finds two peaks at 1.05 and 
20.4 days, which are 1-day$^{-1}$ aliases of each other. The strongest of the two has 
a $0.9\%$ FAP that it just at our detection threshold of 1\% FAP . However, the BIC 
values of the models with three or four Keplerians ($\Delta BIC_{K3,K4}=4.4$) do not 
favor this more complex model. In addition, a periodicity of 20.4 days is half the 
rotation of GJ~3138. We therefore interpret that this low RV signature is possibly 
provoked by stellar activity.

\begin{figure}[t]
\centering
\includegraphics[scale=0.47]{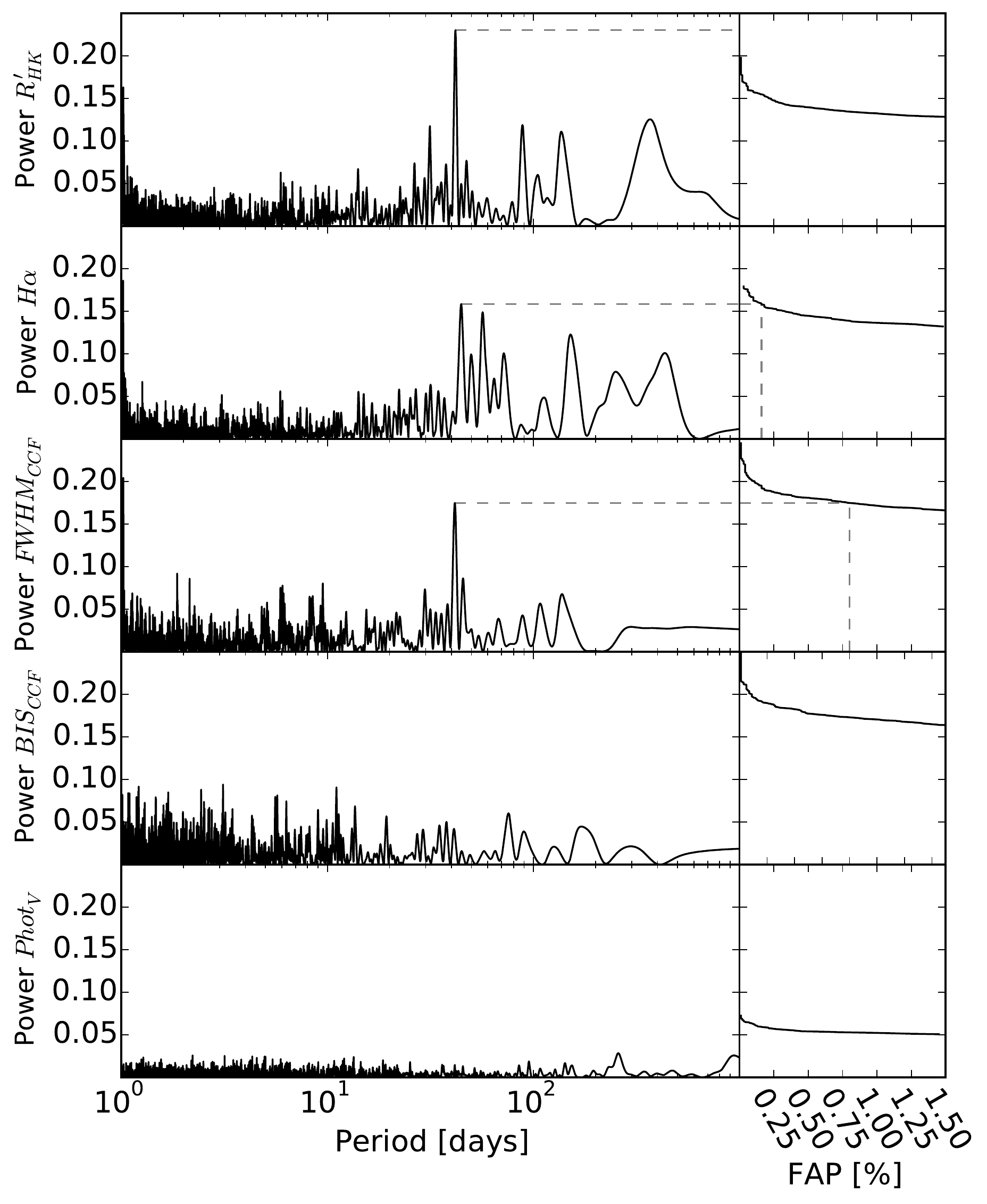}
\caption{\small Periodograms of the $R^\prime_{HK}$, H$\alpha$, and FWHM$_{CCF}$ 
measurements of GJ~3138. Their low-FAP peaks at respectively 41.8~days, 
44.6~days, and 41.2~days all reflect the stellar rotation period. 
}
\label{fig:GJ3138_act_period}
\end{figure}

\begin{figure}[t]
\centering
\includegraphics[scale=0.469]{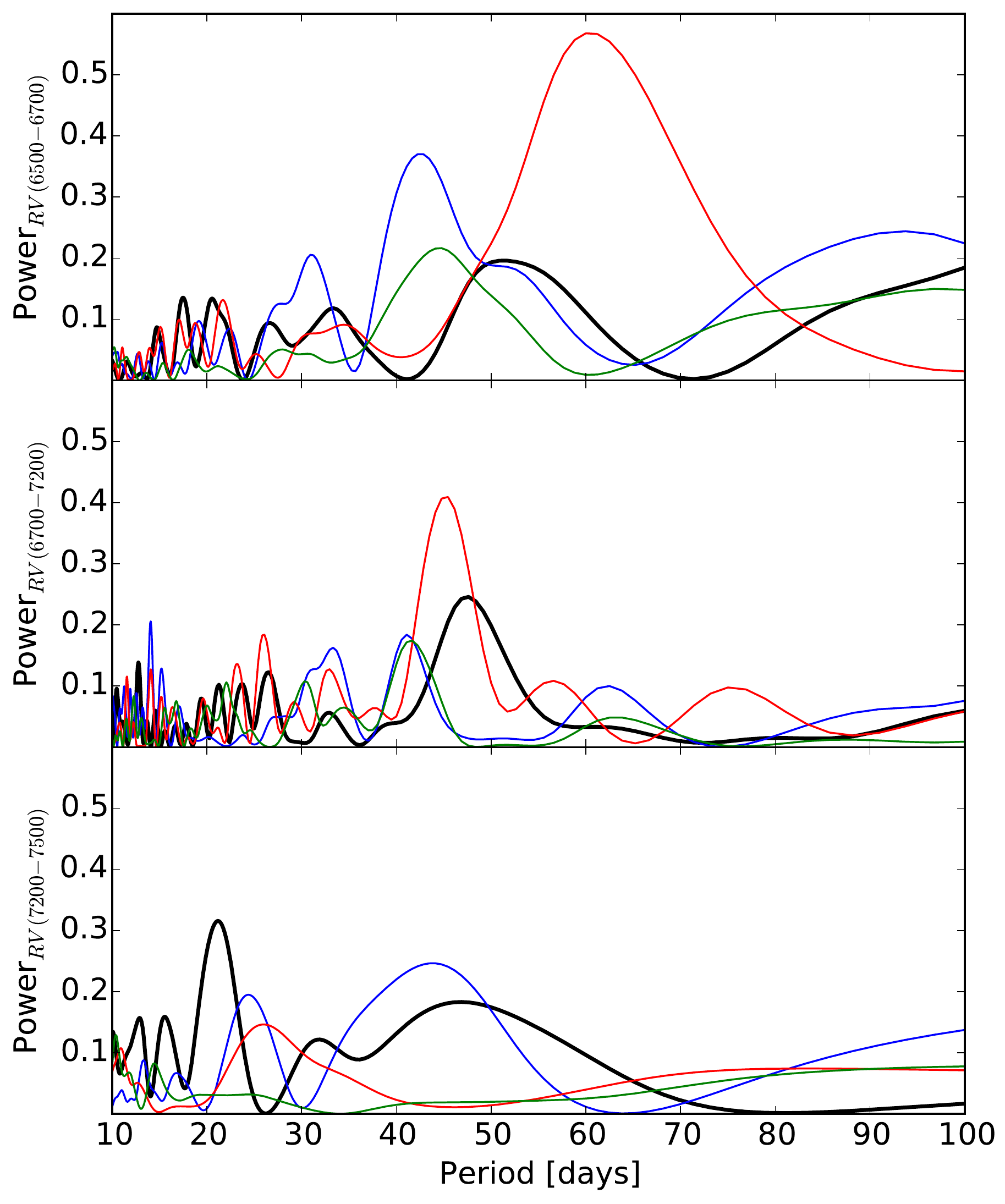}
\includegraphics[scale=0.469]{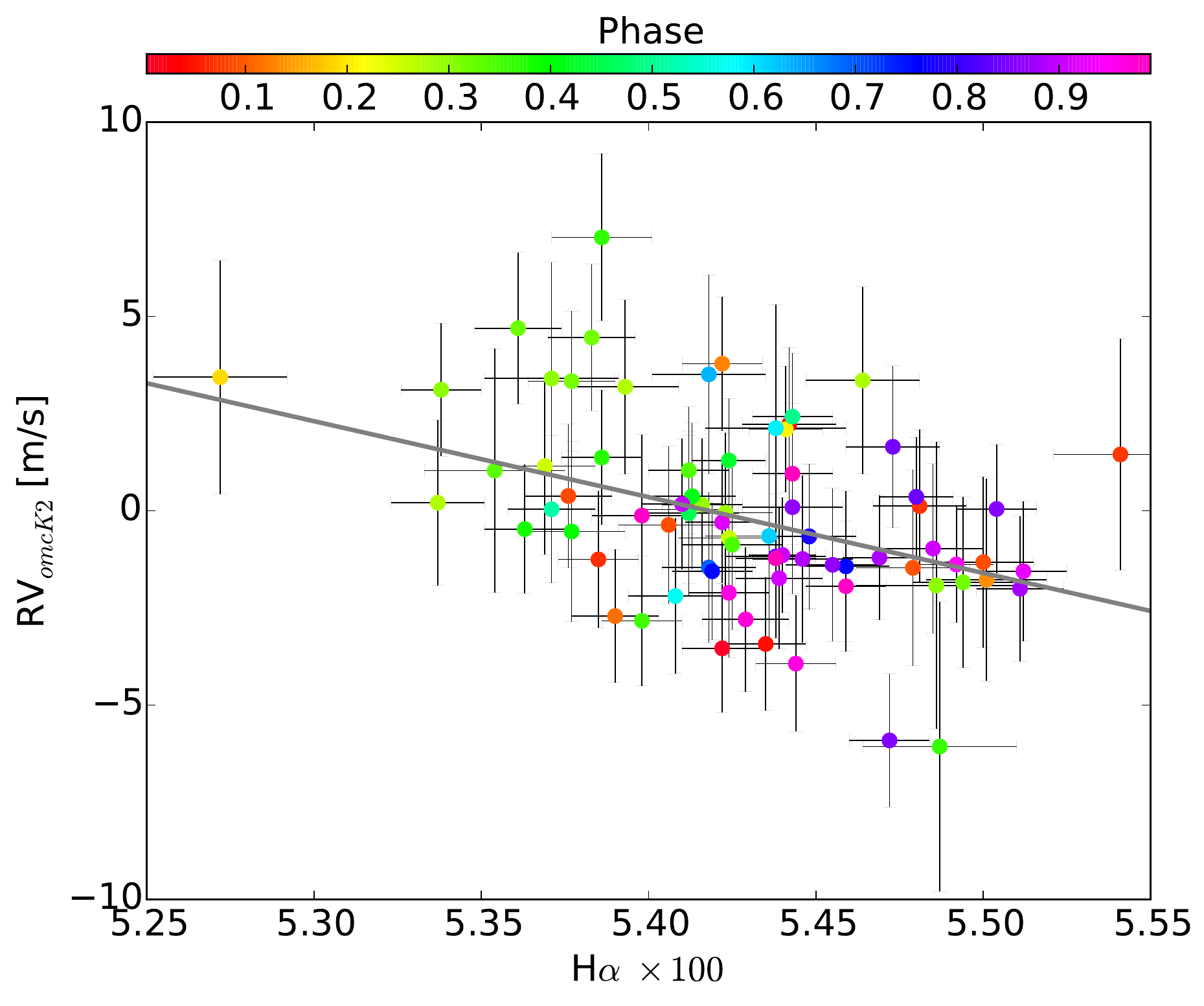}
\caption{\small Top panel:  Black, red, blue and green curves depict the periodogram of 
the GJ~3138 RVs after removing the 1.22 and 6.0 days periodicities, H$\alpha$, $R^\prime_{HK}$, 
and FWHM$_{CCF}$, respectively. The first row shows periodograms for BJD-2450000 between 
6500 and 6700, second row for 6700-7200, and third row for 7200-7500. The second row 
shows that RVs and H$\alpha$ have a power excess at the same periodicity. 
RVs--H$\alpha$ anti-correlation is subtracted (see text). Bottom panel: The 77 radial 
velocities as a function of the H$\alpha$ for BJD from 2456700 to 2457200. The phase for a periodicity of 48.2 days is color-coded. The linear fit is used to correct 
the RV interval.}
\label{fig:GJ3138_rvomck2P1_6_Halpha}
\end{figure}

\subsubsection{Validating the planet interpretation}
\label{sec:GJ3138_orb_act}

\paragraph{The 48-day signal as stellar rotation}
The stellar rotation period estimated from the average R$^\prime_{HK}$ is
34~days (Table~\ref{tab:targetsProperties}).
This is safely remote from the 1.2-day, 6.0-day, and 257-day periods, but
within its likely error bar of the 48-day period.
We therefore search the time series of the activity indicators for 
periodicities (Fig.~\ref{fig:GJ3138_act_period}) and for correlation with 
the measured radial velocity (Fig.~\ref{fig:GJ3138_rvomck2P1_6_Halpha}, 
bottom). Since the fiber upgrade introduced an offset in both the 
FWHM$_{CCF}$ and BIS$_{CCF}$ time series, we restrict this analysis to the 
pre-upgrade measurements as they represent the bulk of the data set. The 
$R^\prime_{HK}$, H$\alpha$, and FWHM$_{CCF}$ periodograms show significant peaks 
at periods of respectively 41.8, 44.6, and 41.2 days, while neither the periodogram of 
the bisector nor that of the ASAS photometry (497 points over 9.0 years) 
show significant periodicities. The periodograms for the seasonal 
photometry do not show evidence of  stellar rotation.
The three identified signals have similar 
periods, which are also close to the 34-day stellar rotation period estimate
from R$^\prime_{HK}$ and to the 48.1-day periodicity in the RV time series. We 
surmise that these periods all reflect modulation of photospheric inhomogeneities 
by the stellar rotation;  their dispersion is due to some combination of differential
rotation and measurement noise.

We turn to subsamples of the time series to confirm this suspicion. We subtract 
the 1.2- and 6.0-day signals from the RV series to focus on the 48-day signal,
and divide the activity diagnostic and RV time series into three BJD$-$2450000 
intervals: 6500$-$6700, 6700$-$7200, and 7200$-$7500, with 57, 77, and 41 
measurements. Figure~\ref{fig:GJ3138_rvomck2P1_6_Halpha} shows the corresponding
periodograms.  The periodograms of the RV residuals and H$\alpha$ for the 
second epoch ([6700$-$7200]) are strikingly similar;  both
have significant power excess at $\sim$43 days. The RV residuals 
clearly anti-correlate with H$\alpha$ for that epoch 
(Fig.~\ref{fig:GJ3138_rvomck2P1_6_Halpha}, bottom), with a Pearson coefficient 
of $-0.42$. The same figure also shows H$\alpha$ phased to a 48.2-day period, 
with points for phases between 0.2-0.5 mostly in the upper left corner and points for
phases between 0.7-0.9 mostly in the lower right part of the plot. 
We fit the anti-correlation and obtain 

\begin{equation}
\label{eq:Halpha_cor}
RV =(-1.953\pm0.262) \times H \alpha +(0.106 \pm 0.001)
.\end{equation}

We use this relation to correct the activity effect over the central interval. Without
evidence for a similar relation over the other two intervals, we do not correct the
rest of the RV time series. This partial correction reduces the strength of the 
48-day signal very significantly, as demonstrated by a comparison of the second 
and third periodograms in Figure~\ref{fig:GJ3138_RVs_period}. This adds to the
evidence for stellar activity, and we classify the 48-day periodicity as an activity 
signal rather than a planet.

\begin{figure}[t]
\centering
\includegraphics[scale=0.47]{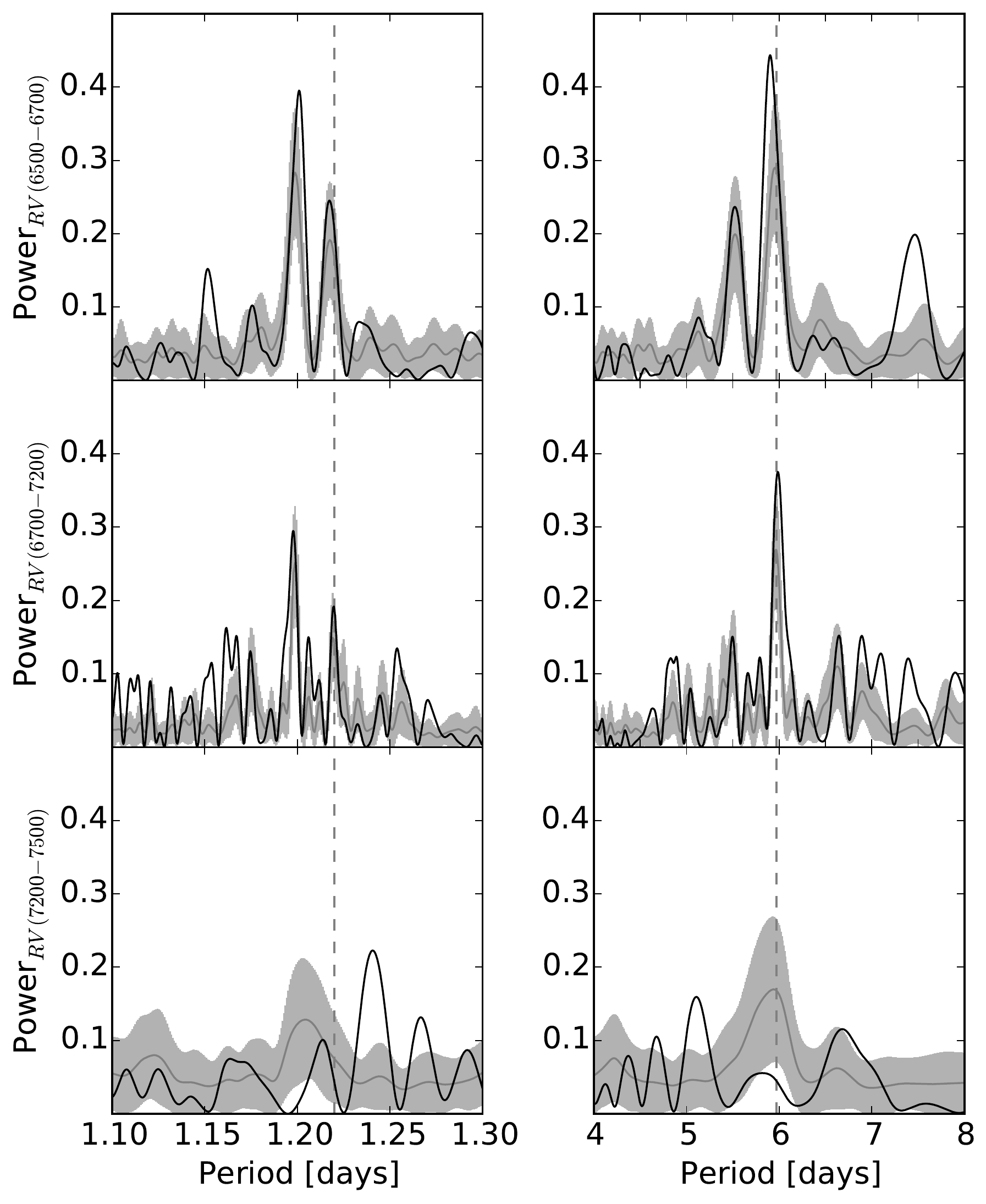}
\caption{\small Periodograms of the tree subsets of RVs. The left and right columns are zooms 
around the 1.2 and 6 days zones of interest, respectively. The black 
curves depict the periodograms for the observed RVs (raw), the gray 
curves show the average periodograms of the 100 synthetic RVs, while the
light gray areas represent the 1$\sigma$ zones. The top and middle 
rows of the left column show separately the 1-day alias of the 6.0 day 
peak (located at 1.20 days) and the 1.22-day signal (vertical dashed line). 
The bottom row  in the right column shows that the lack of power for the subsample 
with BJD-2450000 between 7200 and 7500 is consistent (very close to the 
1$\sigma$ level) with the constrained coverage in phase of the 6.0-day 
signal.}
\label{fig:GJ3138_rv_split}
\end{figure}

We inquire if the RV signals for the two shorter periodicities (1.22 and 6 days) 
are present along the entire data set -- favoring the scenario where RV variations 
are due to the presence of an exoplanet -- or concentrate in a particular subsample 
-- challenging the planetary interpretation. We thus divide in three our RV 
time series, with subsamples that satisfy BJD-2450000 in the ranges 6500-6700, 6700-7200, and 
7200-7500 and compute their periodograms. Figure~\ref{fig:GJ3138_rv_split} shows those 
periodograms specifically around 1.22 and 6 days, where both the 1.22- and 6~-~day signals 
seen as power excess in most epochs.

In parallel, we need to evaluate if the power excess are actually expected given the sampling 
of each one of the three subsamples. To that intend, we make additional synthetic time series 
(100) by drawing random RV points with a normal distribution centered around the best Keplerian 
fit and with a $2\ m/s$ standard deviation. For each one of the synthetic time series we make 
a new periodogram, and the gray areas in Fig~\ref{fig:GJ3138_rv_split} represent their $\pm$1 
sigma power distribution.

Around both 1.22 and 6 days, the periodogram of the original time series (black lines in 
 Fig~\ref{fig:GJ3138_rv_split}) is never found below the $\pm$1 sigma region (gray areas). 
This means both signals are indeed detected every time the sampling allows so, which in turn 
means the signals seem coherent rather than transitory.

Overall, we conclude on the planetary nature of both the 1.22- and 6-day signals.

\subsubsection{Orbital parameters}~\\

\begin{figure}[t]
\centering
\includegraphics[scale=0.495]{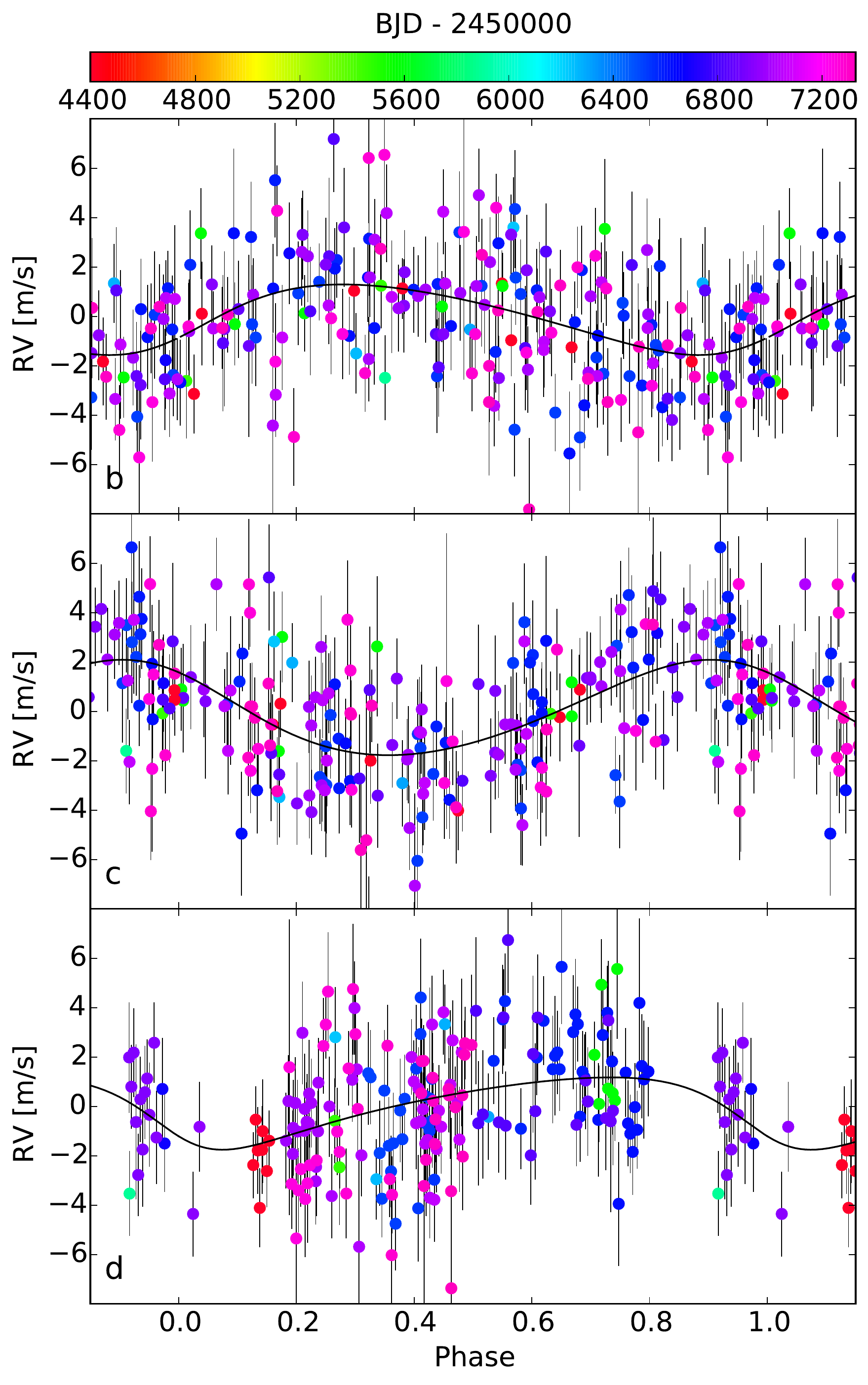}
\caption{\small Phase folded GJ~3138 RVs for the 1.22-day (top), 5.97-day (middle), 
and 258.5-day (bottom) orbits. The black solid curve and rainbow colors depict the Keplerian 
solution and BJD of observations, respectively.}
\label{fig:GJ3138_rv_sol}
\end{figure}

\begin{table}[h]
\centering
\caption{\small Parameters for the three Keplerians fitted to GJ~3138 RVs.}
\label{tab:GJ3138_k2}
\begin{tabular*}{\hsize}{@{\extracolsep{\fill}} l  l  c  c c}
\noalign{\smallskip}
\hline\hline
\noalign{\smallskip}
N$_{\rm Meas}$&&\multicolumn{2}{c}{ 198}\\
$\sigma_{\rm ext}$ &[m/s]&\multicolumn{2}{c}{ 0.39/1.37}\\
\noalign{\smallskip}
$\sigma_{(O-C)}$&[m/s]&\multicolumn{2}{c}{ 2.05/2.60}\\
\noalign{\smallskip}
$\Delta V_{21}$&[m/s]&\multicolumn{2}{c}{-0.16$_{ -0.53}^{+  0.52}$}\\
\noalign{\smallskip}
BJD$_{\rm ref}$&[days]&\multicolumn{2}{c}{56684.6682638567}\\
\noalign{\smallskip}
$\gamma$&[km/s]&\multicolumn{2}{c}{ 13.5973$\pm$0.0002 }\\
\noalign{\smallskip}
\hline
\noalign{\smallskip}
    &          & GJ~3138c & GJ~3138b& GJ~3138d \\
\noalign{\smallskip}
\hline
\noalign{\smallskip}
 P         &[days]                & 1.22003$_{ -0.00004}^{+  0.00006}$ & 5.974$_{ -0.001}^{+  0.001}$ & 257.8$_{ -3.5}^{+  3.6}$ \\
\noalign{\smallskip}
$K_1$      &[$ms^{-1}$]         & 1.43$_{ -0.26}^{+  0.27}$ & 1.93$_{ -0.26}^{+  0.26}$ & 1.47$_{ -0.30}^{+  0.35}$ \\
\noalign{\smallskip}
e          &                   & 0.19$_{ -0.13}^{+  0.18}$ &  0.11$_{ -0.07}^{+  0.11}$ &  0.32$_{ -0.21}^{+  0.20}$  \\
\noalign{\smallskip}
$\lambda_{0}$ at BJD$_{\rm ref}$  & [deg]& 239.4$_{-10.3}^{+ 10.5}$& 42.5$_{ -7.2}^{+7.2}$& 128.8$_{ -13.6}^{+13.0}$\\
\noalign{\smallskip}
\hline
\noalign{\smallskip}
$m\, sin(i)$ &[M$_\oplus$]       & 1.78$_{ -0.33}^{+  0.34}$ & 4.18$_{ -0.59}^{+  0.61}$& 10.5$_{ -2.1}^{+  2.3}$\\
\noalign{\smallskip}
a & [AU]&    0.0197$_{ -0.0005}^{+  0.0005}$&  0.057$_{ -0.001}^{+  0.001}$&  0.698$_{ -0.019}^{+  0.018}$\\
\noalign{\smallskip}
$S/S_\oplus$ && 118.1 & 13.9 & 0.1 \\
\noalign{\smallskip}
Transit prob. & [\%] & 10.5 & 3.7& 0.3\\
\noalign{\smallskip}
 BJD$_{\rm Trans}$-54000&[days]  & 2685.407$_{ -0.072}^{+ 0.092}$& 2685.47$_{ -0.24}^{+  0.25}$&2897.8$_{ -27.0}^{+  26.0}$\\

\noalign{\smallskip}
\hline

\end{tabular*}
\end{table}

Considering the above analysis, we detect 4 periodicities and conclude to the detection of 3 
planets and the stellar rotation.
Our Keplerian solution is illustrated in Figure~\ref{fig:GJ3138_rv_sol} and
presented in Table~\ref{tab:GJ3138_k2}, which reports separate noise estimates and rms 
residuals around the solution for the data acquired before and after the HARPS fibers 
upgrade. For the stellar mass reported in Table~\ref{tab:targetsProperties}, 
the semi-amplitudes convert to minimum masses of 1.9, 4.4 and 10.9~M$_\oplus$, 
in the regime of super-Earths and mini-Neptunes. Their irradiance are 116, 14, and 0.09 times 
the terrestrial one, which for extreme Bond albedos of 0.75 and 0.0 correspond 
to equilibrium temperatures ranges of respectively 640-900K, 375-530K, 107-150K.

\subsection{GJ~3323}
\label{sec:LHS1723_analysis}

\subsubsection{Periodicity analysis}

\begin{figure}[t]
\centering
\includegraphics[scale=0.47]{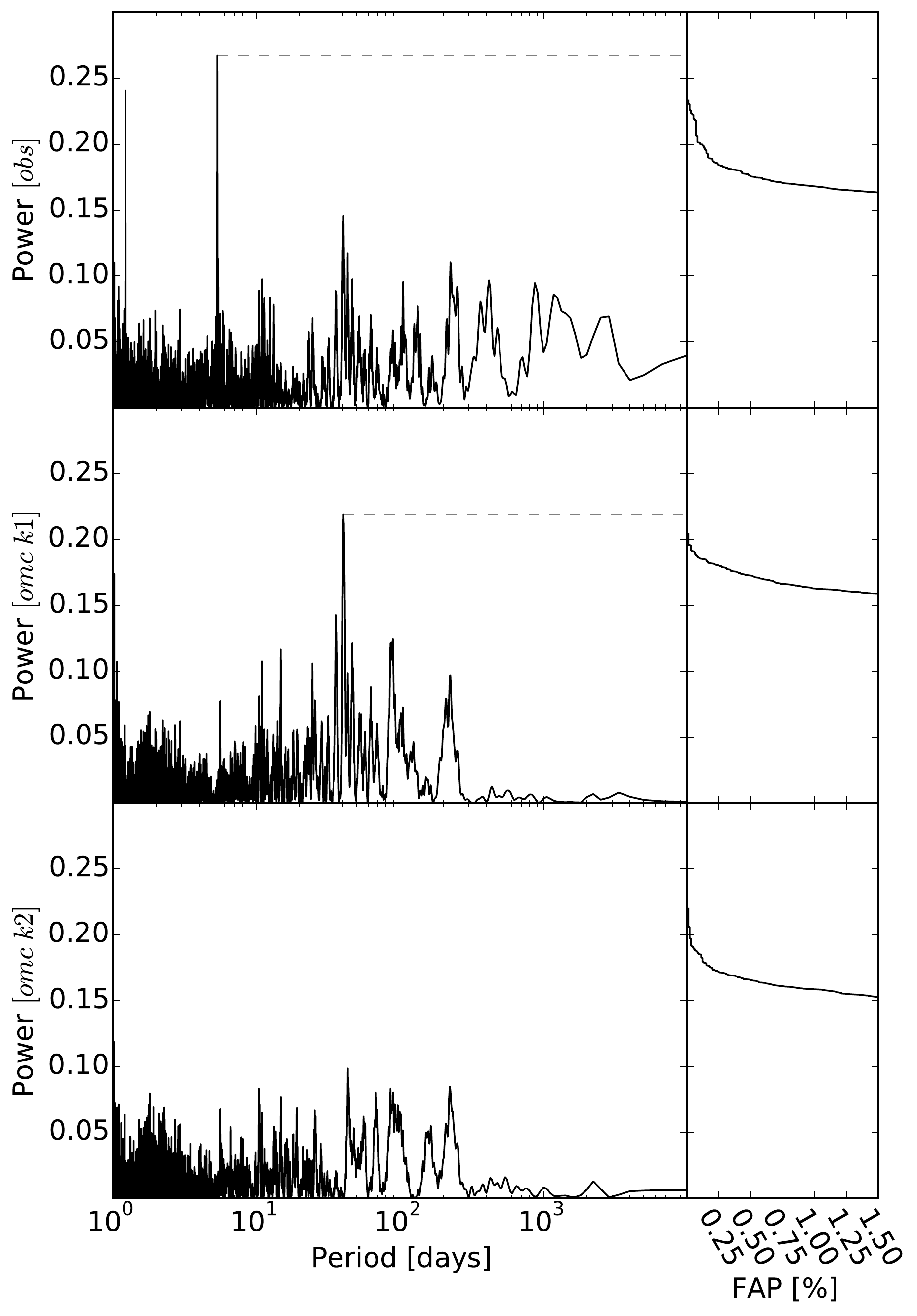}
\includegraphics[scale=0.47]{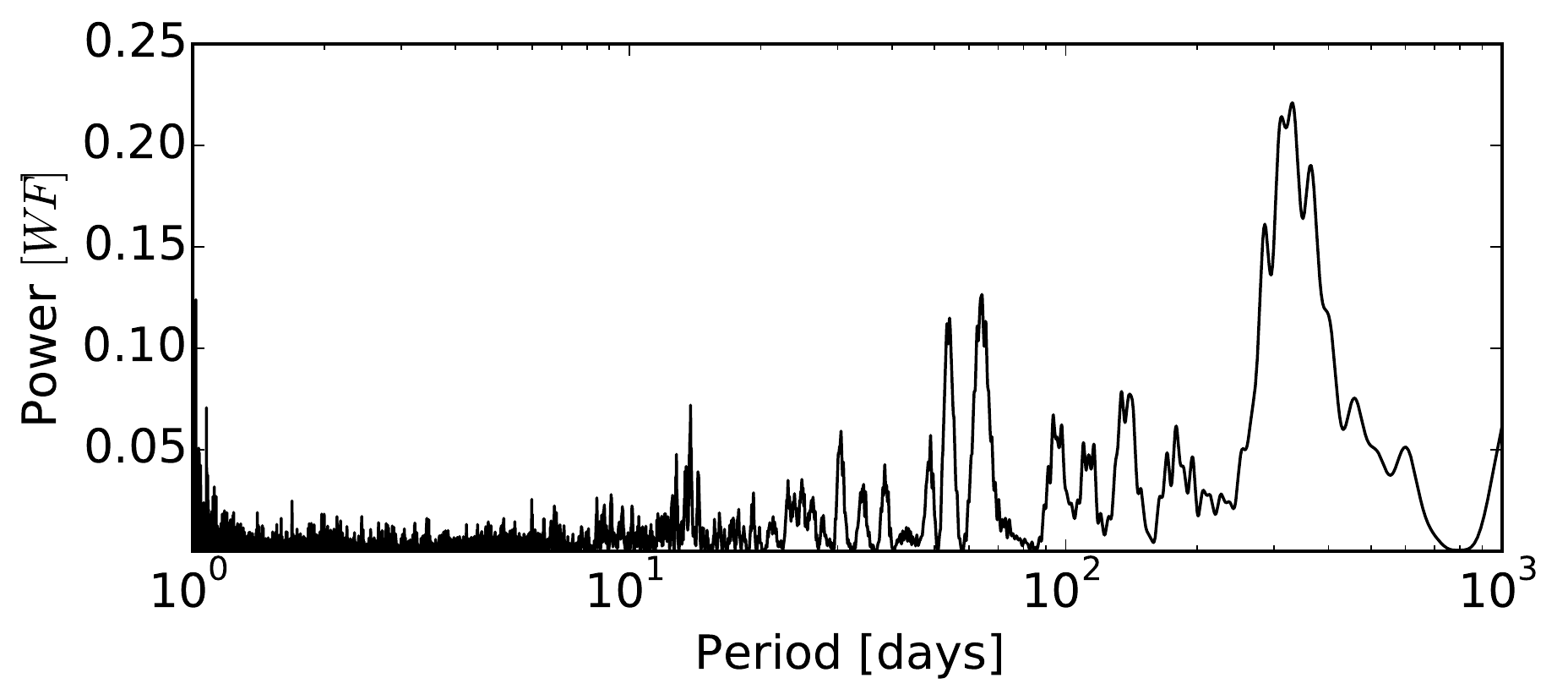}
\caption{\small \textit{Top:} Periodogram of GJ~3323 raw radial velocities, with a 
clear detection (FAP$<10^{-2}\%$) of periodicity at 5.4 days (top row). When subtracting
the Keplerian adjusted to this signal, the periodogram of residuals exhibit 
a very significant peak (FAP$<10^{-2}\%$) at 40 days (bottom row). 
\textit{Bottom:} Window function showing peaks at 1 year, 64 days, and 54 days.}
\label{fig:LHS1723_RVs_period}
\end{figure}

We acquired 157 radial velocities on GJ~3323, including 132 before and 25 after 
the HARPS upgrade. We rejected three points because two have very low SNR ($<$3 at 
550 nm; BJD=2456630.65, 2456926.86) and one is a $>5\sigma$ outlier (BJD=2457258.92). 
The 154 remaining RVs show a standard deviation $\sigma_{(O-C)}=2.85\ ms^{-1}$, 
in excess of estimated uncertainties $\sigma_i=2.18\ ms^{-1}$. The periodogram 
of raw RVs (Fig.~\ref{fig:LHS1723_RVs_period}) shows a powerful 
peak at around 5.4 days with FAP $<<0.01\%$. We fit a Keplerian to that first signal and 
compute the periodogram of the residuals. We found a second periodicity at  
$\sim$40 days with, again, a FAP$<<0.01\%$. We made a third iteration and found no more 
significant periodicity.

\subsubsection{Challenging the planet interpretation}

\label{sec:LHS1723_orb_act}

\begin{figure}[t]
\centering
\includegraphics[scale=0.47]{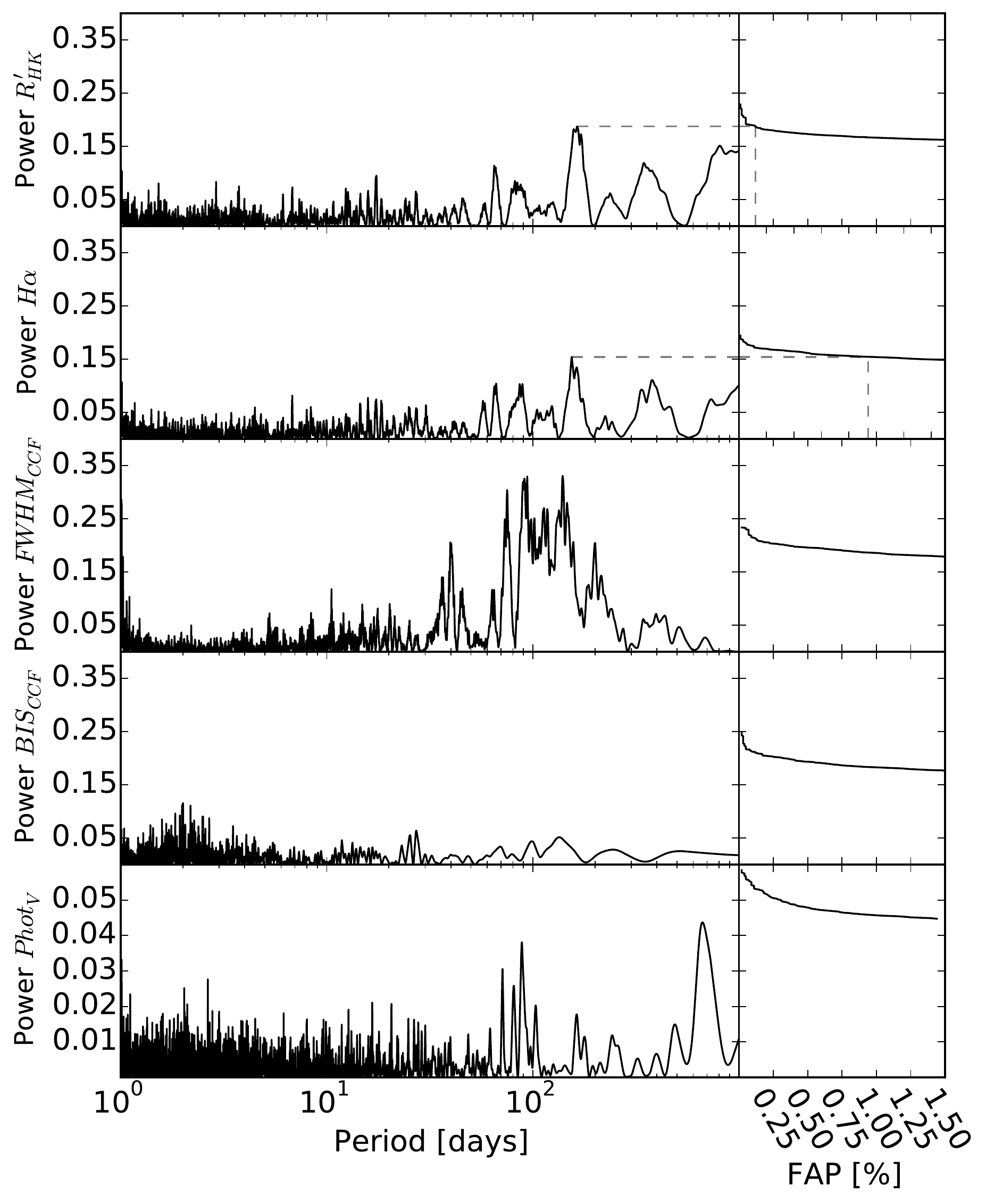}
\caption{\small Periodograms of the $R^\prime_{HK}$, H$\alpha$, FWHM$_{CCF}$, and the 
BIS$_{CCF}$ activity indices of GJ~3323. Significant peaks (FAP$<$1\%) are located 
at $\sim$160 days for $R^\prime_{HK}$ and H$\alpha$. The FWHM$_{CCF}$ periodogram is 
less clean at that periodicity and additionally shows power excess at $\sim$ 40 days. 
No favored periodicity is present in the periodogram of BIS$_{CCF}$, while the photometry 
periodogram shows power excess at about 665 and 88 days with 2\% and 10.2\% FAPs, respectively.}
\label{fig:LHS1723_act_period}
\end{figure}

\begin{figure}[t]
\centering
\includegraphics[scale=0.47]{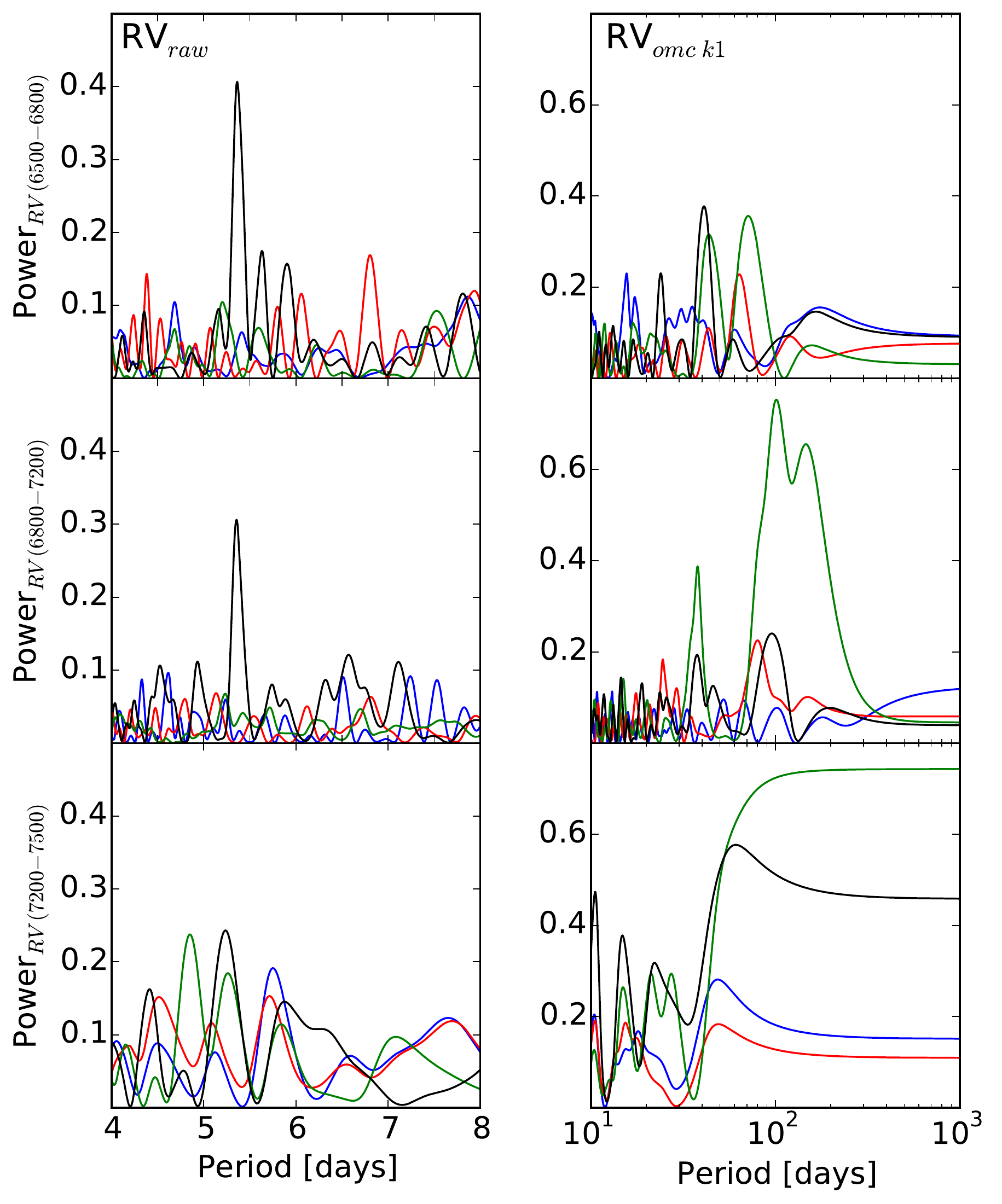}
\includegraphics[scale=0.47]{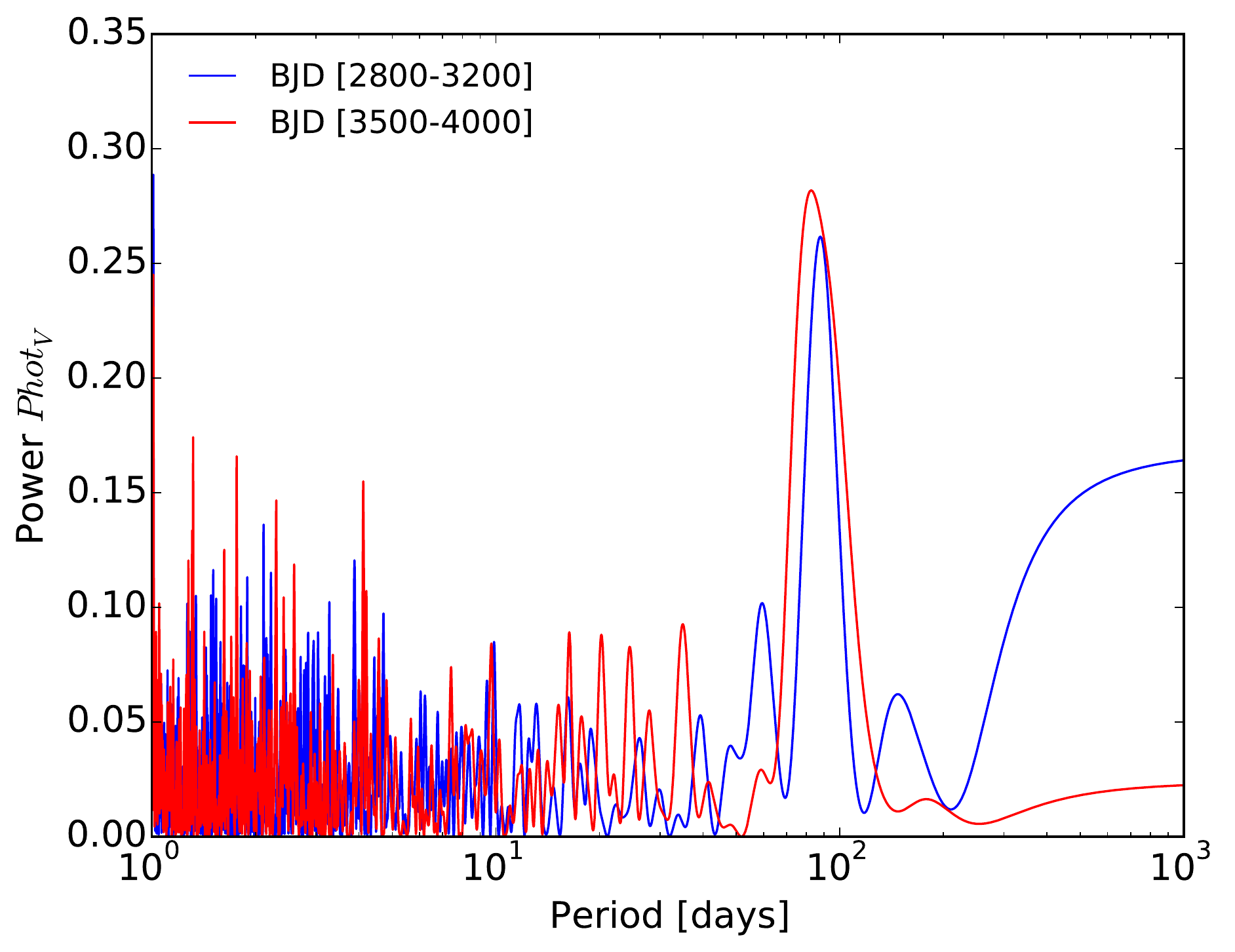}
\caption{\small \textit{Top:} GJ~3323 periodograms of RVs (black curve), $R^\prime_{HK}$ (blue curve), 
H$\alpha$ (red curve), and FWHM$_{CCF}$ (green curve) for subsets satisfying BJD-2450000: 
6500-6800, 6800-7200, and 7200-7500. The left column corresponds to raw RVs and the 
right column to the residues of subtracting the Keplerian adjusted to the 5.4-day 
signal. We note the temporal stability of the shorter RV periodicity and that the 
one at 40 days also shows power excess in the FWHM$_{CCF}$ (first and second rows of 
right column). \textit{Bottom:} Periodogram for the V-band ASAS photometry with 
BJD between 2800-3200 and 3500-4000. Significant power excess arises at about 90 days.}
\label{fig:LHS1723_rv_act_split}
\end{figure}

\paragraph{GJ~3323 b}~\\

GJ~3323 b is found with a period of 5.4 days, which  is significantly shorter than the 
33-day period estimated for the stellar rotation (Table~\ref{tab:targetsProperties}). 
We know that in some configurations, a spot coming in and out of view could produce a 
Doppler signal with a significant power excess at one-half or one-third of the stellar rotation 
period \citep{2011A&A...528A...4B}. It is thus reassuring to know the 5.4-day period is 
also significantly different than any harmonic of the rotation. 

As we did for the GJ\ 3138 RV analysis, we split the RV time series into three time intervals. They 
are chosen such that BJD-2450000 is in the ranges 6500-6800, 6800-7200, and 7200-7500, with 
56, 67, and 24 points, respectively. Figure~\ref{fig:LHS1723_rv_act_split} shows the 
periodograms with, on the left-hand side, a focus on a period of around 5 days. The signal is 
recovered at all epochs.

We can confidently accept that the first signal is a planet.

\paragraph{GJ~3323 (c)}~\\

The periodicity of around 40 days is close to our estimate of the stellar rotation. It  therefore 
requires more attention to the different activity proxies.

Periodograms of $R^\prime_{HK}$, FWHM$_{CCF}$, and $H\alpha$ present significant power excess in 
the period range of $\sim$160 days (Fig.~\ref{fig:LHS1723_act_period}). The FWHM$_{CCF}$ 
also shows additional power excess around 40 days. Conversely, in the same figure, the 
BIS$_{CCF}$ index does not show any significant periodicity and the periodogram for the whole 
ASAS photometry time series (654 points spanning 9.0 years) shows power excess at about 665 
(2\% FAP) and 88 days (10\% FAP). From the seasonal analysis, a periodicity close to 88 days 
is detected for the ASAS seasons between BJD 2800$-$3200 (0.24\% FAP) and 3500$-$4000 (0.42\% FAP). 
This value is compatible with the 88.5 days reported by \citet{2012AcA....62...67K} 
for the rotational period. However, the reliability of the 88.5-day stellar rotation is 
questionable because only 2/10 of ASAS seasons shows the $\sim$90-day periodicity even though 
 spots on M~dwars are usually more stable over time (e.g., GJ~674, GJ~581, GJ~174, GJ~551)
 and the stellar 
rotation period estimated from the $\log(R'_{HK})$ is 33~days. 
Except for the smaller power excess at 40 days in FWHM$_{CCF}$, the variability 
is thus seen in a period range much different than that of the 33-day period estimated for the 
stellar rotation by the $R^\prime_{HK}-P_{\rm rot}$ relation (Table~\ref{tab:targetsProperties}). 
It is possible that these longer timescale changes are related to magnetic cycles instead 
\citep{2011A&A...534A..30G}.

As previously, we subdivide the time series of both RVs and activity proxies into the same three 
time intervals. We subtract the best Keplerian fit to RVs to focus on the 40-day periodicity 
(right-hand side of Fig.~\ref{fig:LHS1723_rv_act_split}).
We note that only the first two intervals have a meaningful 
sampling able to probe the 40-day signal. The third interval only has 24 points spread on just one 
period and a half. A power excess appears at about 40 days for both the RVs and FWHM$_{CCF}$ 
periodograms in the 6500$-$6800 and 6800$-$7200 subsamples.

\begin{figure}[t]
\centering
\includegraphics[scale=0.5]{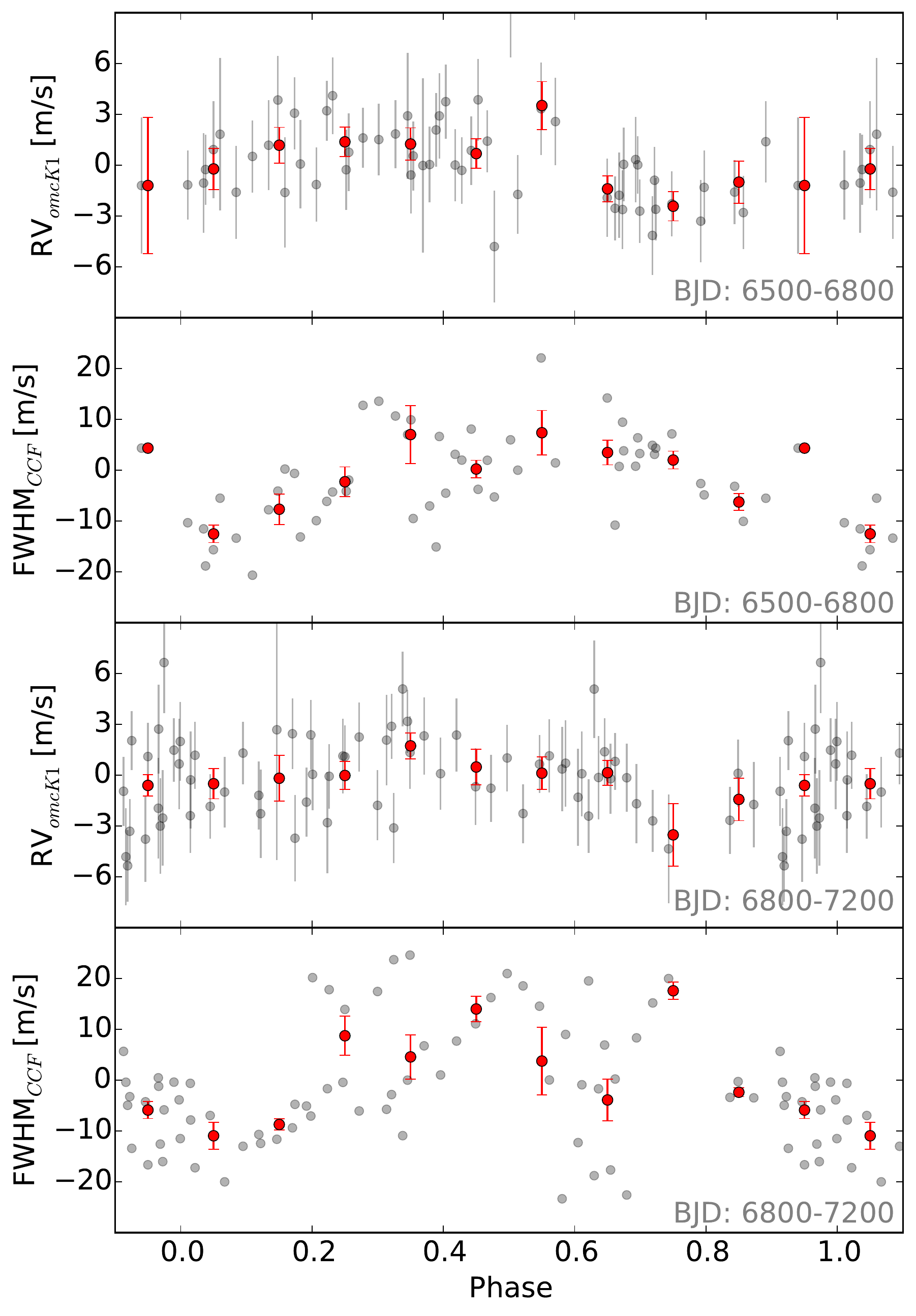}
\caption{\small GJ~3323 RV$_{omc\ k1}$ and FWHM$_{CCF}$ (the median value is subtracted) against 
the phase for the two different epochs where the variables show power excess at about 40 
days in their periodograms (Fig.~\ref{fig:LHS1723_rv_act_split}). Red points depict binned 
data to guide the eye where a possible dependence arises for the BJD interval 6500-6800.}
\label{fig:LHS1723_rv_fwhm_phase}
\end{figure}

In Figure~\ref{fig:LHS1723_rv_fwhm_phase}, we further inspect RVs against FWHM$_{CCF}$ 
for two time intervals and it is hard to exclude possible resemblances, in particular if we 
allow for a shift in phase. However, no other activity indicator shows evidence of 
stellar activity at about 40 days (Fig.~\ref{fig:LHS1723_act_period}) and currently we have  
no example of FWHM$_{CCF}$ variation without \ion{Ca}{ii} H\&K or H$\alpha$ variation; 
additionally, the FWHM$_{CCF}$ periodograms for the 6500$-$6800 and 6800$-$7200 subsamples 
show most of the power centered on periods 80$-$250. We therefore search for other possible 
origins of the power excess seen at about 40 days in the FWHM$_{CCF}$ periodogram.

\begin{table}[h]
\small
\centering
\caption{\small Parameters for the Keplerians fitted to GJ~3323 RVs.}
\label{tab:LHS1723_k2}
\begin{tabular*}{\hsize}{@{\extracolsep{\fill}} l c c c}
\noalign{\smallskip}
\hline\hline
\noalign{\smallskip}
N$_{\rm Meas}$& &\multicolumn{1}{c}{ 154 }\\
\noalign{\smallskip}
$\sigma_{\rm ext}$ &[m/s]&\multicolumn{1}{c}{0.31/0.88}\\
\noalign{\smallskip}
$\sigma_{(O-C)}$&[m/s]&\multicolumn{1}{c}{2.10/2.19}\\
\noalign{\smallskip}
$\Delta V_{21}$&[m/s]&\multicolumn{1}{c}{-2.82$_{ -0.62}^{+  0.63}$}\\
\noalign{\smallskip}
BJD$_{\rm ref}$&[days]&\multicolumn{1}{c}{56800.0918533883}\\
\noalign{\smallskip}
$\gamma$&[km/s]&\multicolumn{1}{c}{42.4508 $\pm$0.0002}\\
\noalign{\smallskip}
\hline
\noalign{\smallskip}
    &          & GJ~3323b& GJ~3323(c) \\
\noalign{\smallskip}
\hline
\noalign{\smallskip}
 P         &[days]                &  5.3636$_{ -0.0007}^{+  0.0007}$ &40.54$_{ -0.19}^{+  0.21}$\\
\noalign{\smallskip}
$K_1$      &[$ms^{-1}$]         &2.55$_{ -0.32}^{+  0.33}$ & 1.49$_{ -0.31}^{+  0.33}$\\
\noalign{\smallskip}
e          &                   &  0.23$_{ -0.11}^{+  0.11}$ & 0.17$_{ -0.12}^{+  0.21}$\\
\noalign{\smallskip}
$\lambda_{0}$ at BJD$_{\rm ref}$  & [deg]&  180.4$_{ -7.0}^{+  7.1}$ &305.3$_{-13.0}^{+ 13.0}$\\
\noalign{\smallskip}
\hline
\noalign{\smallskip}
$m\, sin(i)$ &[M$_\oplus$]       & 2.02$_{ -0.25}^{+  0.26}$ &2.31$_{ -0.49}^{+  0.50}$\\
\noalign{\smallskip}
a& [AU]&  0.03282$_{ -0.00056}^{+  0.00054}$  & 0.1264$_{ -0.0022}^{+  0.0021}$\\
\noalign{\smallskip}
$S/S_\oplus$ && 2.58 & 0.17 \\
\noalign{\smallskip}
Transit prob. & [\%] & 2.1 & 0.5  \\
\noalign{\smallskip}
BJD$_{\rm Trans}$-54000.0&[days]  & 2803.797$_{ -0.218}^{+  0.213}$ &2816.39$_{ -2.72}^{+  2.54}$\\

\noalign{\smallskip}
\hline

\end{tabular*}
\end{table}

\begin{figure}[t]
\centering
\includegraphics[scale=0.47]{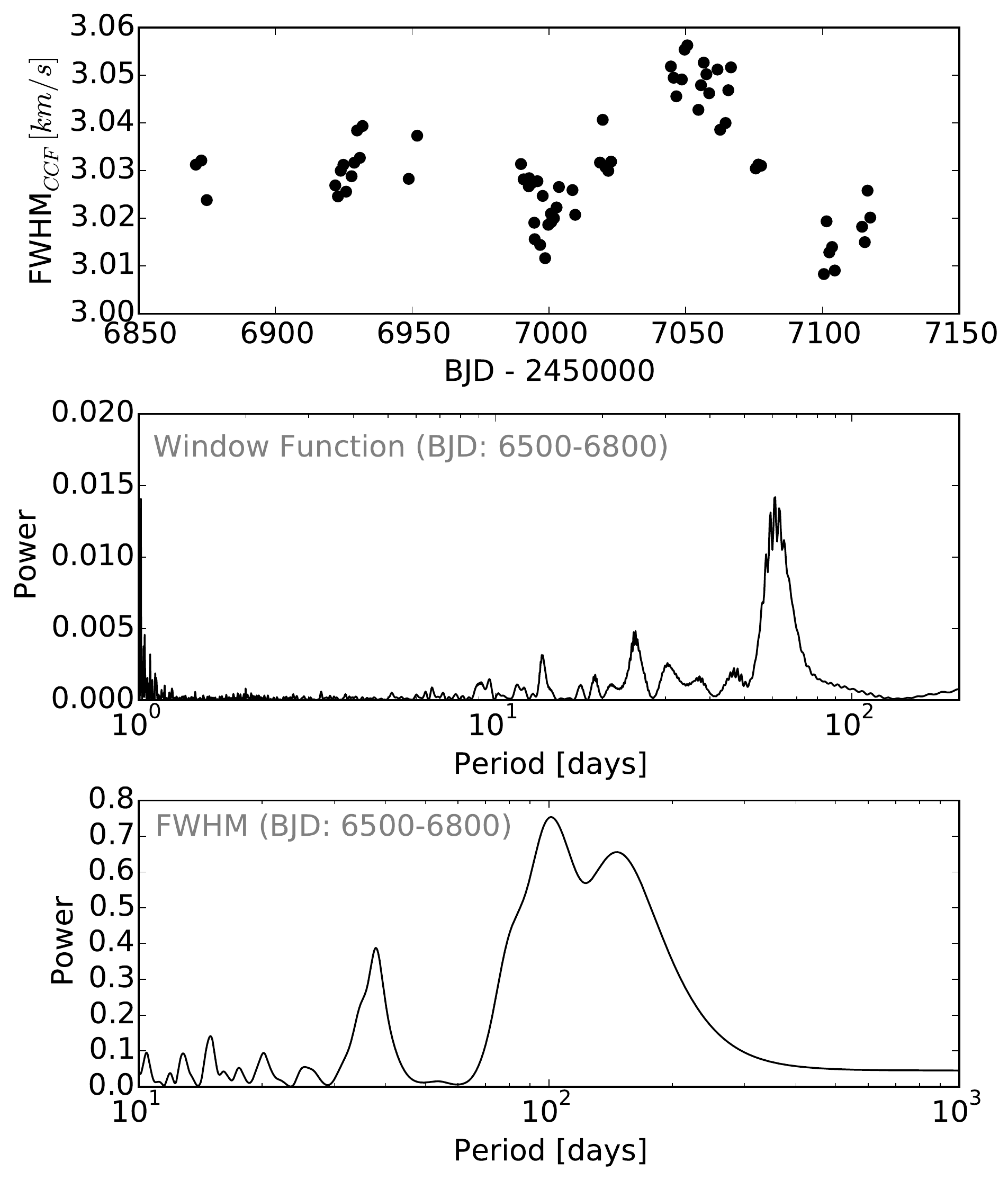}
\caption{\small When focusing on data with BJD between 6800 and 7200, we note that the 
FWHM$_{CCF}$ increases at about BJD 7050, while at BJD 7000 and 7100 it is closer to the 
average value (upper panel) and produces the peak at $\sim$100 days in its periodogram. 
The window function -- with power excess at $\sim$60~days (middle panel) -- for this 
subsample shows that our sampling clearly explains that the 40-day peak is an alias of 
the stronger 100-day peak (bottom panel).}
\label{fig:LHS1723_fwhm_6800-7200}
\end{figure}

We show in Figure~\ref{fig:LHS1723_fwhm_6800-7200} the time series for the FWHM$_{CCF}$ 
subsample satisfying BJD-2450000 $\epsilon$ [6800$-$7200]. It is evident that this activity 
tracer is higher at about BJD-2450000=7050. In the same figure, we note that our sampling 
generates a peak at $\sim$60 days in the window function. This means that the power 
excess at about 40 days is an alias of the strongest peak located at about 100 days 
(1/37.5= 1/100+1/60); when subtracting a 100-day periodicity, power excess is no longer 
present at about 40 days. However, this peak at $\sim$100 days and its $\sim$40-day alias remains unexplained because the other activity indicators show a power excess concentrated 
at $\sim$150 days for the $R^\prime_{HK}$ and H$\alpha$,  and at 665 days for the photometry 
(Fig~\ref{fig:LHS1723_act_period}), while a stellar rotation period is expected to be  about 
33 days (Table~\ref{tab:targetsProperties}). Taking these data all together, and before more data could be acquired, 
we think GJ~3323c should be considered a planet candidate.

\subsubsection{Keplerian analysis}

\begin{figure}[t]
\centering
\includegraphics[scale=0.5]{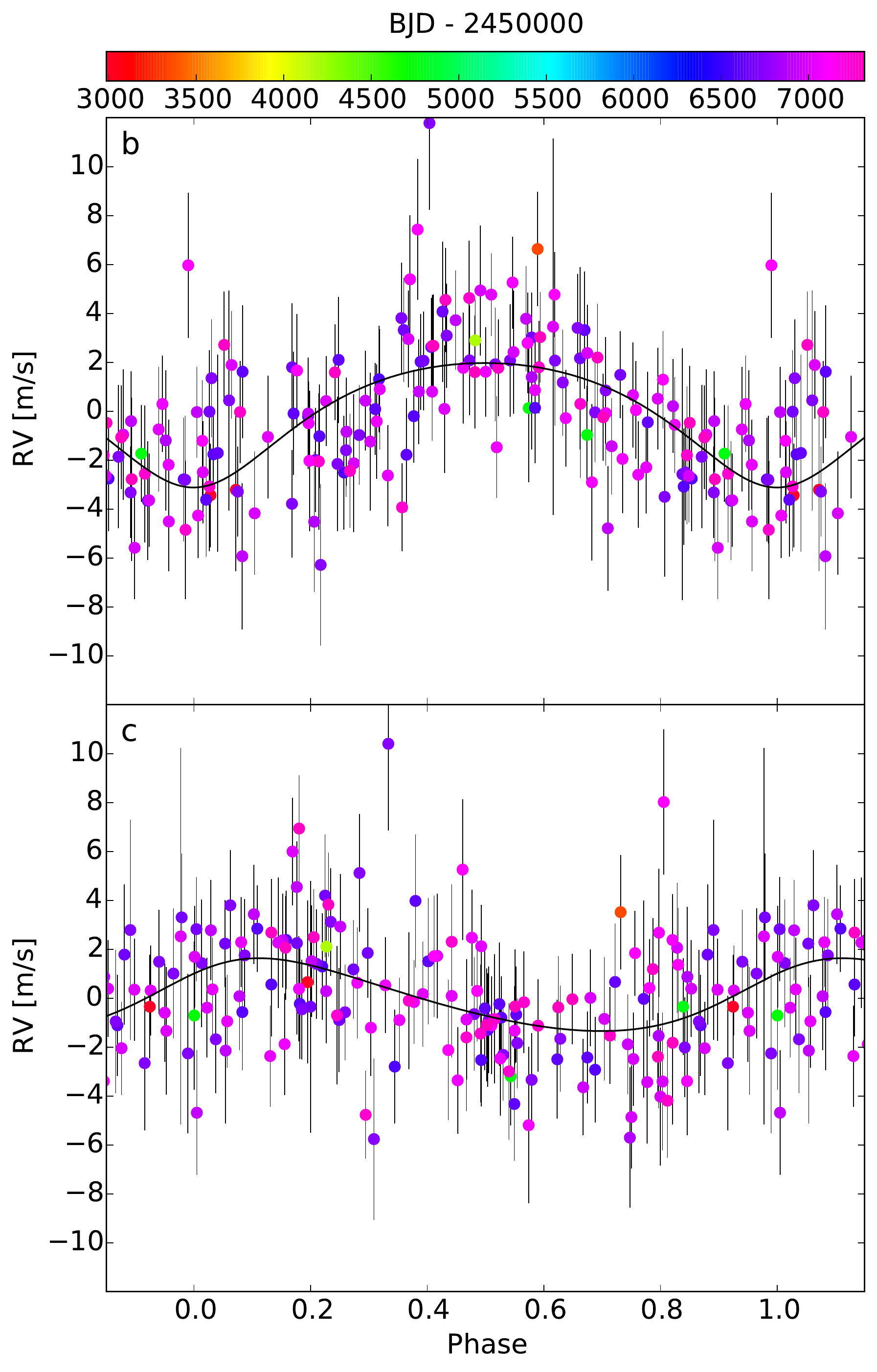}
\caption{\small Phase folded -- 5.36 (top) and 40.5  days (bottom) -- GJ~3323 RVs. Solid 
black curve and rainbow color-coding represent the Keplerian solution and the BJD, respectively.}
\label{fig:LHS1723_rv_sol}
\end{figure}

We use \textrm{\small YORBIT} to fit a model with two Keplerians  
with semi-amplitudes of 2.55$\pm$0.32 and 1.49$\pm$0.32~m/s. We estimated a stellar mass of 
0.164~M$_\odot$ (Table~\ref{tab:targetsProperties}). This converts the semi-amplitudes to 
2.02$\pm$0.25 and 2.31$\pm$0.50~M$_\oplus$ planets and makes GJ~3323 one of the lowest-mass 
stars (0.16M$_\odot$) known with planets discovered by radial velocity. 
GJ~3323b and GJ~3323(c) receive 2.51 and 0.17 times the Earth irradiance and have an 
equilibrium temperature in the range of 420-595K and 214-303K, respectively, if   bond 
albedos between 0 and 0.75 are assumed.
Details of the derived parameters 
are given in Table~\ref{tab:LHS1723_k2}, and Fig.~\ref{fig:LHS1723_rv_sol} shows the phase 
folded RVs.

\subsection{GJ~273}
\label{sec:Gl273_analysis}

\subsubsection{Periodicity analysis}
We intensively monitored GJ~273, recording 280 spectra between December 2, 2003, and 
September 30, 2016. From this 12.8 years of RV monitoring, 43 of them were acquired after 
the HARPS fibers upgrade. The RVs have a standard deviation of $\sigma_{(O-C)}=2.75\ ms^{-1}$, 
a value which  is well above the estimated uncertainties of $\sigma_i=0.94\ ms^{-1}$.

\begin{figure}[t]
\centering
\includegraphics[scale=0.47]{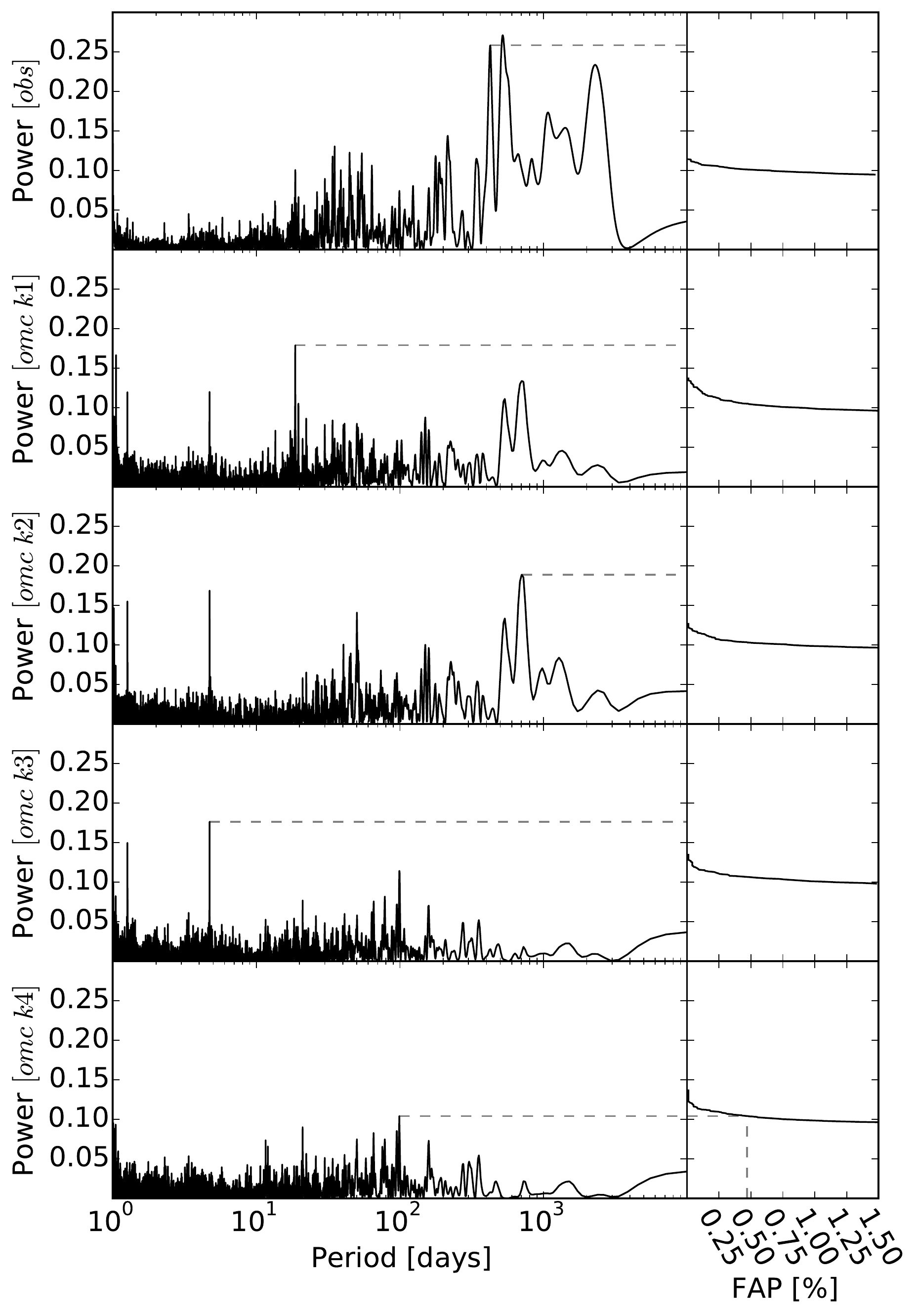}
\includegraphics[scale=0.47]{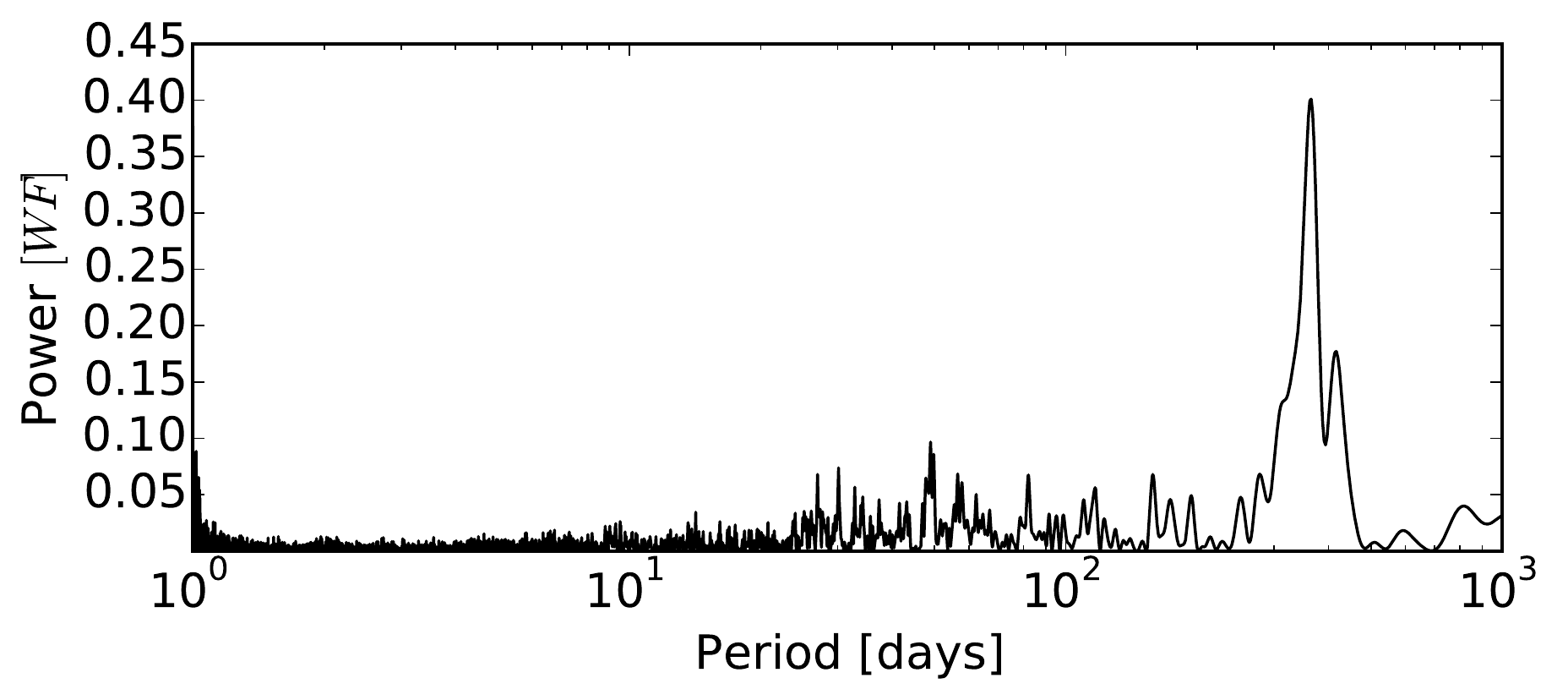}
\caption{\small textit{Top:} Periodograms for the radial velocities of GJ~273 and their residues after sequentially 
subtracting the models accounting for the main periodicity. From top to bottom, up to four RV 
signals are seen with periodicities of 420, 20, 700, and 5 days. 
\textit{Bottom:} Window function showing peaks located at 1 year and at
49 days.}
\label{fig:Gl273_rv_perio}
\end{figure}

Figure~\ref{fig:Gl273_rv_perio} shows the periodograms, first for the raw time series, 
and then for the residuals of each iteration.  Most of the power in the raw RV periodogram 
is concentrated at low frequencies with FAP$<<0.01$, and \textrm{\small YORBIT} converges 
to a solution with a periodicity of $\sim$420 days. The periodogram of the residues shows a 
powerful peak largely above our 1\% FAP threshold at about 20 days, modeled by a second Keplerian. 
After subtraction of this model, the two highest peaks in the periodogram of the residues are 
located at about 700 days and 5 days (both with FAP$<<0.01$). When adding a third Keplerian 
to the model, the favored solution contains the two previous periodicities plus the one at 
700 days. Now the 5-day peak and its 1-day alias dominate the periodogram of the residues. 
Although the residues still have power excesses at about 100 days (0.5\% FAP), our final model 
consists of four Keplerian because this periodicity is  explained well by a signal generated 
by the stellar rotation (99 days; Table~\ref{sec:stellar_properties}). As we  describe below, 
we suspect that both the two longest periodicities (with P$\sim$ 420 and $\sim$ 700 days) 
also have a stellar origin.

\begin{figure}[t]
\centering
\includegraphics[scale=0.47]{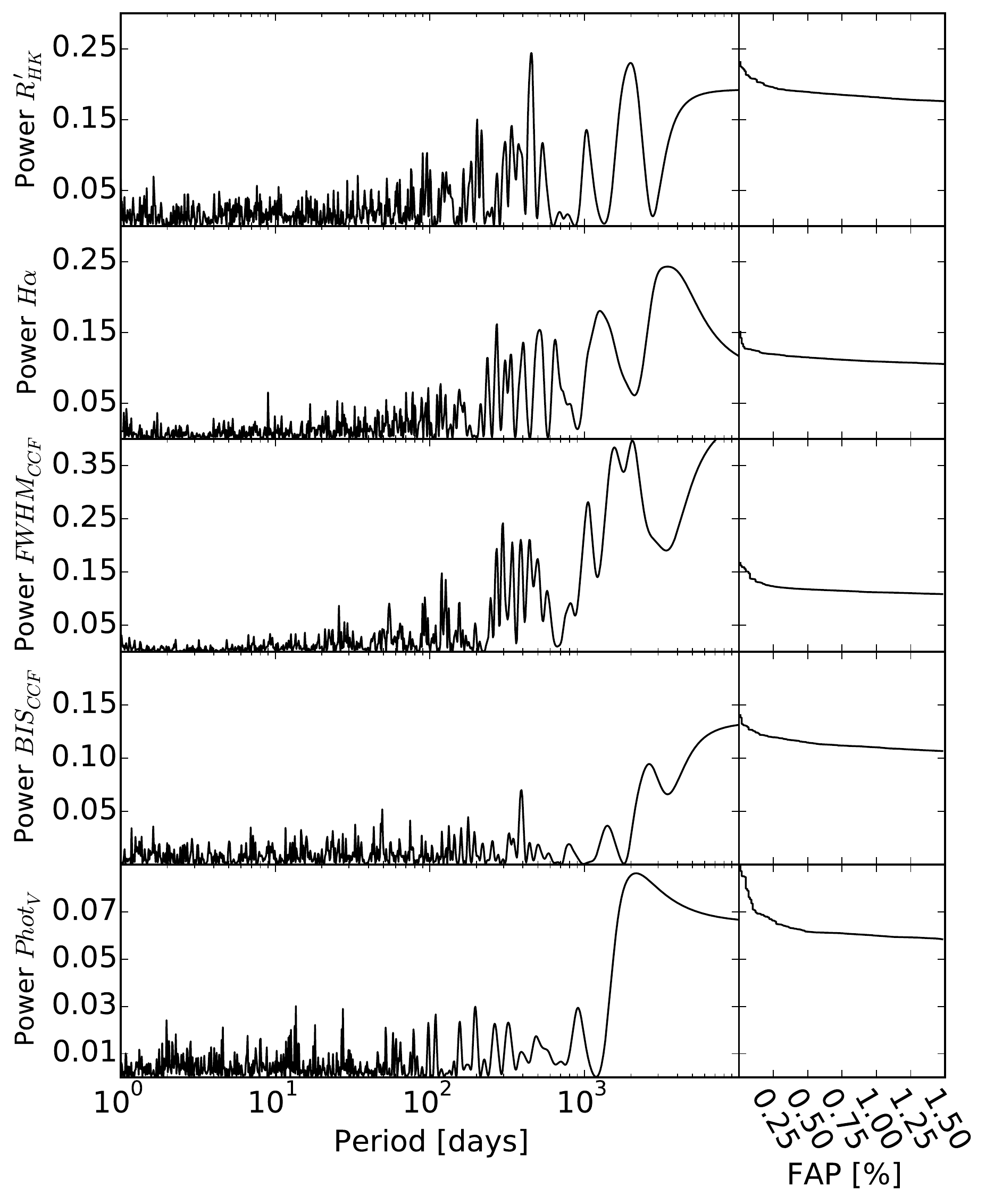}
\includegraphics[scale=0.47]{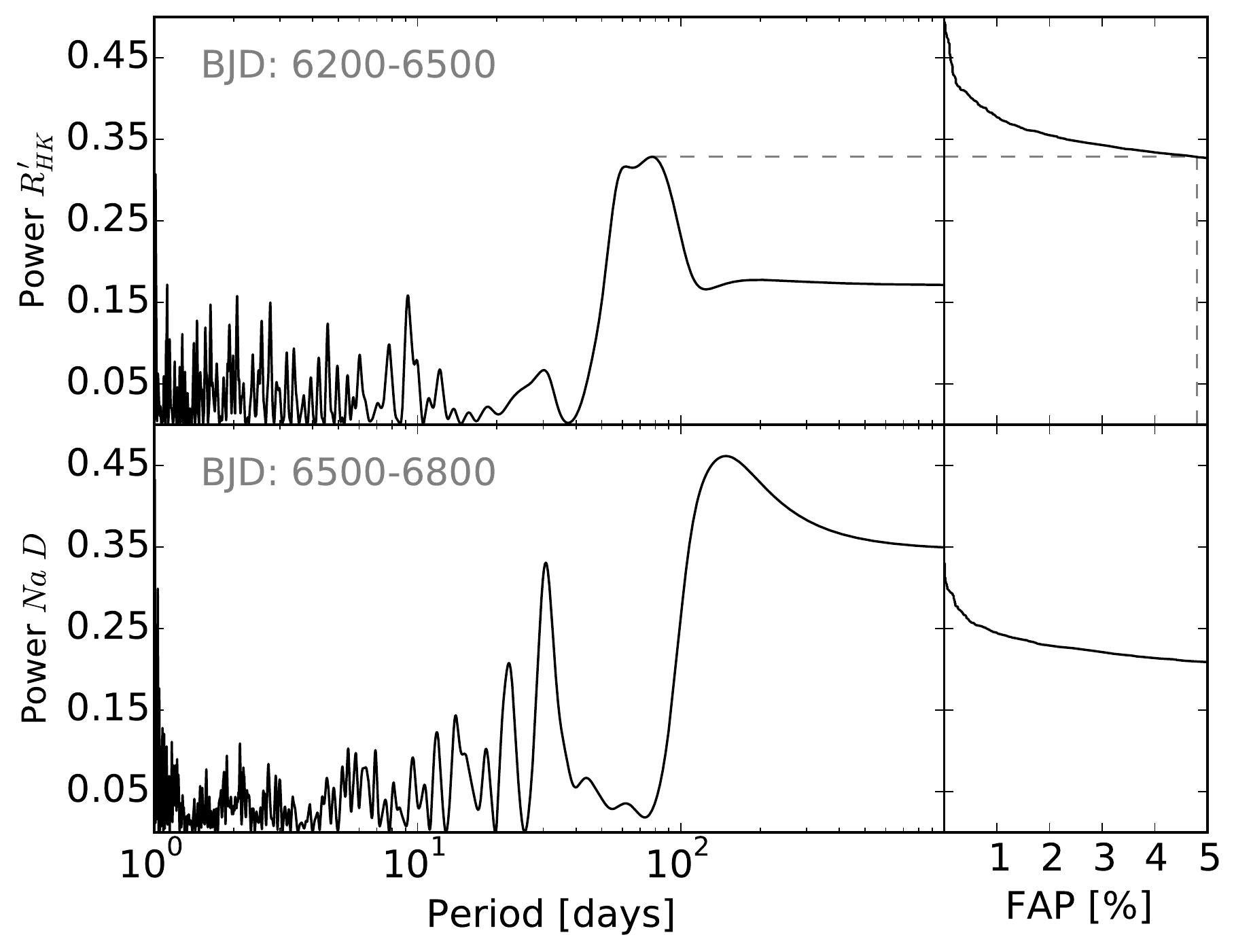}
\caption{\small \textit{Top:} GJ~273 stellar activity indicators show a long-period modulation 
($\sim$2000 days), and the periodogram of $R^\prime_{HK}$ that has a prominent peak located at 
about 400 days. \textit{Bottom:} Periodograms for seasonal $R^\prime_{HK}$ and Na D 
showing clues of the stellar rotation (estimated at about 100 days).}
\label{fig:Gl273_act}
\end{figure}

\subsubsection{Challenging the planet interpretation}

\paragraph{GJ~273 b, c}~\\

The two RV signals with the shortest periodicities (4.7 and 18.6~days) have periods that are 
very different from that estimated for the stellar rotation ($P_{Rot}\sim99$ days; see 
Table~\ref{tab:targetsProperties}), very different from that measured for the two long-period 
RV signals (which we suspect are caused by stellar activity), and very different from the 
harmonics of those periodicities ($\sim$P/2, $\sim$P/3, ...). They thus already appear as 
robust planet detections.

\begin{figure}[t]
\centering
\includegraphics[scale=0.47]{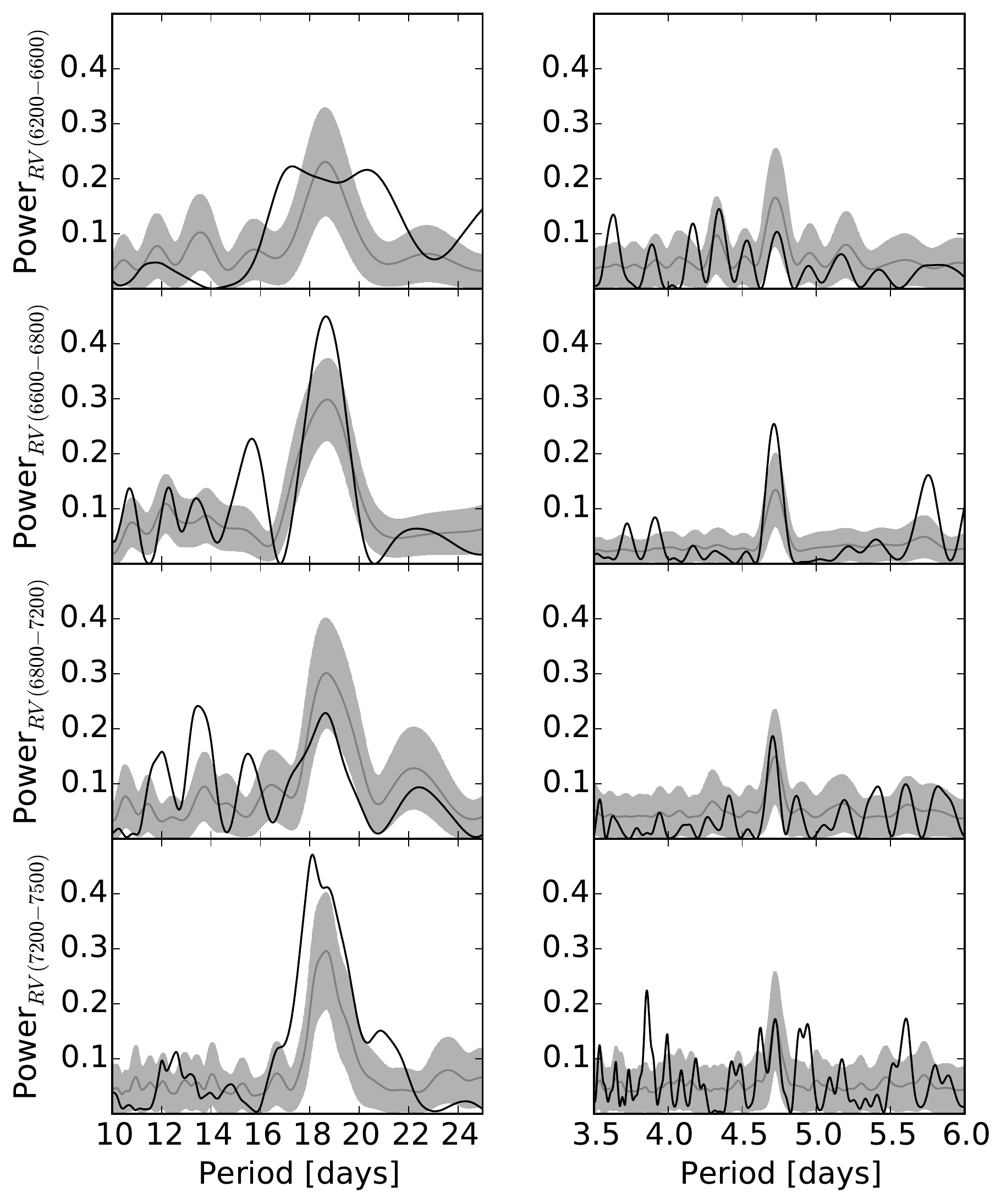}
\caption{\small Analysis for the stability of the two shorter RV signals of GJ~273. Black curves are 
the periodograms of the isolated RV signals with a periodicity of 18 days (left panel) 
and 4.7 days (right panel). The gray curves correspond to the average periodograms of 
the synthetic RVs, and the 1$\sigma$ zones are shown in light gray zones.}
\label{fig:Gl273_rv_split}
\end{figure}

We wanted to check whether the signals are coherent across different epochs. We 
divided our RV time series into four parts, with subsamples satisfying BJD-2450000 within the 
ranges 6200$-$6600, 6600$-$6800, 6800$-$7200, and 7200$-$7500. We computed the periodograms 
for each subsample. As shown in Figure~\ref{fig:Gl273_rv_split}, power excess is seen 
in almost all epochs.

As we did for GJ~3138, we needed to understand whether the signal would be detected at each epoch. We 
proceeded by making a 100 synthetic RV time series with the same sampling as our observation, 
normally distributed ($\sigma=2\ m/s$) and centered on the best Keplerian fit. The periodogram 
for each of the 100 time series was computed, from which we derived the $\pm$1 power distribution. 
Because the periodogram of the original subsamples fell inside the $\pm$1 region (gray zone in 
Fig~\ref{fig:Gl273_rv_sol}), we conclude that there is no evidence that either the 5- or 
the 18-day signal is transitory.

Overall, we accept both the 5- and 18-day signals as bona fide planet detections. 
Figure~\ref{fig:Gl273_rv_sol} shows the phase folded RVs.

\paragraph{Magnetic cycle and rotation}~\\

\begin{figure}[t]
\centering
\includegraphics[scale=0.47]{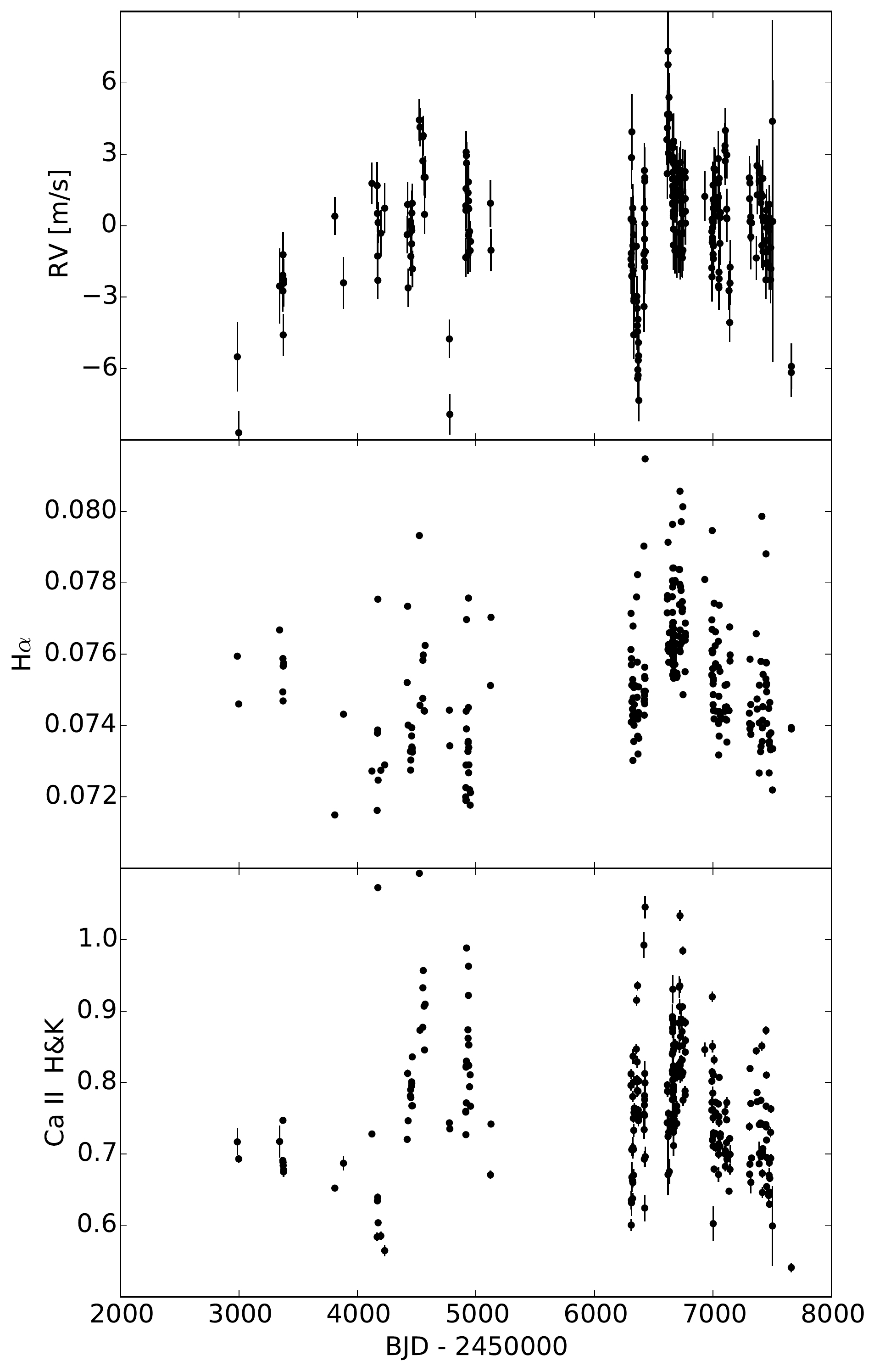}
\caption{\small GJ~273 radial velocities and activity indicator showing resemblances, suggesting that 
RVs are impacted by a stellar activity cycle.}
\label{fig:Gl273_rv_S_Ha}
\end{figure}

\begin{figure}[t]
\centering
\includegraphics[scale=0.5]{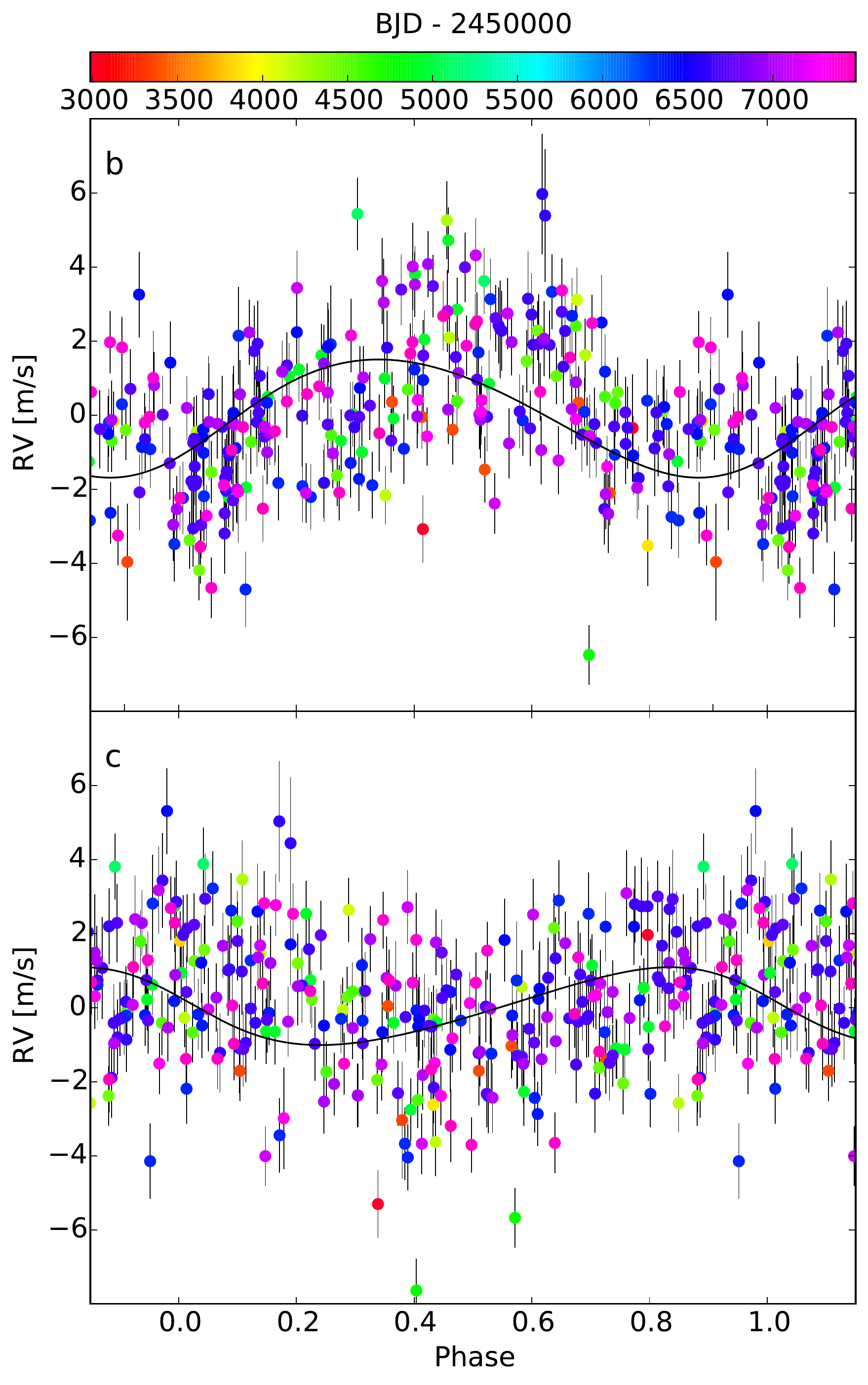}
\caption{\small GJ~273 radial velocities phase folded to 18.7 (top) and 4.7 days (bottom). 
The black curve and rainbow colors represent the Keplerian solution and BJD of observations, respectively.}
\label{fig:Gl273_rv_sol}
\end{figure}

In Figure~\ref{fig:Gl273_rv_perio} we note that the raw RV periodogram exhibits power excess 
for periodicities from 300-4000 days, with peaks at 400-500 and 2000 days. Similarly, the 
periodograms of activity indicators show significant power excess in the same range of 
periodicities (Fig.~\ref{fig:Gl273_act}). Time series for radial velocities and activity 
indicators are shown in Figure~\ref{fig:Gl273_rv_S_Ha}. The correlation and p-value are 0.18 
and 0.002 respectively. 
However, the maximum RV corrections from this correlation are of a few cm/s and are too low for a 
noticeable improvement.

Another way to correct the RV-activity dependency is to proceed as in \citet{2011A&A...535A..55D}. 
They noticed that both RVs and \ion{Ca}{ii} H\&K time series show remarkable 
resemblances. \citet{2011A&A...535A..55D} fit a Keplerian to the $R^\prime_{HK}$ and used the 
resulting parameters to fix a Keplerian model to RVs, but leaving the amplitude free. Proceeding 
this way we fix a model with P$\sim$2000~days, then the periodogram of 
RV residues shows subsequently the signals from GJ~273 b and c. The periodogram from a 
third iteration exhibits a clear peak at $\sim$100 days, the expected stellar rotation period 
(Table~\ref{tab:targetsProperties}). 
Evidence of the stellar rotation appears in the periodograms of the $R^\prime_{HK}$ and the 
Na D-index measured between BJD 6200--6500 and 6500--6800, respectively. 

The $\sim$420-day peak in the RV periodogram could be explained as the 1-year alias 
of the longer period of $\sim$2000 days seen in RVs and activity tracers 
(Figs.~\ref{fig:Gl273_rv_perio},~\ref{fig:Gl273_act},~\ref{fig:Gl273_rv_S_Ha}). The peak 
arising at $\sim$700~days in the RV residues is close to the second harmonic (P/3) of the 
magnetic cycle. Because of that, we suspect these two RV signals (P$\sim$~420, 700~days) 
are the product of a stellar activity cycle that has not been modeled well.

\subsubsection{Keplerian analysis}

GJ~273b is a super-Earth with a minimum mass of $2.89\pm0.26$~M$_\oplus$ with an orbital 
period of $18.650\pm0.006$ days. Orbiting at a distance of $0.09110\pm0.00002$ AU from its parent 
star, 
it is grazing the inner edge of the conservative habitable zone \citep{2013ApJ...765..131K}, 
but with an incident flux of 1.06$S_\oplus$ it is well within the HZ if one assumes the planet 
is surrounded by an atmosphere and accounts for GCM \citep{2016ApJ...819...84K}. 
It has an equilibrium temperature in the range 206-293 K (using albedos of 0.75 and 0, 
respectively). GJ~273c is among the less massive planets detected by radial velocities, 
with a minimum mass of $1.18\pm0.16$~M$_\oplus$. It completes one orbit in $4.7234\pm0.0004$ 
days. At the orbital distance of $0.036467$ AU it receives almost seven times 
the stellar flux received at Earth and has an equilibrium temperature within 327-462 K, 
and thus much too close to the star to be considered in the circumstellar habitable zone.

\begin{table}
\centering
\caption{\small Parameters for the Keplerian fitted to GJ~273  RVs.}
\label{tab:Gl273_orbparam}
\begin{tabular*}{\hsize}{@{\extracolsep{\fill}} l c c c}
\noalign{\smallskip}
\hline\hline
\noalign{\smallskip}
N$_{\rm Meas}$& &\multicolumn{1}{c}{ 280 }\\
\noalign{\smallskip}
$\sigma_{\rm ext}$ &[m/s]&\multicolumn{1}{c}{0.76/0.44}\\
\noalign{\smallskip}
$\sigma_{(O-C)}$&[m/s]&\multicolumn{1}{c}{1.59/1.36}\\
\noalign{\smallskip}
$\Delta V_{21}$&[m/s]&\multicolumn{1}{c}{$-1.98_{ -0.38}^{+0.40}$}\\
\noalign{\smallskip}
BJD$_{\rm ref}$&[days]&\multicolumn{1}{c}{56238.2123938802}\\
\noalign{\smallskip}
$\gamma$&[km/s]&\multicolumn{1}{c}{ $18.4086\pm0.0002$}\\
\noalign{\smallskip}
\hline
\noalign{\smallskip}
    &          & GJ~273b & GJ~273c \\
\noalign{\smallskip}
\hline
\noalign{\smallskip}
 P         &[days]                &$18.6498_{ -0.0052}^{+0.0059}$ &  $4.7234_{ -0.0004}^{+  0.0004}$\\
\noalign{\smallskip}
$K_1$      &[$ms^{-1}$]          & $1.61_{ -0.15}^{+0.15}$ & $1.06_{ -0.15}^{+ 0.15}$\\
\noalign{\smallskip}
e          &                    & $0.10_{ -0.07}^{+0.09}$&  $0.17_{ -0.12}^{+0.13}$\\
\noalign{\smallskip}
$\lambda_{0}$ at BJD$_{\rm ref}$  & [deg] & $229.6_{-5.5}^{+5.3}$&  $75.60_{ -8.4}^{+8.1}$\\
\noalign{\smallskip}
\hline
\noalign{\smallskip}
$m\, sin(i)$ &[M$_\oplus$]        &$2.89_{ -0.26}^{+0.27}$& $1.18_{ -0.16}^{+ 0.16}$\\
\noalign{\smallskip}
a& [AU] & $0.091101_{ -0.000017}^{+ 0.000019}$& $0.036467_{ -0.000002}^{+ 0.000002}$\\
\noalign{\smallskip}
$S/S_\oplus$ && 1.06 & 6.66 \\
\noalign{\smallskip}
Transit prob. & [\%] & 1.6   & 4.3\\
\noalign{\smallskip}
BJD$_{\rm Trans}$-54000.0&[days]  &$2249.295_{ -0.573}^{+ 0.489}$ & $2238.58_{ -0.19}^{+0.26}$\\

\noalign{\smallskip}
\hline

\end{tabular*}
\end{table}

\subsection{GJ~628}
\label{sec:Gl628_analysis}

\begin{figure}[t]
\centering
\includegraphics[scale=0.47]{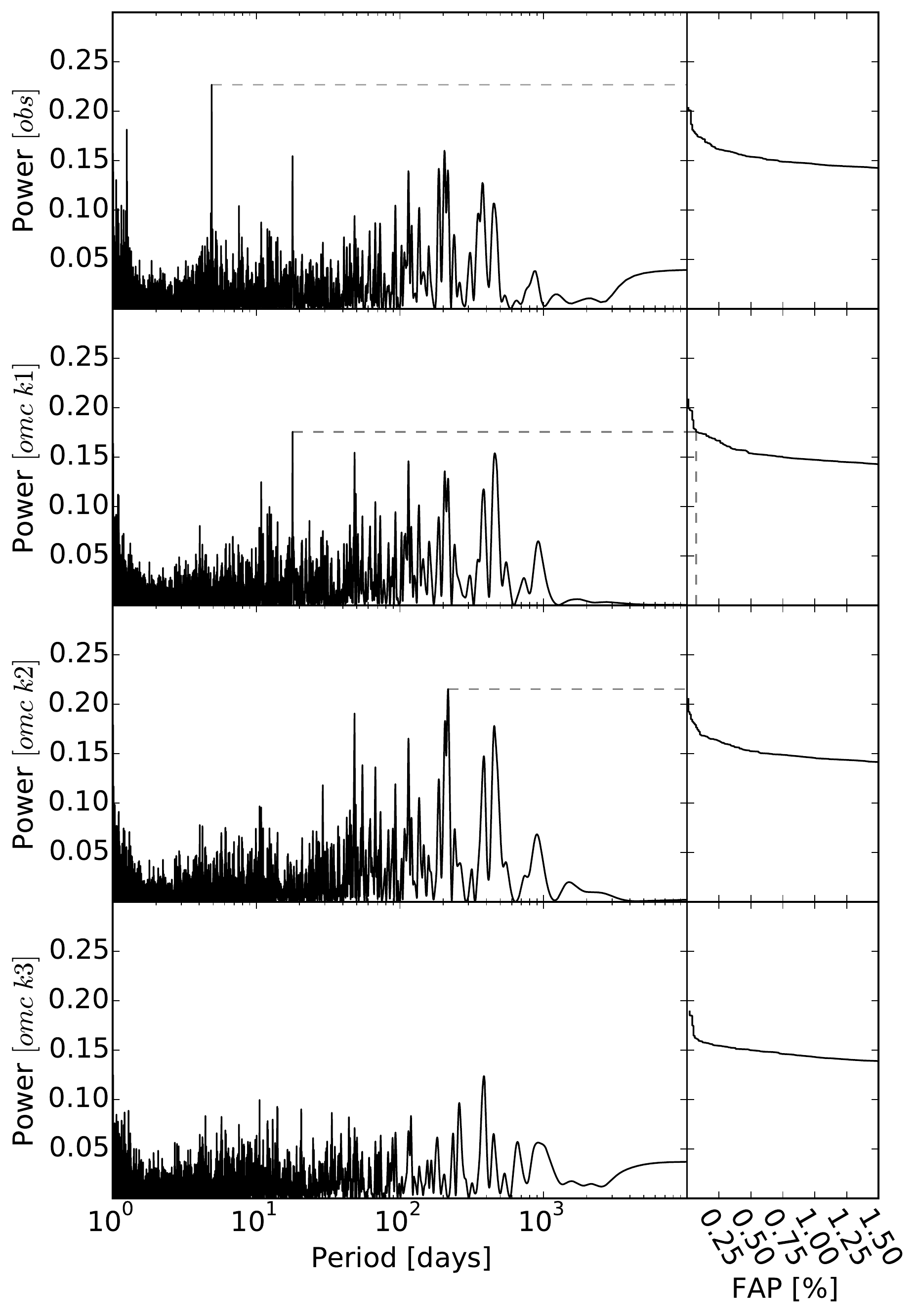}
\includegraphics[scale=0.47]{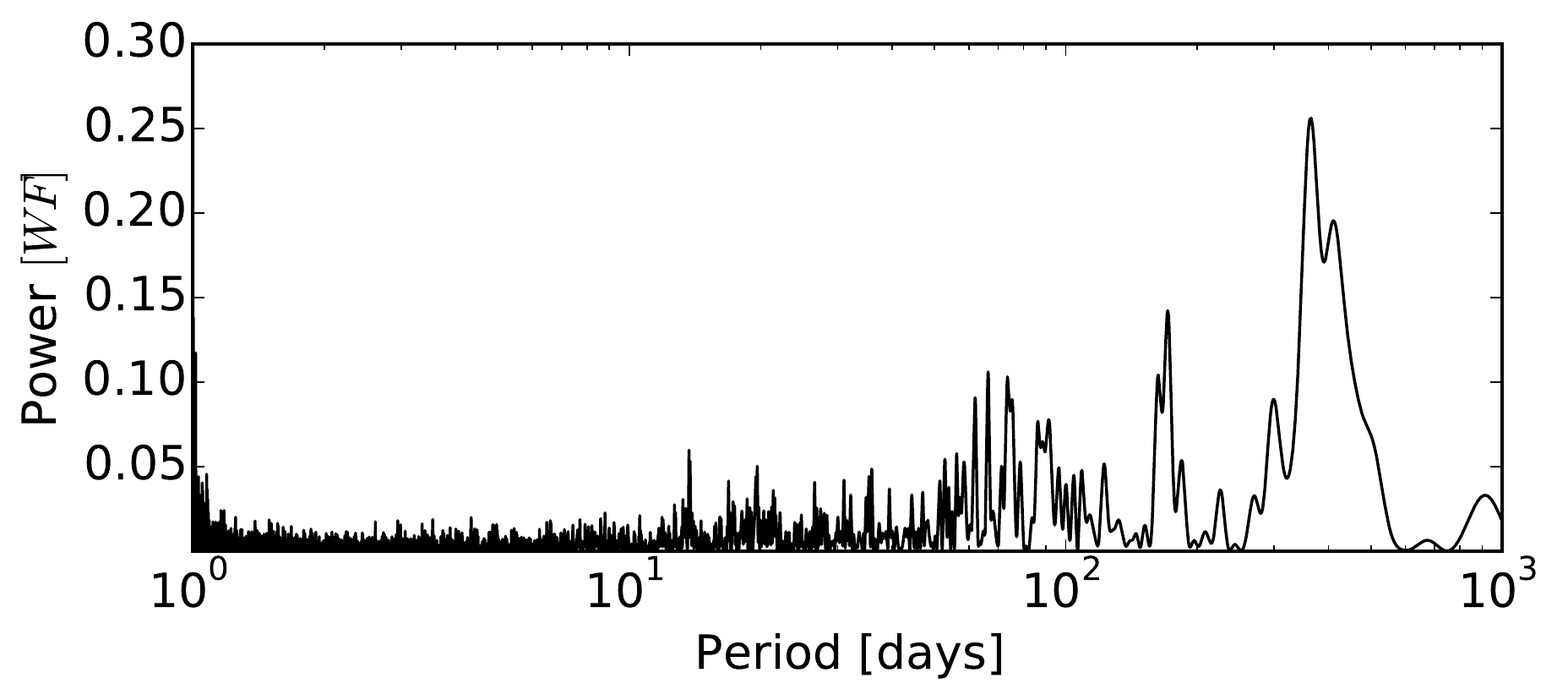}
\caption{\small \textit{Top panel:} GJ~628 periodograms of RVs. The first row shows the periodogram 
of raw RVs and the other  rows show the periodograms of RV 
residues after subtracting the Keplerian adjusted to the periodicity marked with horizontal 
shaded line. \textit{Bottom panel:} Window function showing peaks at 1 year, 0.5 year, 
66 days, 73 days, and 62 days.}
\label{fig:Gl628_rv_perio}
\end{figure}

\begin{figure}[t]
\centering
\includegraphics[scale=0.47]{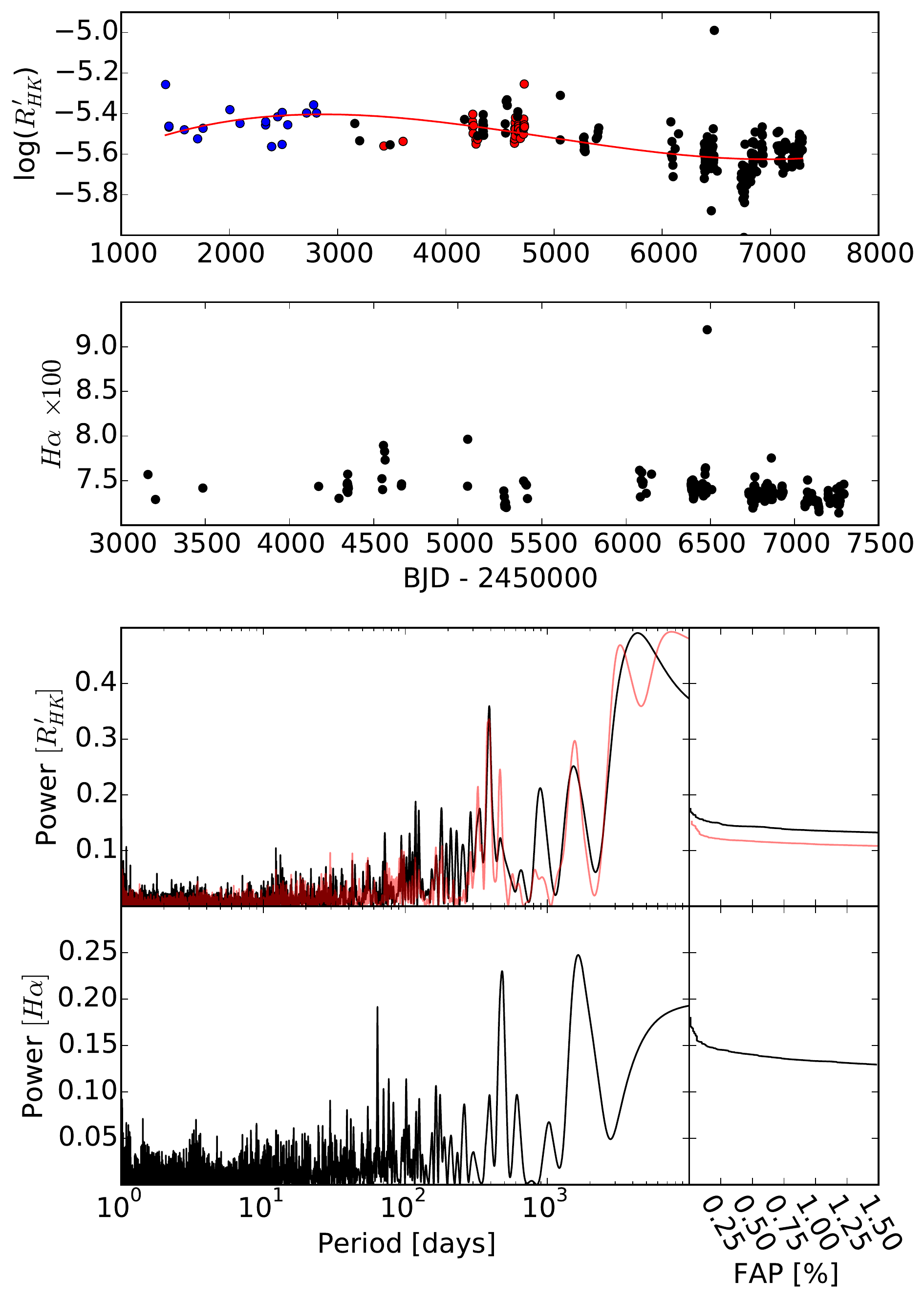}
\caption{\small \textit{Top panel:} GJ~628 measures of the $\log(R^\prime_{HK})$ and the $H\alpha$ activity 
indices against BJD, where the presence of an activity cycle becomes evident: 
the red curve in the top row helps to guide the eyes. 
The black points represent our HARPS measurements, while the blue 
and red points depict measurements from \citet{2004ApJS..152..261W} and \citet{2010ApJ...725..875I}, 
respectively, acquired with HIRES. A flare event occurs at BJD 2456481.6. \textit{Bottom panel:} 
 $R^\prime_{HK}$ and $H\alpha$ periodograms. The $R^\prime_{HK}$ periodogram for HARPS+HIRES 
data (red curve) shows the activity cycle of $\geqslant$7000 days and a peak at about 400 days 
compatible with its yearly alias. Periodogram of the only HARPS $R^\prime_{HK}$ (black curve) shows power excess at about 120 days while the one for $H\alpha$ shows a clear peak at 64 days. }
\label{fig:Gl628_act}
\end{figure}

\begin{figure}[t]
\centering
\includegraphics[scale=0.47]{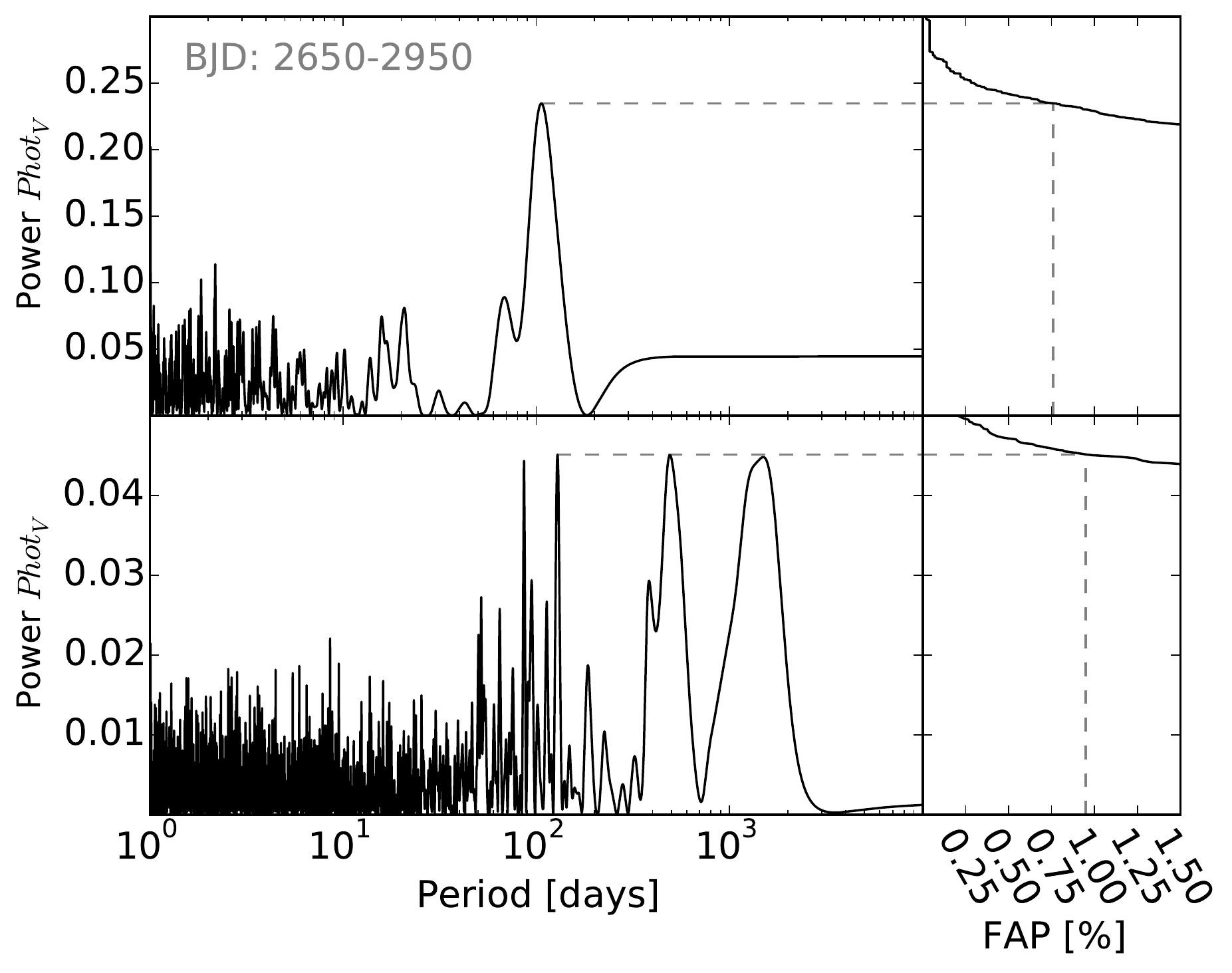}
\caption{\small Periodogram of the photometry for GJ~628. The first row shows the periodogram using data 
for BJD between 2452650 and 2452950, where we see a clear peak associated to the stellar rotation. 
The periodogram using the whole data set is shown in the second row.}
\label{fig:Gl628_phot}
\end{figure}

Between June 2, 2004, and September 28, 2015, we collected 190 spectra 
of GJ~628,  148 of which are publicly available and were previously analyzed by 
\citet{2016ApJ...817L..20W}. Three of the measurements show a difference between GJ~628 
systemic and the Moon RVs $<$1 $kms^{-1}$ and the angle between the objects $<$25$^\circ$. 
We considered these measures (BJD=2454661.6, 2454662.6, and 2454663.7) as contaminated 
by the Moon and they were rejected after the subsequent analysis. As we  show below, we came 
to the same conclusions for the RV signals at shorter periodicity (4.9 and 17.9 days), 
but a different conclusion for the periodicity at 67 days. Here we detect a third RV signal 
at a longer periodicity and identify a stellar activity cycle.

GJ~628 RVs have a dispersion of $\sigma_{(O-C)}=2.72\ ms^{-1}$ and a mean internal 
error of $\sigma_i=1.02\ ms^{-1}$. The dispersion excess comes from several periodic 
variations as is shown in Fig.~\ref{fig:Gl628_rv_perio}. The periodogram of raw RVs 
exhibits two narrow spikes at 4.9 and 17.9 days, with FAP $<10^{-2}\%$ and 0.46\%, 
respectively, and a third one at about 205 days with a 0.29\% FAP. After subtracting 
a two-Keplerian model from the 4.9- and 17.9-day periods, the periodogram of resulting 
residues (third row in Fig~\ref{fig:Gl628_rv_perio}) shows several peaks located at 
217, 48, 454, 115 days, detected with FAPs $<0.01\%$, 0.02\%, 0.07\%, and 0.17\%, 
respectively. We added a third Keplerian to each period with power excess 
(one at a time) and found that the solution including the 217-day signal results in 
significantly lower $\chi^2_\nu$ and BIC (see Table~\ref{tab:model_stats}). We note 
that the 67-day RV signature discussed in \citeauthor{2016ApJ...817L..20W} is not 
significant here, as the weak peak located at this periodicity has a 2.5\% FAP. 
After including the 217-day signal in our model the highest peak in the periodogram of RVs 
residues (P$\sim$380 days) has a FAP of 6.3\%, and therefore no more signatures are detected.

\paragraph{Stellar activity and orbital solution}~\\
\label{sec:Gl628_activity}

The \ion{Ca}{ii} H\&K chromospheric emission reveals that GJ~628 presents an activity 
cycle (Fig.~\ref{fig:Gl628_act}). We combined the literature \ion{Ca}{ii} H\&K measurements 
\citep{2004ApJS..152..261W, 2010ApJ...725..875I} to better understand the cycle identified 
with the HARPS data alone. Although the data time span is increased by about 4 years, the combined 
data does not cover a complete activity cycle, which is at least 19 years. 
Figure~\ref{fig:Gl628_act} 
also shows one flare event detected in both $R^\prime_{HK}$ and $H\alpha$ (BJD=2456481.55). 
Concerning the FWHM$_{CCF}$, in addition to the expected offset due to the HARPS fiber upgrade, 
this activity index -- computed through the DRS -- 
presents other unexplained offsets of about 1 $km/s$ (too large to be of stellar origin) 
and its analysis is omitted here.

\begin{figure}[t]
\centering
\includegraphics[scale=0.47]{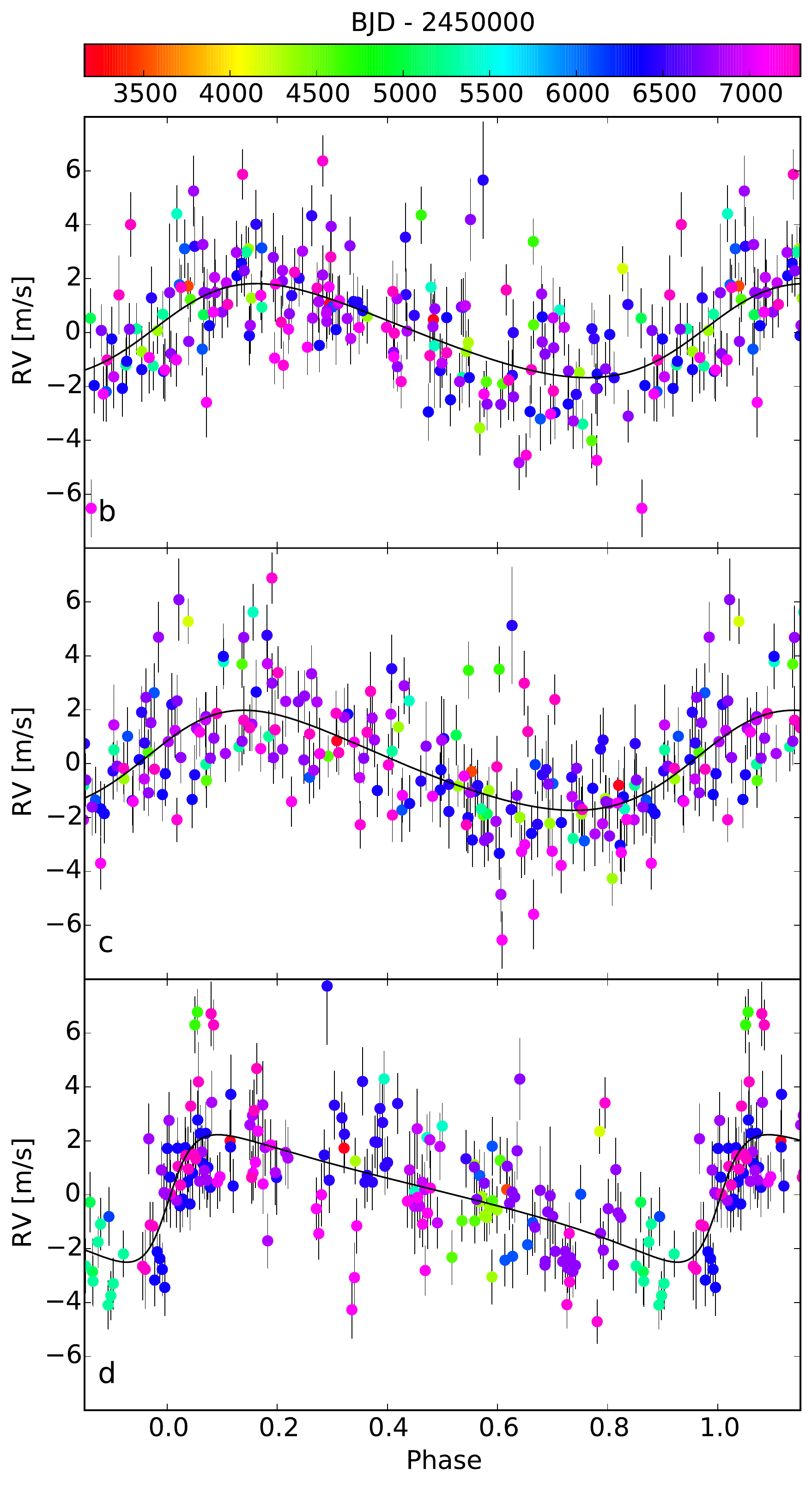}
\caption{\small Radial velocities of GJ~628 folded to 4.89 (top row), 17.87 (middle row), and 217.2 
days (bottom row). Our solution consisting of three Keplerians is shown as the black solid curve, 
while the BJD is  color-coded.}
\label{fig:Gl628_sol}
\end{figure}

The stellar rotation period of GJ~628 is estimated to be 93 days (Table~\ref{tab:targetsProperties}) 
and the $H\alpha$ periodogram (Fig.~\ref{fig:Gl628_act}) presents a significant power excess 
at 64 days; however, we cannot assert that this periodicity is linked to a yearly alias of the 
stellar rotation. 
We do not identify any periodicity linked to the stellar rotation either from the seasonal 
$H\alpha$ or from the seasonal $R^\prime_{HK}$ measurements. 
The peak at about 64 days in the $H\alpha$ periodogram and the 67 days RV signal reported in 
\citet{2016ApJ...817L..20W} are worryingly close. 
The HARPS only $R^\prime_{HK}$ periodogram presents a power excess at 120 days that 
can be explained as the yearly alias of the stellar rotation period. Considering the possible 
aliases, and unlike 
\citet{2016ApJ...817L..20W}, we do not regard the 67-day signal as a planet detection. 
The ASAS photometry, which consists of 600 points 
spread over 8.7 years, shows hints of the stellar rotation. The periodogram for the photometry 
satisfying BJD between 2452650 and 2452950 shows a strong peak at about 100 days, and the entire 
data set shows two peaks  near this periodicity: one peak with 0.95\% FAP at 129 days and 
another with a FAP slightly greater at about 86 days.

Additionally, the third strongest peak of the window function data set is located at the 
66-day periodicity (bottom panel in Fig.~\ref{fig:Gl628_rv_perio}), hence we cannot exclude 
that the 67-day signal seen in RVs in \citeauthor{2016ApJ...817L..20W} and $H\alpha$ are 
generated by the data temporal sampling itself. In the same figure we note the strong 
peak located at one year, showing that one-year aliases are highly expected.

Stellar differential rotation could be inferred from the APT photometry (spanning 6 years) 
presented in \citet{2016arXiv161209324K} plus the ASAS photometry. Minimum and maximum periods 
from the photometry are $P_{min}=84.2\pm3.0$ and $P_{min}=106.3\pm7.3$. Well-determined 
periodicities from the APT photometry shows no evidence of $P_{Rot.} \ge 120$ days, and we 
interpret the 120-day and 129-day peaks in Figures~\ref{fig:Gl628_act} and~\ref{fig:Gl628_phot}, 
respectively, as aliases of the true stellar rotation period.

\begin{table}[t]
\centering
\caption{\small Parameters for the three Keplerians fitted to GJ~628 RVs.}
\label{tab:Gl628_k3}
\begin{tabular*}{\hsize}{@{\extracolsep{\fill}} l  l  c  c c}
\noalign{\smallskip}
\hline\hline
\noalign{\smallskip}
N$_{\rm Meas}$&&\multicolumn{2}{c}{187}\\
$\sigma_{\rm ext}$ &[m/s]&\multicolumn{2}{c}{0.86/ 2.01}\\
\noalign{\smallskip}
$\sigma_{(O-C)}$&[m/s]&\multicolumn{2}{c}{1.71/ 2.34}\\
\noalign{\smallskip}
$\Delta V_{21}$&[m/s]&\multicolumn{2}{c}{0.5093$_{ -0.5267}^{+  0.5348}$}\\
\noalign{\smallskip}
BJD$_{\rm ref}$&[days]&\multicolumn{2}{c}{56309.9363933752}\\
\noalign{\smallskip}
$\gamma$&[km/s]&\multicolumn{2}{c}{-21.03738$_{ -0.00015}^{+  0.00015}$}\\
\noalign{\smallskip}
\hline
\noalign{\smallskip}
    &          & GJ~628b & GJ~628c& GJ~628d \\
\noalign{\smallskip}
\hline
\noalign{\smallskip}
 P         &[days]                & 4.8869$_{ -0.0005}^{+  0.0005}$ & 17.8719$_{ -0.0059}^{+  0.0059}$ & 217.21$_{ -0.52}^{+  0.55}$ \\
\noalign{\smallskip}
$K_1$      &[$ms^{-1}$]         &1.67$_{ -0.19}^{+  0.20}$  & 1.92$_{ -0.19}^{+  0.19}$ & 2.23$_{ -0.29}^{+  0.31}$ \\
\noalign{\smallskip}
e          &                   & 0.15$_{ -0.10}^{+  0.13}$ & 0.11$_{ -0.07}^{+  0.10}$  &  0.55$_{ -0.09}^{+  0.08}$  \\
\noalign{\smallskip}
$\lambda_{0}$ at BJD$_{\rm ref}$  & [deg]& 333.6$_{ -6.9}^{+  7.0}$ & 141.5$_{ -5.9}^{+  5.9}$ & 129.0$_{ -7.8}^{+  7.7}$ \\
\noalign{\smallskip}
\noalign{\smallskip}
\hline
\noalign{\smallskip}
$m\, sin(i)$ &[M$_\oplus$]       & 1.91$_{ -0.25}^{+  0.26}$ & 3.41$_{ -0.41}^{+  0.43}$ & 7.70$_{ -1.06}^{+  1.12}$\\
\noalign{\smallskip}
a& [AU]&  0.0375$_{ -0.0013}^{+  0.0012}$ & 0.0890$_{ -0.0031}^{+  0.0029}$ & 0.470$_{ -0.017}^{+  0.015}$ \\
\noalign{\smallskip}
$S/S_\oplus$ && 7.34 & 1.30 & 0.06 \\
\noalign{\smallskip}
Transit prob. & [\%] & 3.3 & 1.4& 0.2\\
\noalign{\smallskip}
 BJD$_{\rm Trans}$-54000&[days]  & 2311.498$_{ -0.204}^{+  0.191}$  & 2325.333$_{ -0.597}^{+  0.584}$ & 2501.7$_{-16.6}^{+ 16.6}$\\

\noalign{\smallskip}
\hline

\end{tabular*}
\end{table}

\subsubsection{Keplerian analysis}

When adjusting a Keplerian to the 48-day periodicity (second highest peak in the RV$_{omc\ k2}$ 
periodogram) the resulting orbit eccentricity is relatively high (e$\sim$0.7). Then, the 
periodogram of the residues shows the 217-day peak with high significance. Regarding the 
stability of the orbits, a solution including this periodicity is uncomfortable because it 
crosses the orbits with 17.9- and 217-day periodicities, if coplanar orbits are assumed.

The more conservative solution, and that preferred by both minimum $\chi^2$ and $BIC$, consists 
in three Keplerians adjusting the 4.9-, 17.9-, and 217-day periodicities.  The orbital parameters for 
such a solution are listed in Table~\ref{tab:Gl628_k3}, and Fig.~\ref{fig:Gl628_sol} depicts 
the phase folded RVs with the model. Gl~628c falls outside the conservative HZ 
\citep{2013ApJ...765..131K}; however, 
when considering the GCM in synchronous rotating planets \citep{2016ApJ...819...84K}, 
the incident radiation means  that the planet lies within the circumstellar HZ of its parent 
star.

\subsection{GJ~3293}
\label{sec:GJ3293_analysis}

\begin{figure}[t]
\centering
\includegraphics[scale=0.47]{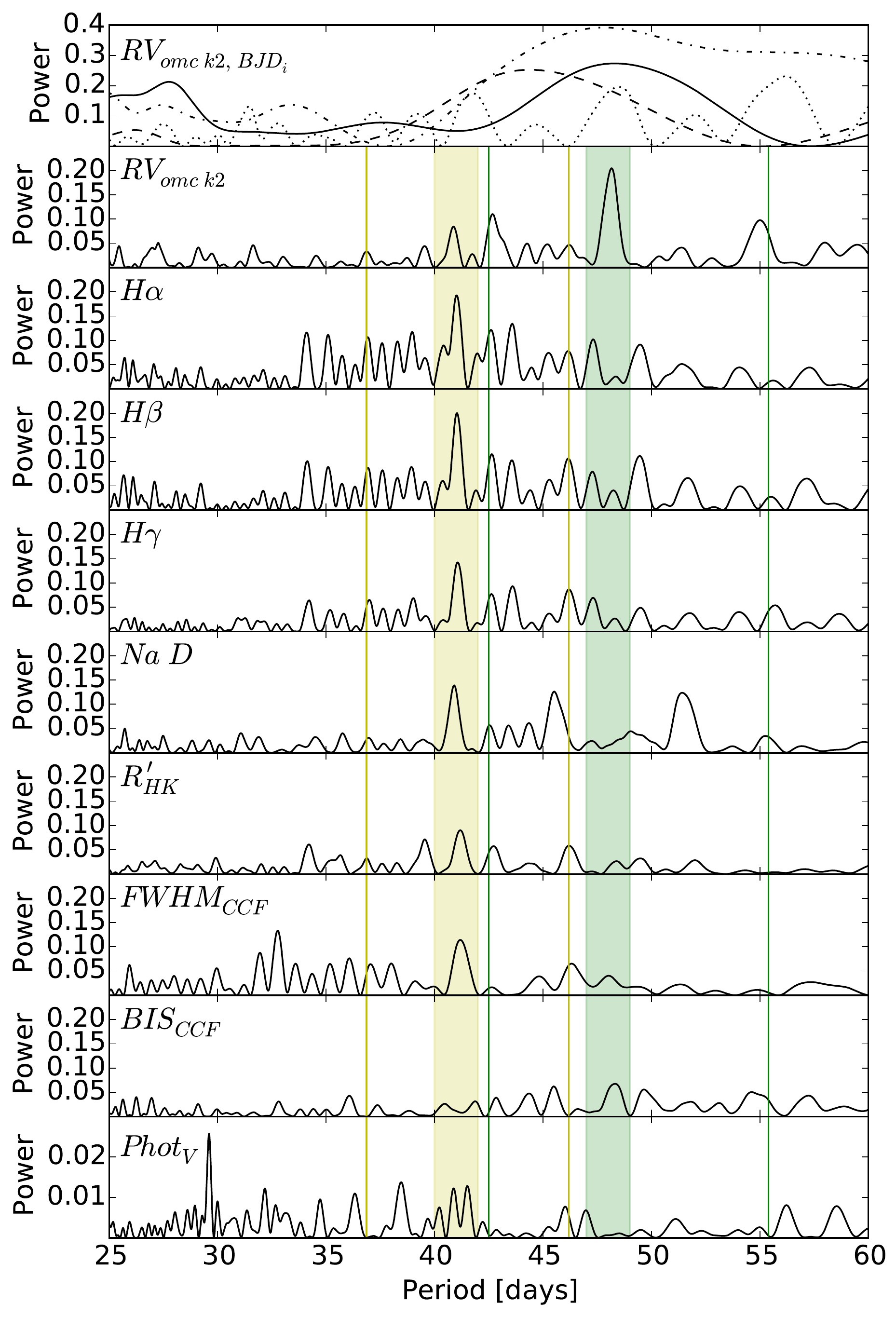}
\caption{\small From top to bottom: The first row shows the periodogram of $RV_{omc\ k2}$ for 
different RV subsets where BJD is within 4790-5750 (dotted), 5750-6100 (solid), 6100-6350 
(dash-dotted), and 6900-7150 (dashed). There is power excess at about 48 days for all the 
subsets. The second row shows the $RV_{omc\ k2}$ periodogram for the entire sample. The third 
to ninth rows show the periodograms of different stellar activity indicators. The 
green and yellow shaded areas represent the planetary and stellar signals, respectively. 
The same color-code is used for the thin vertical lines depicting the yearly aliases. 
The periodicities of the two signals are clearly separated.}
\label{fig:GJ3293_rvomck2_act}
\end{figure}

\citet{2015A&A...575A.119A} reported the detection of two Neptune-like companions 
orbiting GJ~3293 with periodicities of 30.6 and 124 days, and a possible super-Earth 
orbiting with a periodicity of  48.1 days. To shed light on the nature of the 48.1-day 
signal, we gathered 61 new RVs from October 13, 2014, to April 4, 2015. 
In this work we measure a RV dispersion of $\sigma_{(O-C)}=7.73\ ms^{-1}$ and a mean internal 
error of $\sigma_i=2.33\ ms^{-1}$.

\begin{figure}[t]
\centering
\includegraphics[scale=0.47]{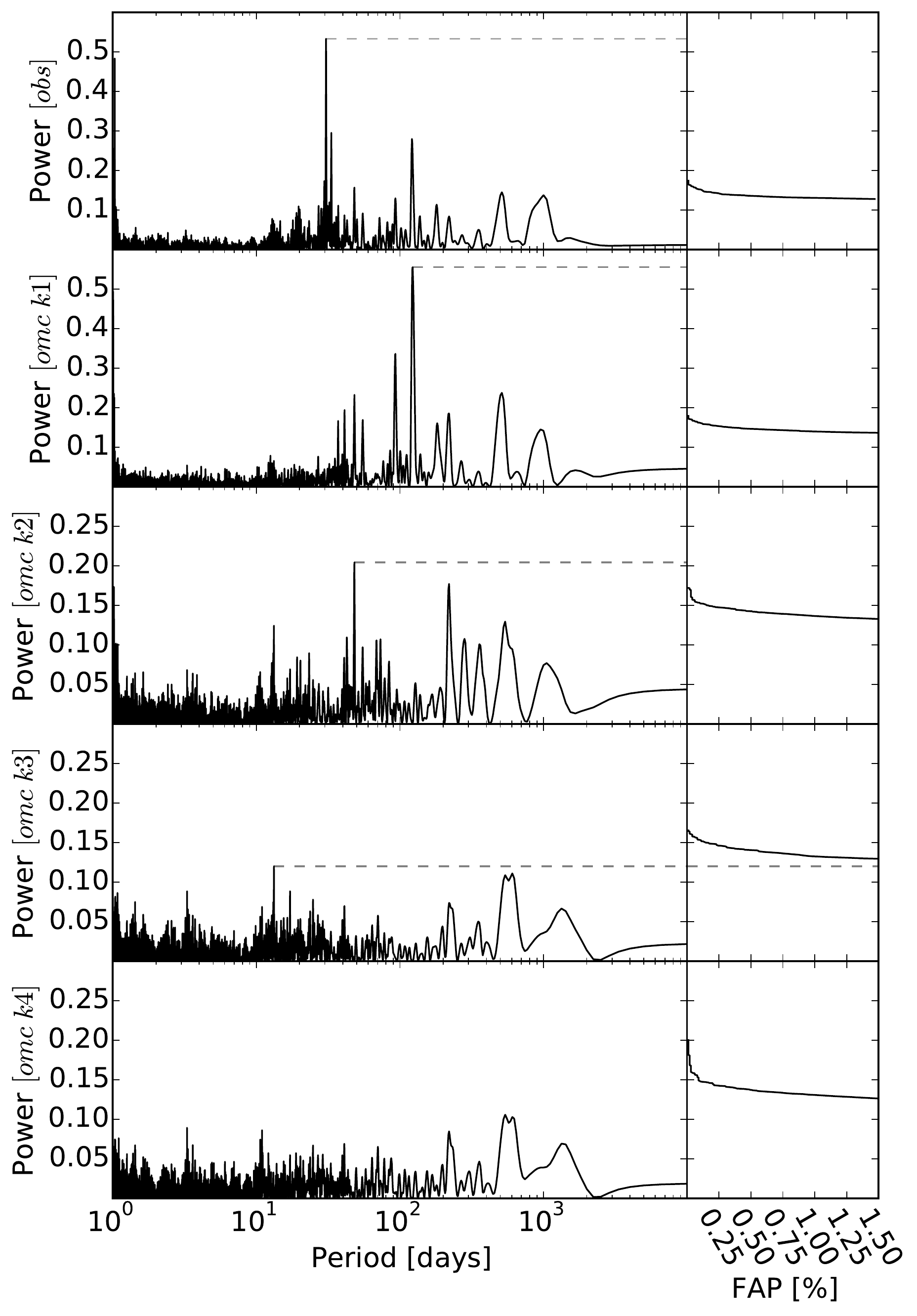}
\includegraphics[scale=0.47]{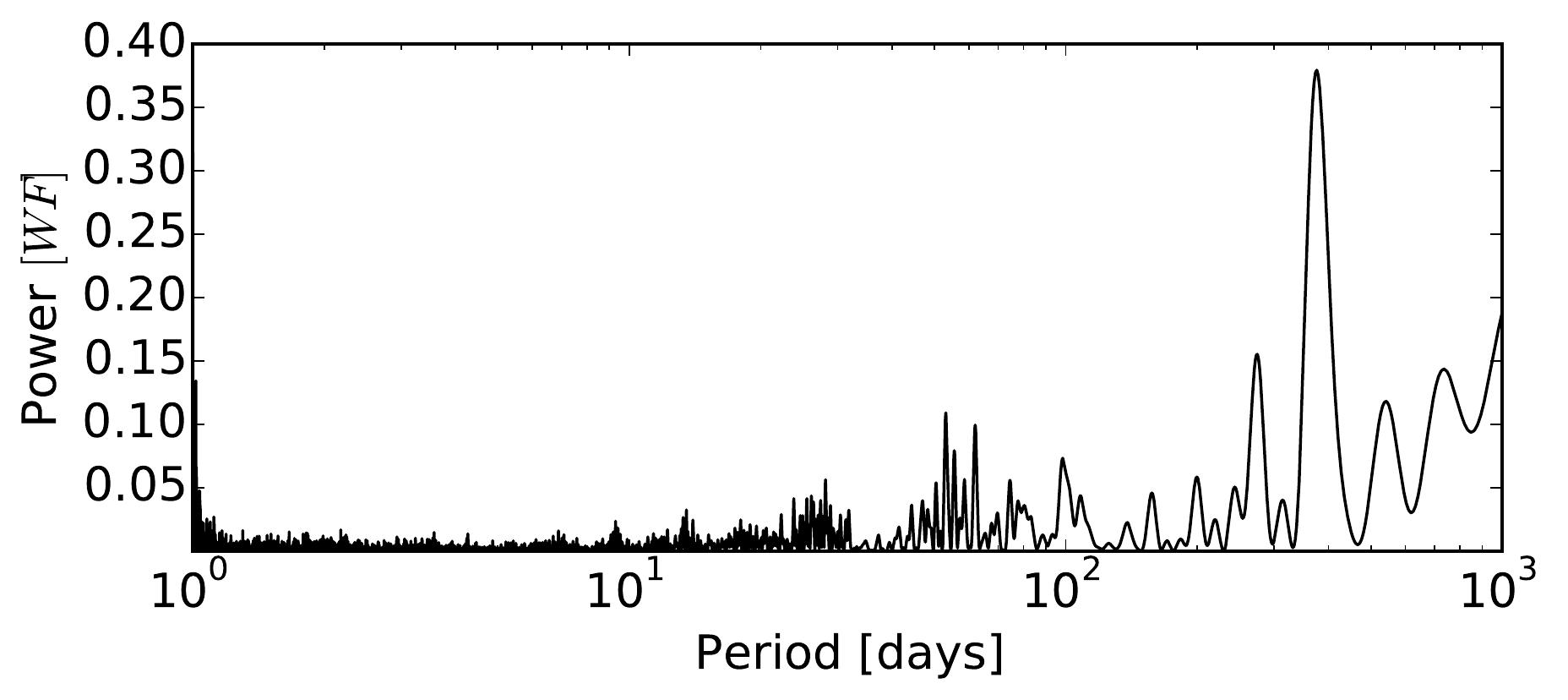}
\caption{\small \textit{Top:} Periodograms of GJ~3293 of RVs showing the successive signals. Since 
\citet[][see their Fig. 4]{2015A&A...575A.119A}, the RV$_{omc\ k3}$ periodogram shows 
an increase of  13.3 days by adding new RV measurements. 
\textit{Bottom:} The window function exhibits peaks at 1 year, 275 days, 53 days, 
and 62 days.}
\label{fig:GJ3293_rv_perio}
\end{figure}

The \citeauthor{2015A&A...575A.119A} $H\alpha$ analysis suggested that the stellar rotation 
period of GJ~3293 is 41 days, and from RVs they argued about an uncertain nature of the 
48.1-day signal  because 48.1 days is close to the stellar rotation and because of the 
lack of such a signal in the 2010-2011 data set. By using several chromospheric activity 
indicators ($H\beta, H\gamma,\ Na\ D,\ R^\prime_{HK},\ FWHM_{CCF}$) and improving the $H\alpha$ 
detection, we confirm that the stellar rotation is   41.0 days. The ASAS photometry 
(599 points spanning 8.7 yr) does not show periodic variability compatible with stellar rotation. 
However, the Moon's synodic period is evident in the photometry periodogram. The synodic periodicity 
appears with 2.3\% FAP for the photometry acquired between BJD 1500 and 2020. 
Neither a 41.0-day signal nor its yearly aliases at 36.9 and 46.2 days appears to affect 
the 48.1-day RV signature in the periodogram. Currently, the 41-day 
period of the stellar rotation and the 48.1-day period seen in RVs seem distinct enough 
to interpret the 48.1-day signal as being due to a planetary companion. Moreover, after subtracting 
 both the 30.6- and 124-day signals and after analyzing the RV residues ($RV_{omc\ k2}$), 
we found that the 48.1-day periodicity is temporally stable when splitting the sample in 
a more optimal way 
(target seasonal visibility instead of calendar year). The four RV subsets used for the 
present analysis satisfies BJD between 4790 and 5750, 5750 and 6100, 6100 and 6350, and  6900 
and 7150. Figure~\ref{fig:GJ3293_rvomck2_act} shows such a stability of the 48.1-day 
signal and the periodograms of a set of activity indicators. This planet is within the 
habitable zone of its parent star, with an equilibrium temperature between 171 K and 241 K.

\begin{figure}[t]
\centering
\includegraphics[scale=0.47]{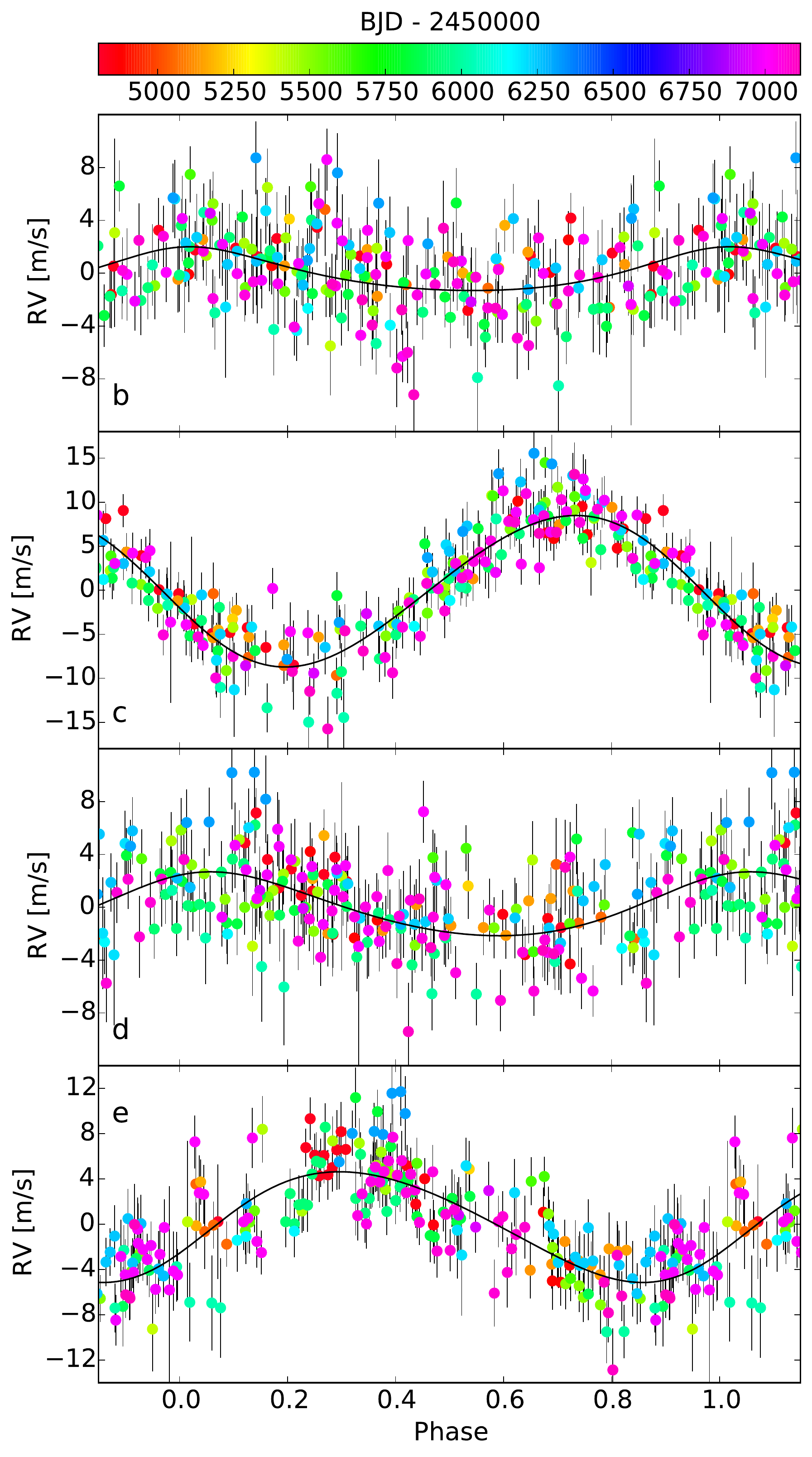}
\caption{\small From top to bottom: Radial velocities of GJ~3293 folded to 13.3 (first row), 30.6 
(second row), 48.1 (third row), and 122.6 days (fourth row). The Keplerian solution 
is represented by the solid black curve and the BJD of observations by the rainbow color-code.}
\label{fig:GJ3293_sol}
\end{figure}

When modeling the 30.6-, 48.1-, and 124-day RV periodicities by three Keplerians, a 
fourth signal appears in the residues. The periodogram (Fig.~\ref{fig:GJ3293_rv_perio}) 
shows power excess at a periodicity of 13.3 days. Such a peak was also present in 
\citet[][see their Fig. 4]{2015A&A...575A.119A}, but without enough significance 
(1$\sigma$). Here the corresponding peak is clearly visible despite its weakness, 
and the FAP decreases to 3.4$\%$ (2$\sigma$). Furthermore, a model consisting of 
four Keplerians is significantly favored over the three-Keplerian model (see the 
BIC difference in Table~\ref{tab:model_stats}). Accordingly, our solution 
for GJ~3293 consists of two super-Earths (P=13.3, 48.1 days) and two Neptune-mass 
planets (P=30.6, 122.6 days) with minimum masses of 3.3~M$_\oplus$, 7.6~M$_\oplus$ and 
23.5~M$_\oplus$, 21.1~M$_\oplus$. At their separation from GJ~3293, the Neptune-mass 
planet b and the super-Earth planet b are within the circumstellar habitable zone. 
Figure~\ref{fig:GJ3293_sol} shows the phase folded RVs 
with the four Keplerians, and Table~\ref{tab:GJ3293_k4} gives the orbital parameters.

\begin{table}[]
\centering
\caption{\small Parameters for the four Keplerians fitted to GJ~3293 RVs.}
\label{tab:GJ3293_k4}
\begin{tabular*}{\hsize}{@{\extracolsep{\fill}} l l c c}
\noalign{\smallskip}
\hline\hline
\noalign{\smallskip}
N$_{\rm Meas}$&&\multicolumn{2}{c}{207}\\
$\sigma_{\rm ext}$ &[m/s]&\multicolumn{2}{c}{1.02}\\
\noalign{\smallskip}
$\sigma_{(O-C)}$&[m/s]&\multicolumn{2}{c}{ 2.78}\\
\noalign{\smallskip}
BJD$_{\rm ref}$&[days]&\multicolumn{2}{c}{56023.1731538393}\\
\noalign{\smallskip}
$\gamma$&[km/s]&\multicolumn{2}{c}{13.29558$\pm$0.00023}\\
\noalign{\smallskip}
\hline
\noalign{\smallskip}
    &          & GJ~3293e & GJ~3293b\\
\noalign{\smallskip}
\hline
\noalign{\smallskip}
 P         &[days]            & 13.2543$_{ -0.0104}^{+  0.0078}$  & 30.5987$_{ -0.0084}^{+  0.0083}$  \\
\noalign{\smallskip}
$K_1$      &[$ms^{-1}$]     &  1.658$_{ -0.321}^{+  0.328}$  &  8.603$_{ -0.323}^{+  0.320}$ \\
\noalign{\smallskip}
e          &                & 0.21$_{ -0.14}^{+  0.20}$  &  0.06$_{ -0.04}^{+  0.04}$   \\
\noalign{\smallskip}
$\lambda_{0}$ at BJD$_{\rm ref}$  & [deg]& 349.5$_{-12.5}^{+ 11.9}$ &  102.7$_{ -2.0}^{+  2.0}$ \\
\noalign{\smallskip}
\hline
\noalign{\smallskip}
$m\, sin(i)$ &[M$_\oplus$]   &  3.28$_{ -0.64}^{+  0.64}$   & 23.54$_{ -0.89}^{+  0.88}$ \\
\noalign{\smallskip}
a& [AU]&  0.08208$_{ -0.00004}^{+  0.00003}$  & 0.14339$_{ -0.00003}^{+  0.00003}$\\
\noalign{\smallskip}
$S/S_\oplus$ && 3.34 & 1.07 \\
\noalign{\smallskip}
Transit prob. & [\%] & 2.3 & 1.3\\
\noalign{\smallskip}
 BJD$_{\rm Trans}$-54000&[days]  & 2026.847$_{ -0.664}^{+  0.538}$ &  2052.320$_{ -0.436}^{+  0.384}$  \\

\noalign{\smallskip}
\hline
\noalign{\smallskip}
    &          & GJ~3293d & GJ~3293c\\
\noalign{\smallskip}
\hline
\noalign{\smallskip}
 P         &[days]           & 48.1345$_{ -0.0661}^{+  0.0628}$  &  122.6196$_{ -0.2371}^{+  0.2429}$ \\
\noalign{\smallskip}
$K_1$      &[$ms^{-1}$]      & 2.420$_{ -0.334}^{+  0.338}$  &  4.891$_{ -0.295}^{+  0.300}$  \\
\noalign{\smallskip}
e          &         &      0.12$_{ -0.09}^{+  0.13}$   &  0.11$_{ -0.08}^{+  0.10}$  \\
\noalign{\smallskip}
$\lambda_{0}$ at BJD$_{\rm ref}$  & [deg]& 333.7$_{ -7.6}^{+  7.5}$ & 242.0$_{ -4.1}^{+  4.0}$  \\
\noalign{\smallskip}
\hline
\noalign{\smallskip}
  $m\, sin(i)$ &[M$_\oplus$]    & 7.60$_{ -1.05}^{+  1.05}$  & 21.09$_{ -1.26}^{+  1.24}$ \\
\noalign{\smallskip}
a& [AU]&  0.19394$_{ -0.00018}^{+  0.00017}$ & 0.36175$_{ -0.00047}^{+  0.00048}$ \\
\noalign{\smallskip}
$S/S_\oplus$ && 0.59 & 0.17 \\
\noalign{\smallskip}
Transit prob. & [\%] &  1.0 & 0.5 \\
\noalign{\smallskip}
 BJD$_{\rm Trans}$-54000&[days]  & 2038.471$_{ -1.827}^{+  1.883}$ &  2090.855$_{ -4.681}^{+  3.682}$  \\
\noalign{\smallskip}
\hline

\end{tabular*}
\end{table}

In comparison to \citet{2015A&A...575A.119A}, where a model with three Keplerians was used, GJ~3293 
RVs are modeled here with four Keplerians. This may involve changes in some of the orbital parameters. 
The only significant change (2$\sigma$)  comes from the eccentricity of the orbit with a 122.7-day 
periodicity, where we found e=0.11$\pm$0.09 instead of the 0.37$\pm$0.06 reported in 
\citeauthor{2015A&A...575A.119A} The other orbital parameters remain unchanged.

\begin{figure}[t]
\centering
\includegraphics[scale=0.4]{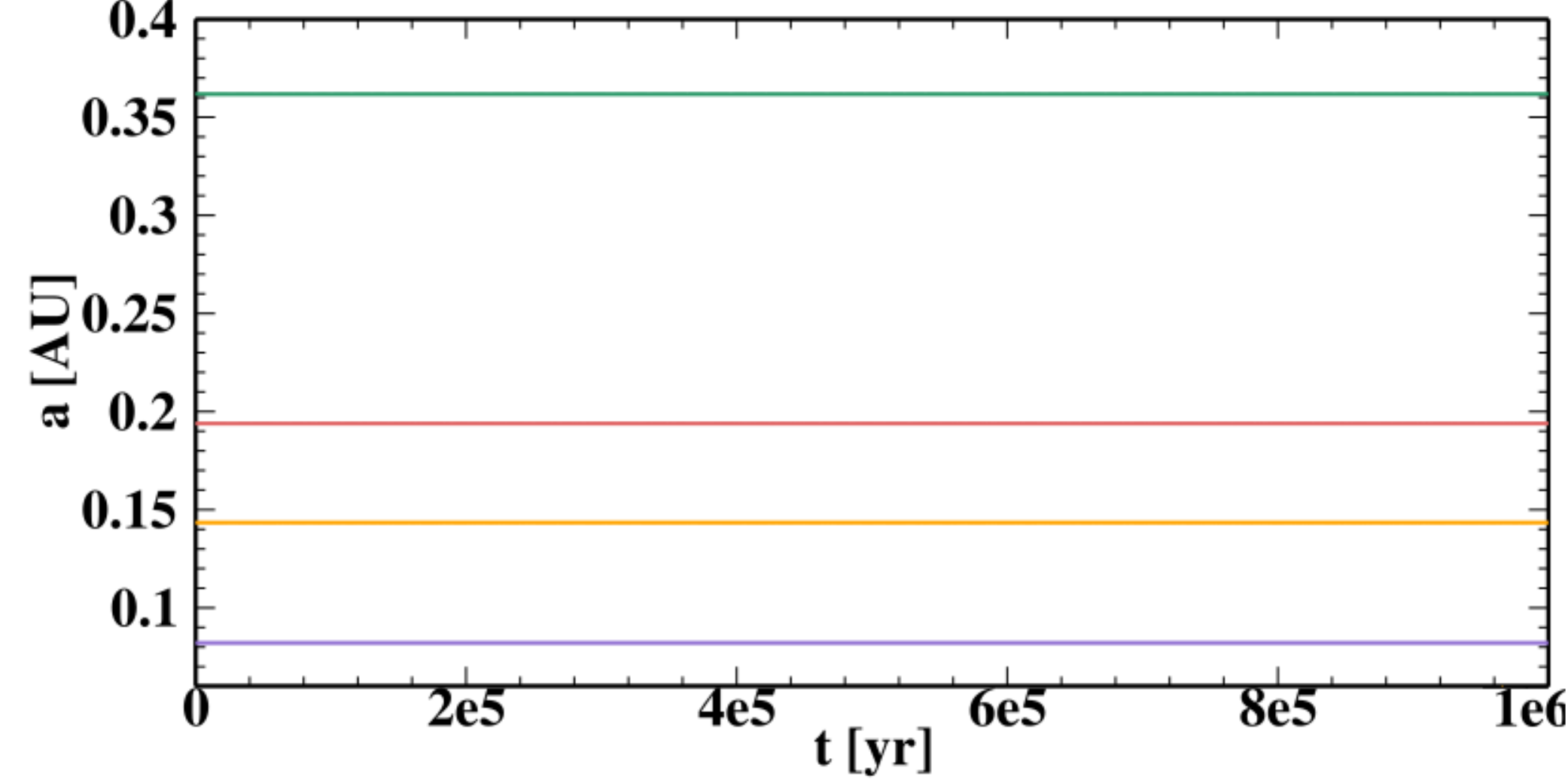}
\includegraphics[scale=0.4]{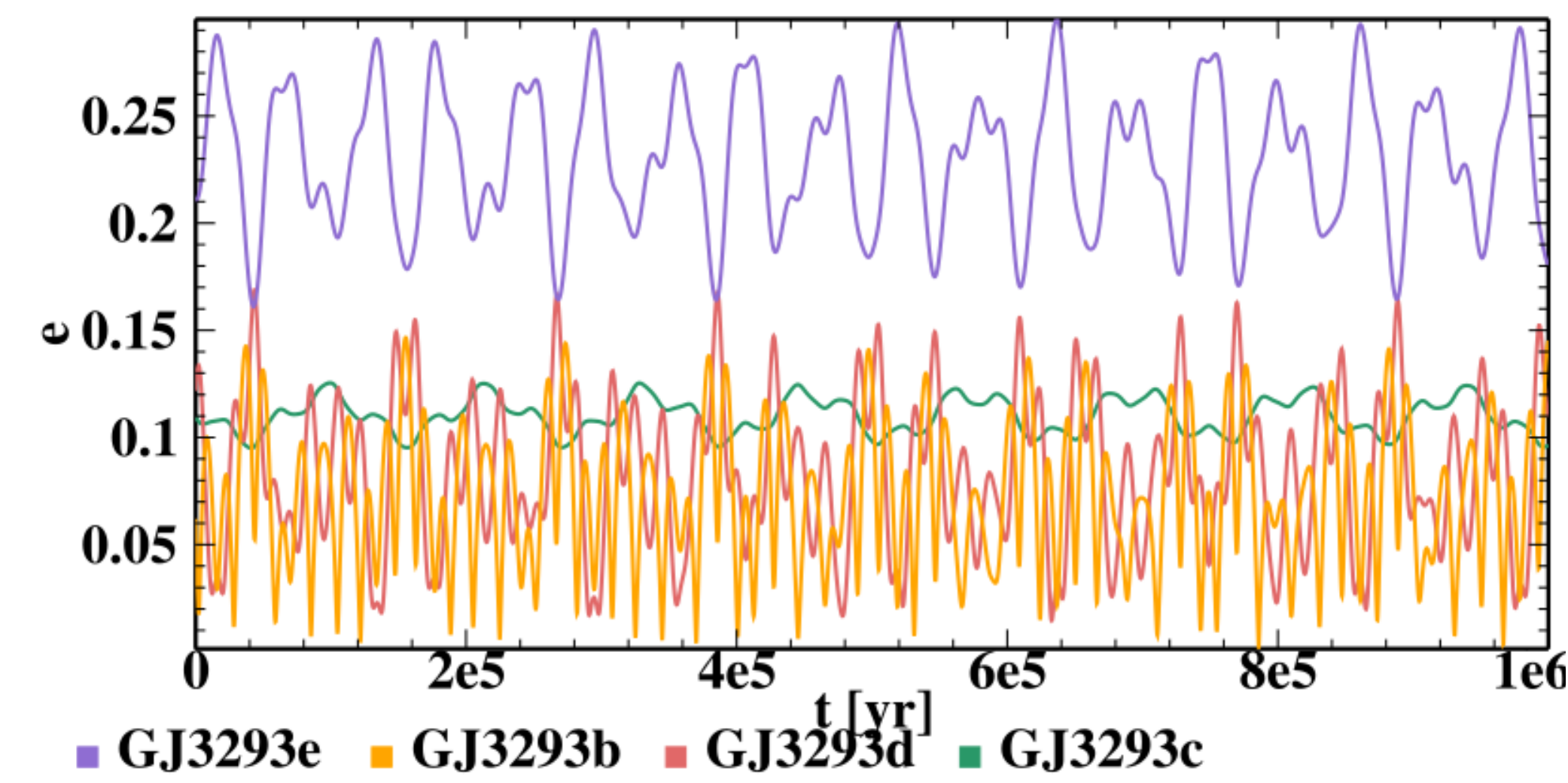}
\caption{\small Results from the \textit{GENGA} simulation showing the stability of the GJ~3293 
system over $10^6$ years.}
\label{fig:GJ3293_genga}
\end{figure}

We performed an additional test on our GJ~3293 solution by running an N-body integrator. 
We ran \textit{GENGA} \citep{2014ApJ...796...23G} through the Data and Analysis Center for 
Exoplanets (DACE\footnote{\url{https://dace.unige.ch/}}). The initial conditions in 
Table~\ref{tab:GJ3293_k4} are used and our simulation lasts for $10^6\ yr$. 
Figure~\ref{fig:GJ3293_genga} shows the stability of the system over this time span, as well as 
the secular variation of the eccentricity.

\begin{table}[]
\centering
\caption{\small Statistical factors for the models where we gradually add (according to the peak 
strength in the periodogram)  a Keplerian to the simplest constant  model. 
}
\label{tab:model_stats}
\begin{tabular*}{\hsize}{@{\extracolsep{\fill}}lccccc}

\hline
\noalign{\smallskip}

Model & Added period & FAP & BIC & $\sigma_{(O-C)}$ & $\chi_\nu^2$\\
 & [days] & [\%] & & $[ms^{-1}]$ & \\
\noalign{\smallskip}
\hline\hline
\noalign{\smallskip}
\multicolumn{6}{c}{GJ~3138}\\
\noalign{\smallskip}
\hline
\noalign{\smallskip}
Const.  & -- & -- & 459.93 & 2.87 & 2.39 \\
K1 & 5.97  & $<$10$^{-2}$ & 379.26 & 2.45 & 1.79 \\
K2 & 1.22  & 0.29 & 349.08 & 2.24 & 1.54 \\
K3 & 258  & 0.17 & 326.52 & 2.04 & 1.31 \\
K4 & 20.4 & 0.9 & 322.17 & 1.90 & 1.17 \\

\noalign{\smallskip}
\hline\hline
\noalign{\smallskip}
\multicolumn{6}{c}{GJ~3323}\\
\noalign{\smallskip}
\hline
\noalign{\smallskip}
Const.  & -- & -- & 293.04 & 2.85 & 1.66 \\
K1 & 5.4  & $<$10$^{-2}$ & 196.02 & 2.27 & 1.09 \\
K2 & 40  & $<$10$^{-2}$ & 184.62 & 2.00 & 0.87 \\

\noalign{\smallskip}
\hline\hline
\noalign{\smallskip}
\multicolumn{6}{c}{GJ~273}\\
\noalign{\smallskip}
\hline
\noalign{\smallskip}
Const.  & -- & -- & 2388.48 & 2.75 & 8.52 \\
K1 & 424 & $<$10$^{-2}$ & 1552.96 & 2.21 & 5.54 \\
K2 & 18.6 & $<$10$^{-2}$ & 1292.05 & 1.99 & 4.57 \\
K3 & 700 & $<$10$^{-2}$ & 952.43 & 1.67 & 3.26 \\
K4 & 4.7 & $<$10$^{-2}$ & 817.24 & 1.50 & 2.69 \\

\noalign{\smallskip}
\hline\hline
\noalign{\smallskip}

\multicolumn{6}{c}{GJ~628}\\
\noalign{\smallskip}
\hline
\noalign{\smallskip}
Const.  & -- & -- & 1346.63 & 2.72  & 7.00 \\
K1 & 4.9  & $<$10$^{-2}$ & 1034.37 & 2.39 & 5.54 \\
K2 &  17.9 & 0.07 & 854.03 & 2.13 & 4.52 \\
K3 &  217 & $<$10$^{-2}$ & 599.18 & 1.71 & 3.00 \\
K3 &  48  & 0.02 & 691.44 & 1.86 & 3.54 \\
K3 &  454 & 0.07 & 645.82 & 1.78 & 3.28 \\
K3 &  115 & 0.17 & 710.05 & 1.88 & 3.65 \\

\noalign{\smallskip}
\hline\hline
\noalign{\smallskip}

\multicolumn{6}{c}{GJ~3293}\\
\noalign{\smallskip}
\hline
\noalign{\smallskip}
Const.  & -- & -- & 2193.79 & 7.73 & 10.62 \\
K1 & 30.6  & $<$10$^{-2}$ & 1025.97 & 5.21 & 4.95 \\
K2 & 122.7  & $<$10$^{-2}$ & 456.62 & 3.3 & 2.03 \\
K3 & 48.1  & $<$10$^{-2}$ & 380.80 & 2.84 & 1.55 \\
K4 & 13.3  & 3.4 & 361.43 & 2.61 & 1.34 \\

\noalign{\smallskip}
\hline
\noalign{\smallskip}
\end{tabular*}
\end{table}

\section{Summary and conclusions}
\label{sec:Conclusion}

We have analyzed between 150 and 280 RVs of GJ~3138, GJ~3323, GJ~273, GJ~628, and GJ~3293. 
To fully exploit the Doppler information contained in M dwarf spectra, the RVs extracted from 
the HARPS data was done by maximizing the likelihood. For each star, a high signal-to-noise reference 
spectrum was built out of all the observed spectra. The RV analysis includes a careful monitoring of 
several spectroscopic activity indicators and photometry, which we used to measure stellar rotation 
periods for three stars, except for GJ~3323. For GJ~628 Ca \textrm{\small II} H\textrm{\small \&}K 
chromospheric emission reveals a long-term variation compatible to a Sun-like magnetic cycle.

Conditioned by a knowledge of the orbital inclination, our analysis results in the detection of 
12 planets (9 new), including Earth-mass planets and super-Earths. The GJ~3138 radial velocities 
reveal four signals: three of them are compatible with planets having minimum masses of 
1.8, 4.2, and 10.5 Earth masses, orbiting with periods of 1.22, 5.97, and 258 days, respectively, 
while a signal with a periodicity of  48 days  is more related to stellar activity as it shows 
dependence on H$\alpha$. We recall that the RV-H$\alpha$ dependence was subtracted and 
that more RVs of this star are needed because the phase for the 258-day signal is not 
fully covered. These orbits are located outside the habitable zone of GJ~3138.

We interpret the RV variations of the fully convective star GJ~3323 as the signal coming from 
two low-mass planets. These planets have minimum masses of 2.0 and 2.3 Earth masses~M$_{\oplus}$ and are 
orbiting with periodicities of 5.36 and 40.5 days, respectively. The planets are orbiting 
outside the conservative boundaries of the HZ. GJ~3323b and GJ~3323c respectively receive 2.5 and 0.17 
times the stellar flux compared to the Sun.

The M~dwarf GJ~273 is orbited by  a super-Earth and by an Earth-mass planet. They have minimum 
masses of 2.89 and 1.18~M$_\oplus$ and orbital periods of 18.65 and 4.72~days, respectively. 
GJ~273b receives 1.06 times as much radiation as our Earth does from our Sun. It thus lies 
well within the habitable zone \citep{2016ApJ...819...84K} and water (if any) may flow on 
its surface. Located at only 3.8 parsec, GJ~273 is also the closest known planetary system 
with a planet in the HZ after Proxima Centauri \citep{2016Natur.536..437A}. Compared to the flaring star Proxima 
Centauri, we highlight the quietness  of GJ~273. This makes GJ~273b even more attractive 
because of its lower atmospheric erosion, which translates into a more favorable environment 
for habitability.

We found that GJ~628 is orbited by at least three planets. We found a Sun-like magnetic cycle and 
clues that the stellar rotation is about 128 days. From this, it follows that the 67-day 
RV signal is more likely produced by stellar activity, and not by a planetary companion as 
reported by \citeauthor{2016ApJ...817L..20W} Here we found that a model including three 
Keplerians with P=4.89, 17.87, 217.2 days is favored. This solution corresponds to planets 
with $m \sin(i)$=1.9, 3.4, 7.7~M$_\oplus$. For GJ~628c we derive a mean orbital distance 
slightly larger than the \citeauthor{2016ApJ...817L..20W} value (using our uncertainty we differ by $1.67\sigma$;  their uncertainty for $a$ is disturbingly low). At this distance and compared to the Sun-Earth 
system, GJ~628c receives 1.29 times the flux received at Earth.

New radial velocities of GJ~3293 allow us to confirm the detection of the planet previously 
suggested in \citet{2015A&A...575A.119A}. Additionally, we report the detection of an extra 
RV signal. The planet we confirm has a minimum mass of 7.6~M$_\oplus$ and orbital period of 48.1 days; 
it is therefore inside the habitable zone of GJ~3293. The additional planet is a super-Earth of 
$m \sin(i)$=3.3~M$_\oplus$ and an orbit of P=13.25 days. The previously detected planets are 
at P=30.6, 122.6 days with $m \sin(i)$=23.5, 21.1~M$_\oplus$; here we report a slightly lower 
eccentricity for the farthest planet.

\begin{acknowledgements}
This publication makes use of data products from the Two Micron All Sky Survey, which is a 
joint project of the University of Massachusetts and the Infrared Processing and Analysis 
Center/California Institute of Technology, funded by the National Aeronautics and Space 
Administration and the National Science Foundation. X.B., X.D., and T.F. acknowledge 
the support of the French Agence Nationale de la Recherche (ANR), under the program 
ANR-12-BS05-0012 Exo-atmos and of PNP (Programme national de plan\'etologie). This work 
has been partially supported by the Labex OSUG@2020. X.B. acknowledges funding from the 
European Research Council under the ERC Grant Agreement n. 337591-ExTrA. 

N.C.S. acknowledges the support by Funda\c{c}\~ao para a Ci\^encia e a Tecnologia (FCT) 
(project ref. PTDC/FIS-AST/1526/2014) through national funds and by FEDER through COMPETE2020 
(ref. POCI-01-0145-FEDER-016886), and through grant UID/FIS/04434/2013 
(POCI-01-0145-FEDER-007672). N.C.S. was also supported by FCT through the Investigador FCT 
contract reference IF/00169/2012 and POPH/FSE (EC) by FEDER funding through the program 
``Programa Operacional de Factores de Competitividade - COMPETE''.

We are very thankful to Jean-Baptiste Delisle for his precious comments about the dynamics 
of planetary systems. We thank the anonymous referee for the comments that 
improved the manuscript. 

This research 
made use of the databases at the Centre de Données astronomiques de Strasbourg 
(\url{http://cds.u-strasbg.fr}), NASA’s Astrophysics Data System Service 
(\url{http://adsabs.harvard.edu/abstract_service.html}), and the \url{http://arXiv.org} 
paper repositories. This work made use of {\small SCIPY} \citep{scipy}, {\small IPYTHON} 
\citep{4160251}, and {\small MATPLOTLIB} \citep{4160265}.
\end{acknowledgements}

\bibliographystyle{aa}
\bibliography{GJ3138_Gl628_LHS1723}


\begin{appendix}
\section{Radial velocities}
\label{sec:RV_appendix}
We list here the radial velocities in the barycentric frame -- without subtracting the 
secular acceleration -- and activity indicators. The uncertainty on RVs  accounts for an instrumental 
error of $0.60\ ms^{-1}$ (added quadratically). A maximum likelihood approach is used to 
compute RVs, using a high signal-to-noise stellar template.
\end{appendix}

\begin{table*}
\caption{\small Not corrected for stellar activity (Eq.~\ref{eq:Halpha_cor}), radial velocity time series for GJ~3138 (minimal; full version available at the CDS).}
\label{tab:gj3138_rv}
\begin{tabular*}{\hsize}{@{\extracolsep{\fill}}lllllllll}

\noalign{\smallskip}
\hline\hline
\noalign{\smallskip}
BJD - 2400000 & RV &  $\sigma_{RV}$ & FWHM & Contrast & BIS & S-index & $H\alpha$\\
 & $[kms^{-1}]$ & $[kms^{-1}]$ & [$kms^{-1}$] & & [$kms^{-1}$] &&  \\

\noalign{\smallskip}
\hline
\noalign{\smallskip}

54396.706036    &       13.59614        &       0.00134 &       3.35441 &       13.46537        &       -16.61000       &       0.94071 &       0.05333 \\
54397.781533    &       13.59639        &       0.00135 &       3.35775 &       13.56172        &       -9.54700        &       0.94202 &       0.05302 \\
54398.695400    &       13.59508        &       0.00176 &       3.35078 &       13.53711        &       -17.41900       &       0.96591 &       0.05326 \\
54399.583478    &       13.59125        &       0.00160 &       3.35057 &       13.43478        &       -6.81600        &       0.84540 &       0.05318 \\
54400.615060    &       13.59404        &       0.00178 &       3.34159 &       13.45599        &       -12.88400       &       0.90862 &       0.05344 \\

\noalign{\smallskip}
\hline

\end{tabular*}
\end{table*}

\begin{table*}
\caption{\small Radial velocity time series for GJ~3323 (minimal; full version available at the CDS).}
\label{tab:gj3323_rv}
\begin{tabular*}{\hsize}{@{\extracolsep{\fill}}lllllllll}
\noalign{\smallskip}
\hline\hline
\noalign{\smallskip}

BJD - 2400000 & RV &  $\sigma_{RV}$ & FWHM & Contrast & BIS & S-index & $H\alpha$\\
 & $[kms^{-1}]$ & $[kms^{-1}]$ & [$kms^{-1}$] & & [$kms^{-1}$] &&  \\

\noalign{\smallskip}
\hline
\noalign{\smallskip}

52986.721748    &       42.44749        &       0.00212 &       3.03195 &       27.57384        &       -12.03000       &       5.92003 &       0.13107 \\
52997.681842    &       42.44901        &       0.00333 &       3.03125 &       26.59910        &       -21.51800       &       3.46817 &       0.11630 \\
53343.728670    &       42.45623        &       0.00234 &       3.01471 &       27.21297        &       -10.98700       &       4.29847 &       0.15122 \\
54174.512651    &       42.45516        &       0.00167 &       3.00959 &       27.37128        &       -10.46400       &       3.26599 &       0.09855 \\
54732.819933    &       42.45234        &       0.00303 &       3.02512 &       26.90758        &       -12.95900       &       4.35294 &       0.17680 \\

\noalign{\smallskip}
\hline

\end{tabular*}
\end{table*}

\begin{table*}
\caption{\small Radial velocity time series for GJ~273 (minimal; full version available at the CDS).}
\label{tab:gj273_rv}
\begin{tabular*}{\hsize}{@{\extracolsep{\fill}}lllllllll}

\noalign{\smallskip}
\hline\hline
\noalign{\smallskip}

BJD - 2400000 & RV &  $\sigma_{RV}$ & FWHM & Contrast & BIS & S-index & $H\alpha$\\
 & $[kms^{-1}]$ & $[kms^{-1}]$ & [$kms^{-1}$] & & [$kms^{-1}$] &&  \\

\noalign{\smallskip}
\hline
\noalign{\smallskip}

52986.769625    &       18.40354        &       0.00145 &       2.97871 &       27.91234        &       -5.50100        &       0.71683 &       0.07591 \\
52998.774768    &       18.40073        &       0.00091 &       2.96841 &       27.67013        &       -8.65700        &       0.69299 &       0.07457 \\
53343.769216    &       18.40549        &       0.00158 &       2.96227 &       27.41423        &       -3.31700        &       0.71733 &       0.07665 \\
53370.806664    &       18.40891        &       0.00081 &       2.97153 &       27.80775        &       -7.43500        &       0.69075 &       0.07491 \\
53371.752065    &       18.40807        &       0.00086 &       2.97149 &       27.63837        &       -6.45600        &       0.74705 &       0.07585 \\

\noalign{\smallskip}
\hline

\end{tabular*}
\end{table*}

\begin{table*}
\caption{\small Radial velocity time series for GJ~628 (minimal; full version available at the CDS).}
\label{tab:gj628_rv}
\begin{tabular*}{\hsize}{@{\extracolsep{\fill}}lllllllll}

\noalign{\smallskip}
\hline\hline
\noalign{\smallskip}

BJD - 2400000 & RV &  $\sigma_{RV}$ & FWHM & Contrast & BIS & S-index & $H\alpha$\\
 & $[kms^{-1}]$ & $[kms^{-1}]$ & [$kms^{-1}$] & & [$kms^{-1}$] &&  \\

\noalign{\smallskip}
\hline
\noalign{\smallskip}

53158.680215    &       -21.03311       &       0.00115 &       3.77357 &       28.86726        &       -15.36500       &       0.88246 &       0.07568 \\
53203.583611    &       -21.03738       &       0.00094 &       3.81234 &       30.43251        &       -9.57700        &       0.78018 &       0.07288 \\
53484.842215    &       -21.03704       &       0.00086 &       3.80200 &       30.29858        &       -9.77500        &       0.75919 &       0.07416 \\
54172.850539    &       -21.03481       &       0.00084 &       3.80888 &       30.15304        &       -11.23400       &       0.90972 &       0.07435 \\
54293.651400    &       -21.03805       &       0.00087 &       3.81178 &       30.19980        &       -14.18100       &       0.80749 &       0.07301 \\

\noalign{\smallskip}
\hline

\end{tabular*}
\end{table*}

\begin{table*}
\caption{\small Radial velocity time series for GJ~3293 (minimal; full version available at the CDS).}
\label{tab:gj3293_rv}
\begin{tabular*}{\hsize}{@{\extracolsep{\fill}}lllllllll}

\noalign{\smallskip}
\hline\hline
\noalign{\smallskip}

BJD - 2400000 & RV &  $\sigma_{RV}$ & FWHM & Contrast & BIS & S-index & $H\alpha$\\
 & $[kms^{-1}]$ & $[kms^{-1}]$ & [$kms^{-1}$] & & [$kms^{-1}$] &&  \\

\noalign{\smallskip}
\hline
\noalign{\smallskip}

54805.684241    &       13.28713        &       0.00505 &       3.60254 &       25.79888        &       -17.14600       &       1.43841 &       0.06417 \\
54825.640268    &       13.30974        &       0.00168 &       3.60782 &       25.14589        &       -5.88700        &       1.32105 &       0.06840 \\
54826.634453    &       13.31091        &       0.00189 &       3.60512 &       25.07722        &       -3.29900        &       1.18284 &       0.06751 \\
54827.648451    &       13.30608        &       0.00191 &       3.60538 &       25.03949        &       -10.48400       &       1.11380 &       0.06634 \\
54828.660596    &       13.30306        &       0.00231 &       3.60655 &       24.93891        &       -3.71900        &       1.35938 &       0.06757 \\

\noalign{\smallskip}
\hline

\end{tabular*}
\end{table*}

\end{document}